\documentclass[twocolumn]{aastex62}
\pdfoutput=1 
\usepackage{epsfig}
\usepackage{amsmath,amstext}
\usepackage[T1]{fontenc}
\usepackage{footmisc}
\usepackage{natbib}
\usepackage[figure,figure*]{hypcap}

\newcommand{\Sp}{{\it Spitzer\/}}
\newcommand{\WISE}{{\it WISE\/}}

\shorttitle{IR Lightcurves of NEOs}
\shortauthors{Hora et al.}
\accepted{22 August 2018 to the Astrophysical Journal Supplement Series}

\begin{document}
\title{Infrared Lightcurves of Near Earth Objects}

\author{Joseph L. Hora}
\affiliation{Harvard-Smithsonian Center for Astrophysics, 60 Garden St., MS-65, Cambridge, MA 02138, USA} 
\author{Amir Siraj}
\affiliation{Harvard University, Cambridge, MA 02138, USA}
\author{Michael Mommert}
\affiliation{ Lowell Observatory, 1400 W Mars Hill Rd, Flagstaff, AZ 86001, USA}
\affiliation{Department of Physics and Astronomy, PO Box 6010, Northern Arizona University, Flagstaff, AZ 86011, USA}
\author{Andrew McNeill}
\affiliation{Department of Physics and Astronomy, PO Box 6010, Northern Arizona University, Flagstaff, AZ 86011, USA}
\author{David E. Trilling}
\affiliation{Department of Physics and Astronomy, PO Box 6010, Northern Arizona University, Flagstaff, AZ 86011, USA}
\author{Annika Gustafsson}
\affiliation{Department of Physics and Astronomy, PO Box 6010, Northern Arizona University, Flagstaff, AZ 86011, USA}
\author{Howard A. Smith}
\affiliation{Harvard-Smithsonian Center for Astrophysics, 60 Garden St., MS-65, Cambridge, MA 02138, USA}
\author{Giovanni G. Fazio}
\affiliation{Harvard-Smithsonian Center for Astrophysics, 60 Garden St., MS-65, Cambridge, MA 02138, USA}
\author{Steven Chesley}
\affiliation{Jet Propulsion Laboratory, California Institute of Technology, 4800 Oak Grove Drive, Pasadena, CA 91109, USA}
\author{Joshua P. Emery}
\affiliation{Department of Earth \& Planetary Science, University of Tennessee, 306 EPS Building, 1412 Circle Drive, Knoxville, TN 37996, USA}
\author{Alan Harris}
\affiliation{German Aerospace Center (DLR), Institute of Planetary Research, Rutherfordstrasse 2, 12489, Berlin, Germany}
\author{Michael Mueller}
\affiliation{Kapteyn Astronomical Institute, Rijksuniversiteit Groningen, PO Box 800, 9700AV Groningen, The Netherlands}
\affiliation{SRON, Netherlands Institute for Space Research, PO Box 800, 9700AV Groningen, The Netherlands}
\correspondingauthor{Joseph L. Hora}
\email{jhora@cfa.harvard.edu}

\begin{abstract}
We present lightcurves and derive periods and amplitudes for a subset of 38 near earth objects (NEOs) observed at 4.5 \micron\ with the IRAC camera on the the \Sp\ Space Telescope, many of them having no previously reported rotation periods. This subset was chosen from about 1800 IRAC NEO observations as having obvious periodicity and significant amplitude.  For objects where the period observed did not sample the full rotational period, we derived lower limits to these parameters based on sinusoidal fits. Lightcurve durations ranged from 42 to 544 minutes, with derived periods from 16 to 400 minutes.

We discuss the effects of lightcurve variations on the thermal modeling used to derive diameters and albedos from \Sp\ photometry. We find that both diameters and albedos derived from the lightcurve maxima and minima
agree with our previously published results, even for extreme objects, 
showing the conservative nature of the thermal model uncertainties. We also evaluate the NEO rotation rates, sizes, and their cohesive strengths. 
\end{abstract}

\tighten

\section{Introduction}

Near Earth Objects (NEOs) are small Solar System bodies whose orbits bring them close to the Earth's orbit. NEOs are compositional and dynamical tracers from elsewhere in the Solar System. The study of NEOs allows us to probe environmental conditions throughout the Solar System and the history of our planetary system, and provides a template for analyzing the evolution of planetary disks around other stars. NEOs are the parent bodies of meteorites, one of our key sources of detailed knowledge about the development of the Solar System, and so studies of NEOs are essential for understanding the origins and evolution of our Solar System and others. 

As of 2018 June there are over 18,000 known NEOs. Roughly 2000 new NEOs are being discovered each year, primarly by the Catalina Sky Survey \citep{leonard17} and Pan-STARRS \citep{veres15}, and the rate will significantly increase when LSST begins operations \citep{veres17}. However, little is known about most NEOs after their discovery, beyond their orbits and optical magnitudes.  The size of objects that pass close to Earth can be measured with radar, for example using the Arecibo or Goldstone facilities. Over 750 NEOs have been observed\footnote{https://echo.jpl.nasa.gov/asteroids/index.html}, at a rate of $\sim$75 -- 100 objects per year during the past three years. This rate cannot be easily scaled up, however, and is not keeping pace with the rate of new NEO discoveries.  Optical or near-IR spectra of NEOs can determine the surface properties and allow their taxonomic classification \citep{bus99,bus02a,bus02b,demeo09}. However, currently less than 2\% of the NEOs in the JPL Small-Body 
Database\footnote{https://ssd.jpl.nasa.gov/sbdb\_query.cgi} have assigned taxonomic types.  Small NEOs are especially difficult to characterize: for example, \citet{perna18} recently conducted a 30-night GTO program at the NTT and obtained spectra of 147 NEOs, focusing on smaller (<300m) objects. With 24 usable nights, they were able to observe $\sim$ 6 objects per night on this moderately-sized telescope. It would  take a major effort on large telescopes to increase the fraction of spectrally-classified objects.

The IRAC instrument\citep{2004ApJS..154...10F} on the \Sp\ Space Telescope \citep{2004ApJS..154....1W} is a powerful NEO characterization system. NEOs typically have daytime temperatures $\sim$250 K, hence their thermal emission at 4.5 \micron\ is almost always significantly larger than their reflected light at that wavelength. We can therefore use a thermal model using the optical and IR fluxes to derive NEO properties, including diameters and albedos (see \citealt{2010AJ....140..770T,2016AJ....152..172T}). Measuring the size distribution, albedos, and compositions for a large fraction of all known NEOs will allow us to understand the scientific, exploration, and civil-defense-related properties of the NEO population. 

After an initial pilot study to verify our observing  techniques and analysis methods with the \Sp\ data \citep{2008ApJ...683L.199T}, our team has conducted three major surveys of NEOs with \Sp/IRAC in the Warm/Beyond Mission phases: the ExploreNEOs program \citep{2010AJ....140..770T}, the NEO Survey \citep{2016AJ....152..172T}, and the NEO Legacy Survey \citep{Trilling2017}. As of 2018 March, \Sp\ has completed a total of over 1800 NEO observations, with an expected total of over 2100  observations by the time that the NEO Legacy program has completed in early 2019. Our initial NEO survey results are summarized in \citet{2010AJ....140..770T,2016AJ....152..172T} and \citet{2011AJ....141...75H}. Since then we have examined the albedo distribution and related them to taxonomic classifications \citep{2011Icar..212..158T}, performed a physical characterization of NEOs in our sample \citep{2014Icar..228..217T}, and examined the physical properties of subsets of the sample, including low-$\Delta\nu$ NEOs \citep{2011AJ....141..109M} and dormant short-period comets\citep{2015AJ....150..106M}. We examined individual objects more closely, such as in our discovery of cometary activity associated with the NEO Don Quixote \citep{2014ApJ...781...25M}. We have also performed additional observations on specific NEOs of interest, including the small (<10 m) NEOs 2009~BD \citep{2014ApJ...786..148M} and 2011~MD \citep{2014ApJ...789L..22M}, and the Hayabusa-2 mission target 162173~Ryugu \citep{2017A&A...599A.103M}. 
One part of our  \Sp\ observations of 162173~Ryugu consisted of repeated integrations during its full period to obtain an IR lightcurve to help to constrain the object's shape and size. This led us to conclude that we could perhaps extract similar lightcurves for objects in the survey programs, which were designed only to obtain a single flux measurement from the mosaic image averaging over all of the exposures in the observation. We found that our predicted NEO fluxes were fairly conservative in many cases, and that we could detect most of the NEOs in the individual IRAC exposures.

The Wide-field Infrared Survey Explorer \citep[\WISE;][]{wright10} has similarly used infrared observations to characterize a large sample of main-belt asteroids and NEOs. This Explorer-class mission obtained images in four broad infrared bands at 3.4, 4.6, 12 and 22 \micron. \WISE\ conducted its 4-band survey of the sky starting in 2010 January, and after the cryogen was depleted later that year, it continued to operate with its 3.4 and 4.6 \micron\ bands until  2011 February. The spacecraft was reactivated in 2013 December as {\it NEOWISE} \citep{mainzer14} and has since been conducting a sky survey in the 3.4 and 4.6 \micron\ bands to focus on NEO discovery and characterization, using a thermal modeling technique similar to what we have employed with \Sp\ as described above. Over its lifetime, {\it NEOWISE} has observed over 860 NEOs\footnote{https://neowise.ipac.caltech.edu/} and published their estimated diameters and albedos \citep[e.g.,][]{masiero17}. The \WISE\ data can also be used to derive lightcurves of asteroids \citep[e.g.,][]{sonnett15}. However, the cadence is quite different; the \WISE\ survey typically provides repeated observations separated by 3 hr over a 1.5 day period, making it useful for sampling periodicities on the order of 1 -- 2 days. The \Sp\ data  samples cadences from a few minutes to hours, making it ideal for small and fast-rotating NEOs, and complementary to the data that \WISE\ provides. Also, since \Sp\ has a larger primary mirror, and it can track the observatory to follow the apparent motion of the NEO, we can integrate for longer periods on each NEO and therefore are more sensitive, detecting objects at the level of a few $\mu$Jy.

In this paper we present the results from an analysis of a sample of the available \Sp\ lightcurve data. Section \ref{ObsRed} describes the observations and the reduction techniques. Section \ref{periods} describes the analysis techniques used to derive periods and amplitudes of the lightcurves and presents those results. Section \ref{NEATM} discusses the effects of rotation-induced brightness variability on the thermal modeling results.

\section{Observations and Data Reduction}\label{ObsRed}
\subsection{The \Sp\ NEO Survey Programs}
Observations were obtained with \Sp/IRAC in the ExploreNEOs program  (\Sp\ Program IDs 60012, 61010, 61011, 61012, 61013), the NEO Survey (Program ID 11002), and the NEO Legacy Survey (Program ID 13006). The observations were conducted in a similar manner for these three large survey programs, taking frames while tracking the NEO motion and dithering during the observations to eliminate instrument systematics such as bad pixels or array location-dependent scattered light effects. In ExploreNEOs, we used the ``Moving Cluster'' target mode with custom offsets to perform the dithers, alternating between the 3.6 and 4.5 \micron\ fields of view. For the other programs, we used the ``Moving Single'' target mode and used a large cycling dither pattern with the source in the 4.5 \micron\ field of view only.  

In order to provide the required scheduling flexibility of the observations, we specified an observing window during which a fixed set of integrations would provide adequate signal-to-noise for the object in the total integration time. This was typically chosen to be near the time when the NEO would have its peak flux as seen by \Sp, in order to minimize the time necessary to detect the source.  The frame time was set to keep the NEO below saturation levels on the IRAC detectors based on the maximum expected NEO flux, and ranged from 12 to 100 seconds. When the uncertainty in the NEO flux was such that we could possibly be close to saturation in the long frames, we used the High Dynamic Range option, which adds little additional overhead but protects against an unexpectedly bright NEO saturating the detectors. We also required a minimum apparent motion of the source relative to the background during the observation, to make it possible to separate the NEO from background objects and isolate the NEO flux. For ones with slow apparent motions from \Sp, we increased the number of frames, or added a second epoch of observations to ensure adequate motion to enable successful background-subtraction and photometry of the object.

The total exposure time was chosen such that the source would be detected at a 10$\sigma$ level in the final mosaic after combining all observations. To assess and schedule each potential target, we predicted the reflected+emitted flux density at 4.5 \micron\ as a function of time. Our flux predictions are based on the Solar System absolute optical magnitude $H$, as reported by $Horizons$\footnote{\citet{giorgini96}; https://ssd.jpl.nasa.gov/?horizons}. $H$ magnitudes for NEOs are of notoriously low quality and tend to be skewed bright  \citep{Juric2002, Romanishin2005, veres15}. We assume an $H$ offset ($\Delta H$) of [+0.6, +0.3, 0.0] mag for [faint, nominal, bright] fluxes, respectively, so that the observations will achieve or exceed the required signal-to-noise ratio. 	 We predicted thermal fluxes using the Near Earth Asteroid Thermal Model\citep[NEATM,][see Section \ref{NEATM}]{1998Icar..131..291H}. We assume albedos ($p_V$) of [0.4, 0.2, 0.05] for [low, nominal, high] thermal fluxes. The nominal $\eta$ value (the infrared beaming parameter) was determined from the solar phase angle using the linear relation given by \citet{Wolters2008}, which is generally in 	 
agreement with the newer results of \citet{Mainzer2011a} and \citet{2016AJ....152..172T}; 0.3 was 	 
[added, subtracted]for [low, high] fluxes to capture the scatter in the empirical relationship 	 
derived in \citet{Wolters2008}. The resulting NEATM fluxes were convolved with the  	 
IRAC passbands \citep{hora08} to yield ``color-corrected'' in-band fluxes. Optical fluxes were calculated from $H + \Delta H$ together with the observing geometry and the solar flux at IRAC wavelengths. 	 
Asteroids were assumed to be 1.6 times more reflective at IRAC wavelengths than in the $V$ 	 
band \citep{2008ApJ...683L.199T, Harris2011,Mainzer2011a}; color-corrections for the 	 
5800 K reflected component are negligible. After removing all dates where an NEO's bright 
predicted flux could saturate the detector, we identified a five day window centered on the peak brightness during the observing cycle and used it as the timing constraint for the AOR. Our experience with these programs have shown that a window of this size allows good scheduling flexibility while enabling us to use the shortest possible integration times. 

There are some slight differences between the observing modes in the survey programs. In the ExploreNEOs program, we obtained near-simultaneous data for 575 sources in the 3.6 and 4.5 \micron\  channels by alternating between the two bands during the observation period. However, we found that the 3.6 \micron\ data was not a significant constraint in the NEATM fitting process, since  the flux in that
band is an  unknown mix of reflected light and thermal emission. in addition, most NEOs
are significantly fainter at 3.6 \micron\ than at 4.5 \micron, and therefore the sensitivity in that band was driving the total integration time requirements. We therefore observed only in the 4.5 \micron\ band in the NEOSurvey and NEOLegacy programs, reducing the required integration time for each NEO and allowing us to observe many more sources in the time awarded. Another change that was done in the NEOLegacy program was to set our minimum total observation time to $\sim$30 minutes, in order to ensure we have sufficient frames for background subtraction and elimination of systematic effects. The maximum time for objects in the survey was chosen to be $\sim$3 hours, to keep our total time request within the range allowed by the \Sp\ program and maximize the number of objects we could characterize.  Our group maintains a web page\footnote{http://nearearthobjects.nau.edu/spitzerneos.html} where we provide the IRAC photometry and the results of the NEATM fitting for each object shortly after it is observed.

\begin{figure*}\label{LSplotsFull}
\centering
	\includegraphics[height=0.223\textwidth]{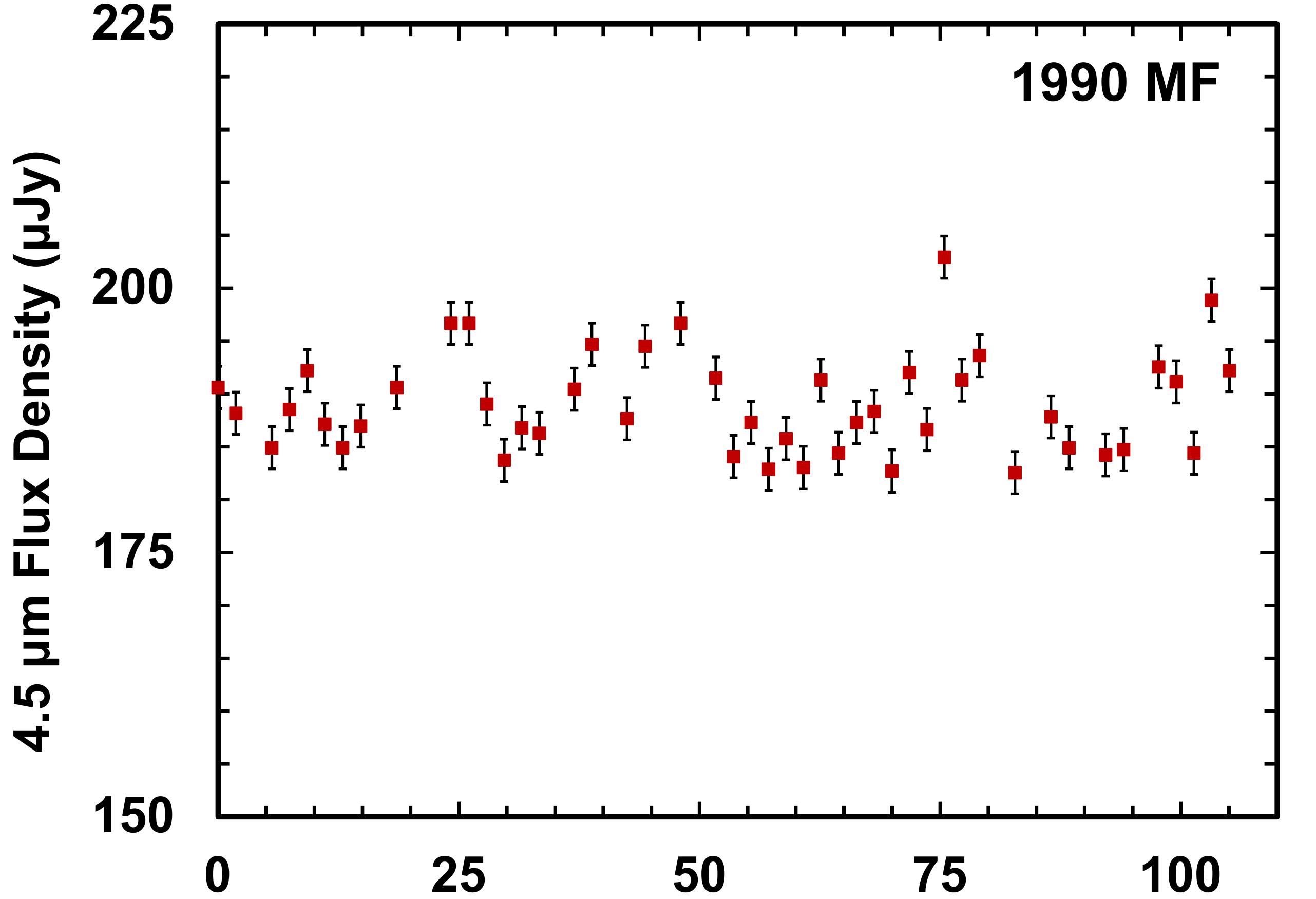}
	\includegraphics[height=0.223\textwidth]{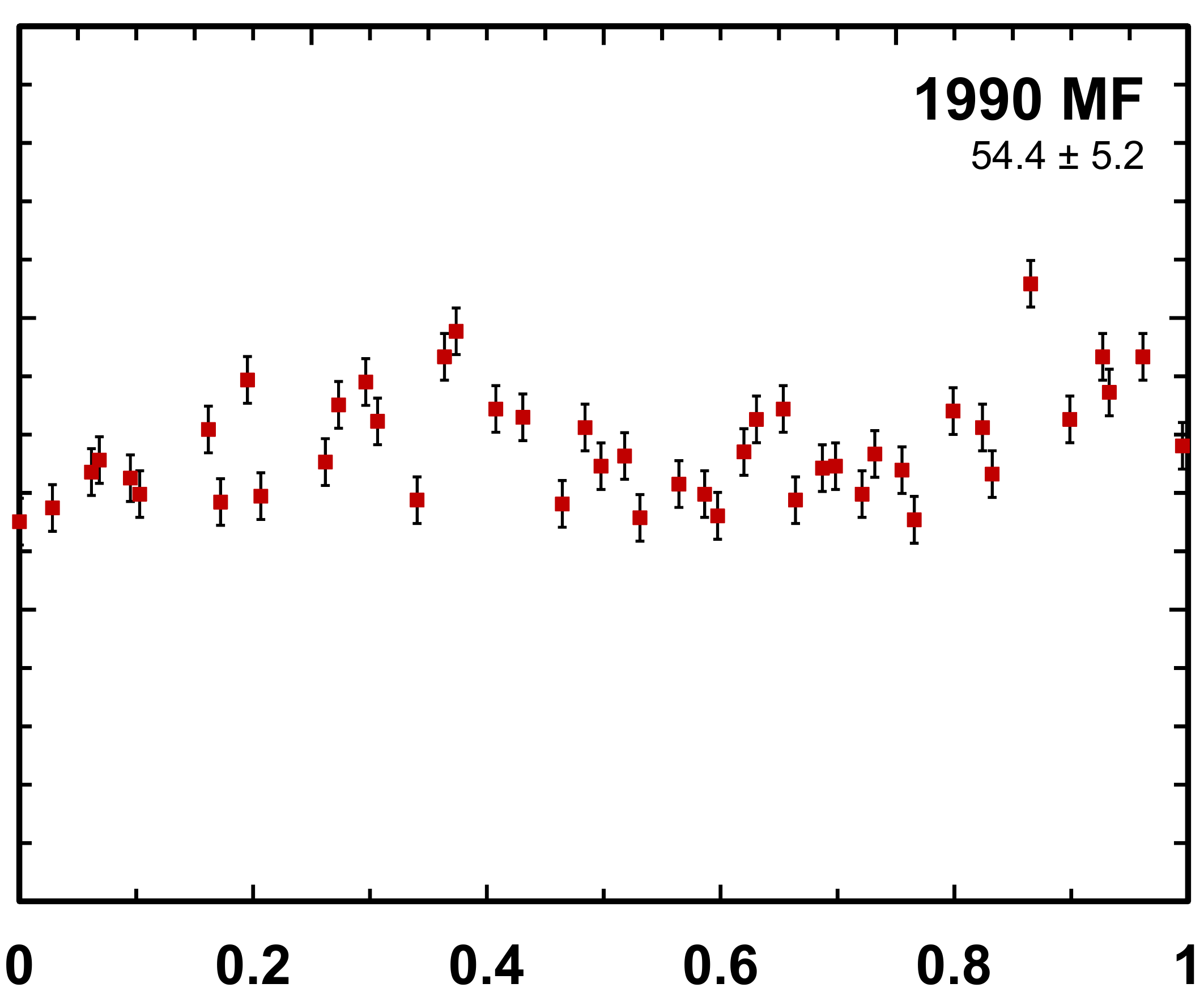}\\ \vskip 5pt
	\includegraphics[height=0.223\textwidth]{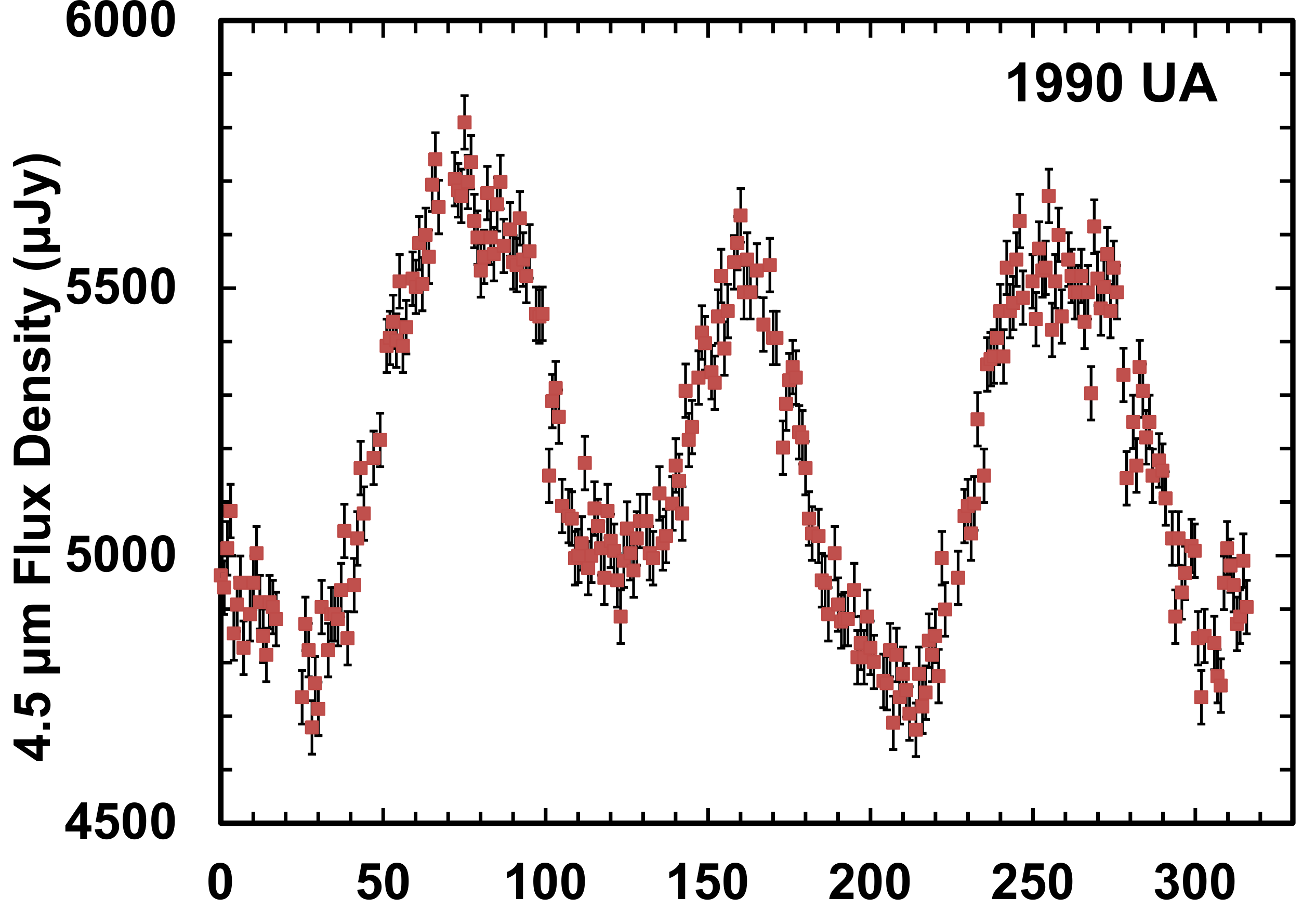}
	\includegraphics[height=0.223\textwidth]{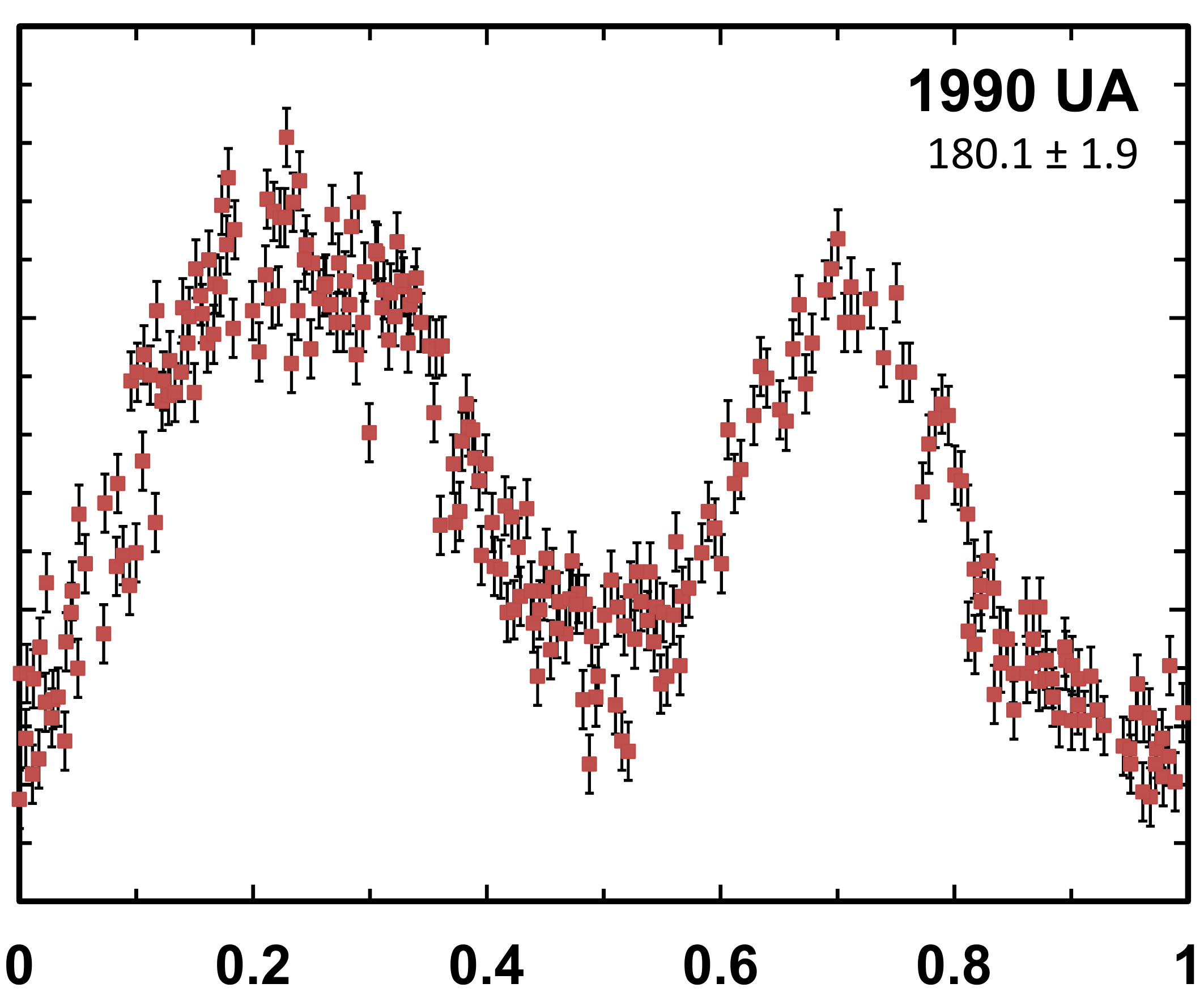}\\ \vskip 5pt
      \includegraphics[height=0.223\textwidth]{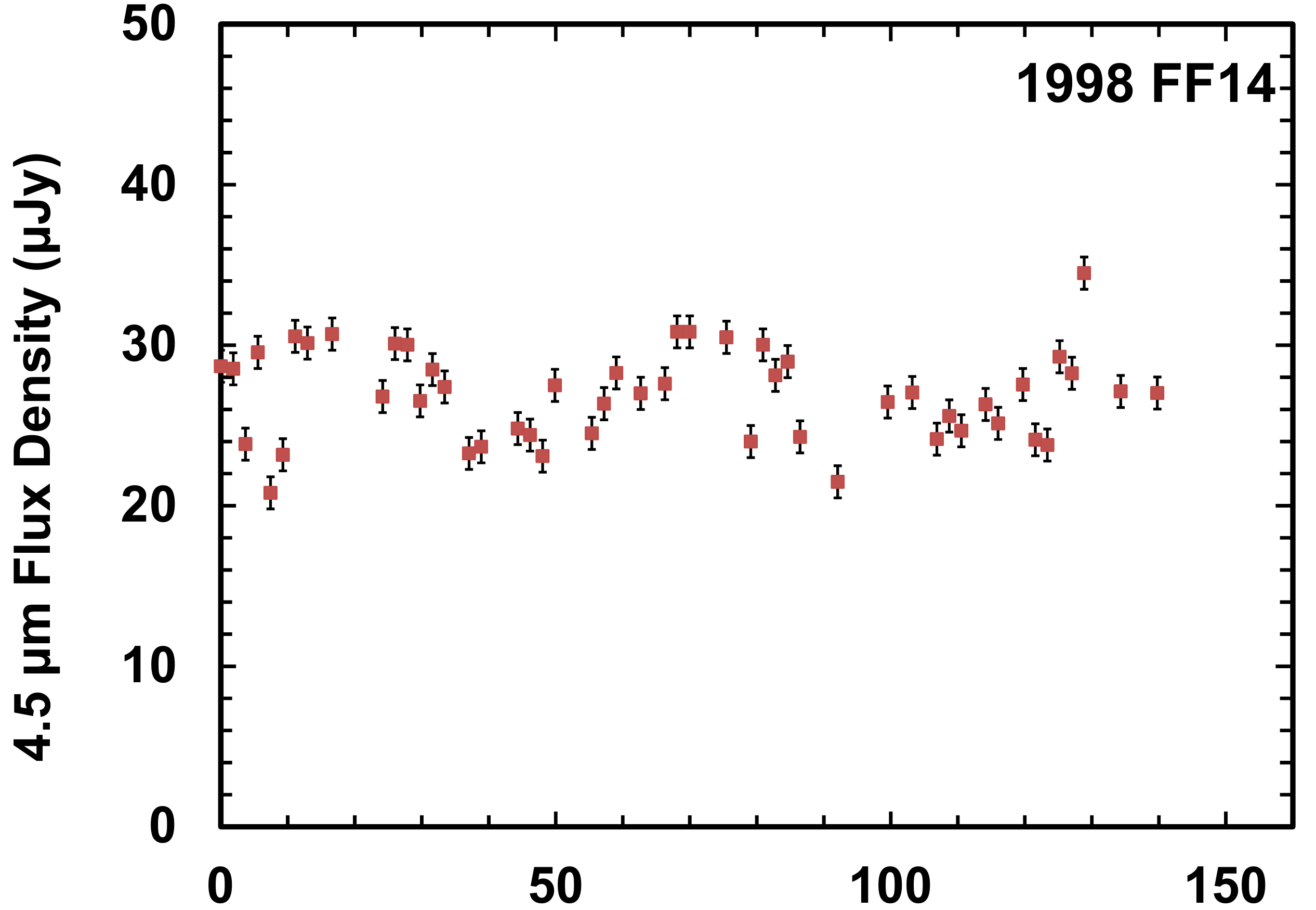}
      \includegraphics[height=0.223\textwidth]{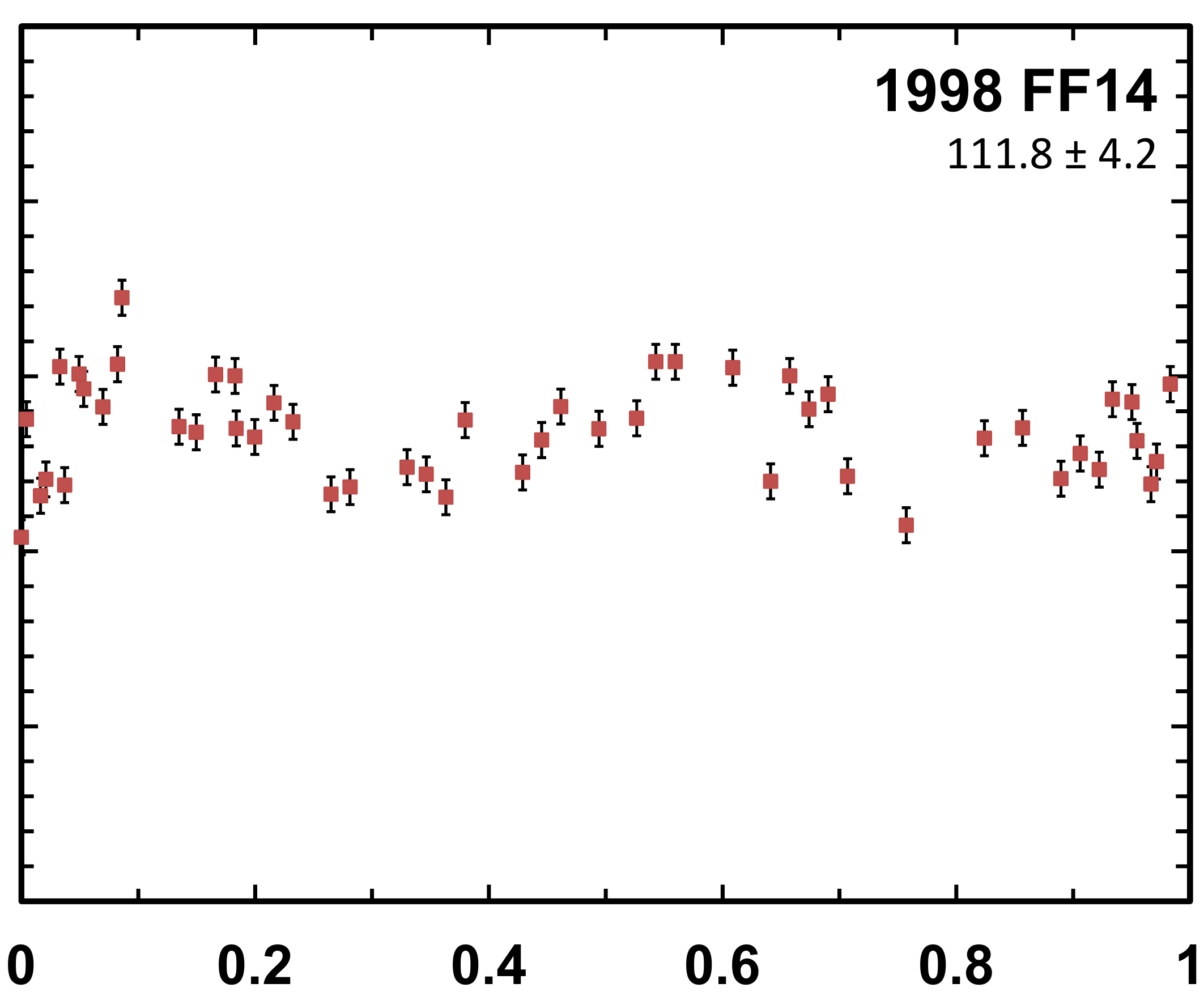}    \\ \vskip 5pt
	\includegraphics[height=0.223\textwidth]{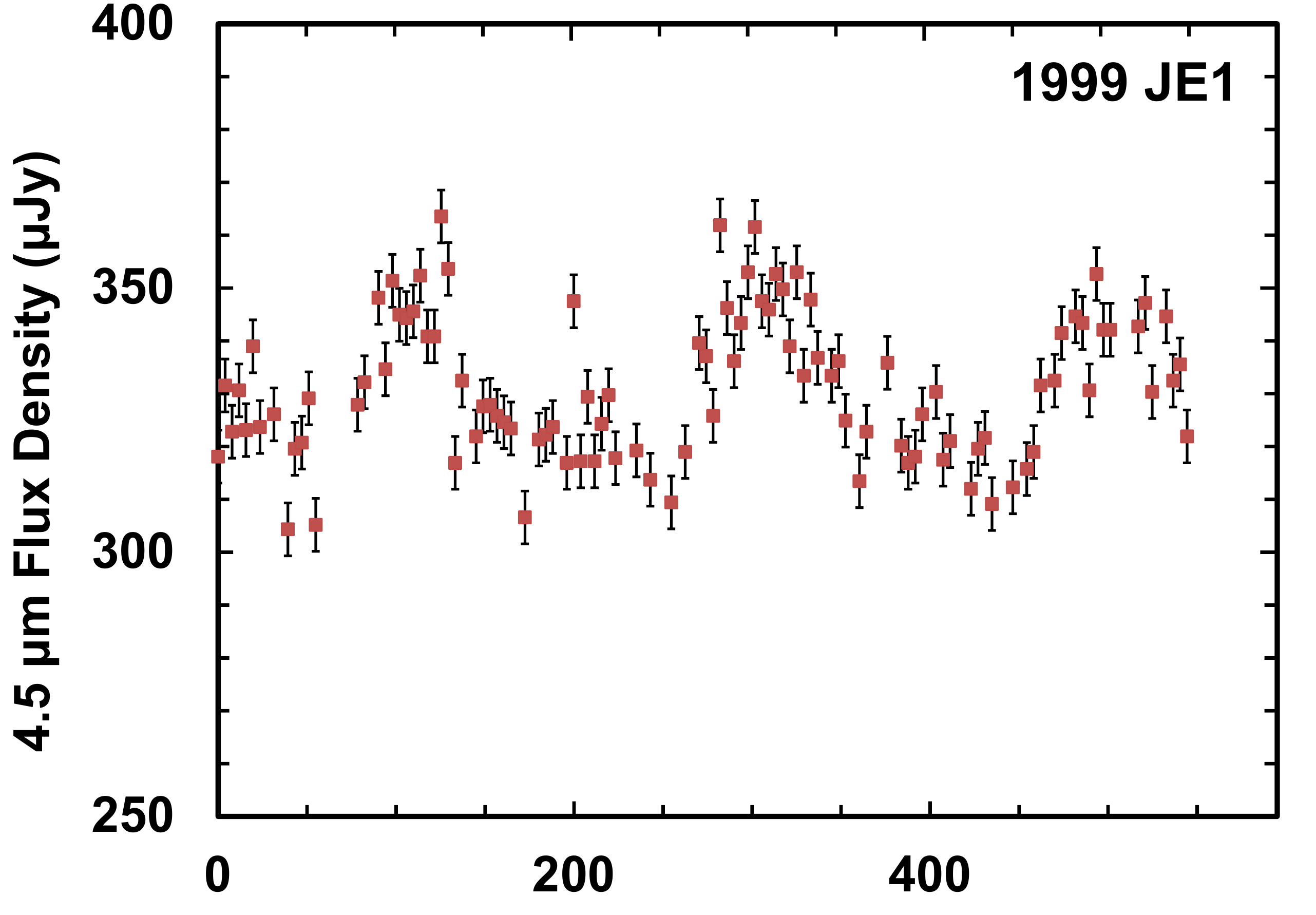}
	\includegraphics[height=0.223\textwidth]{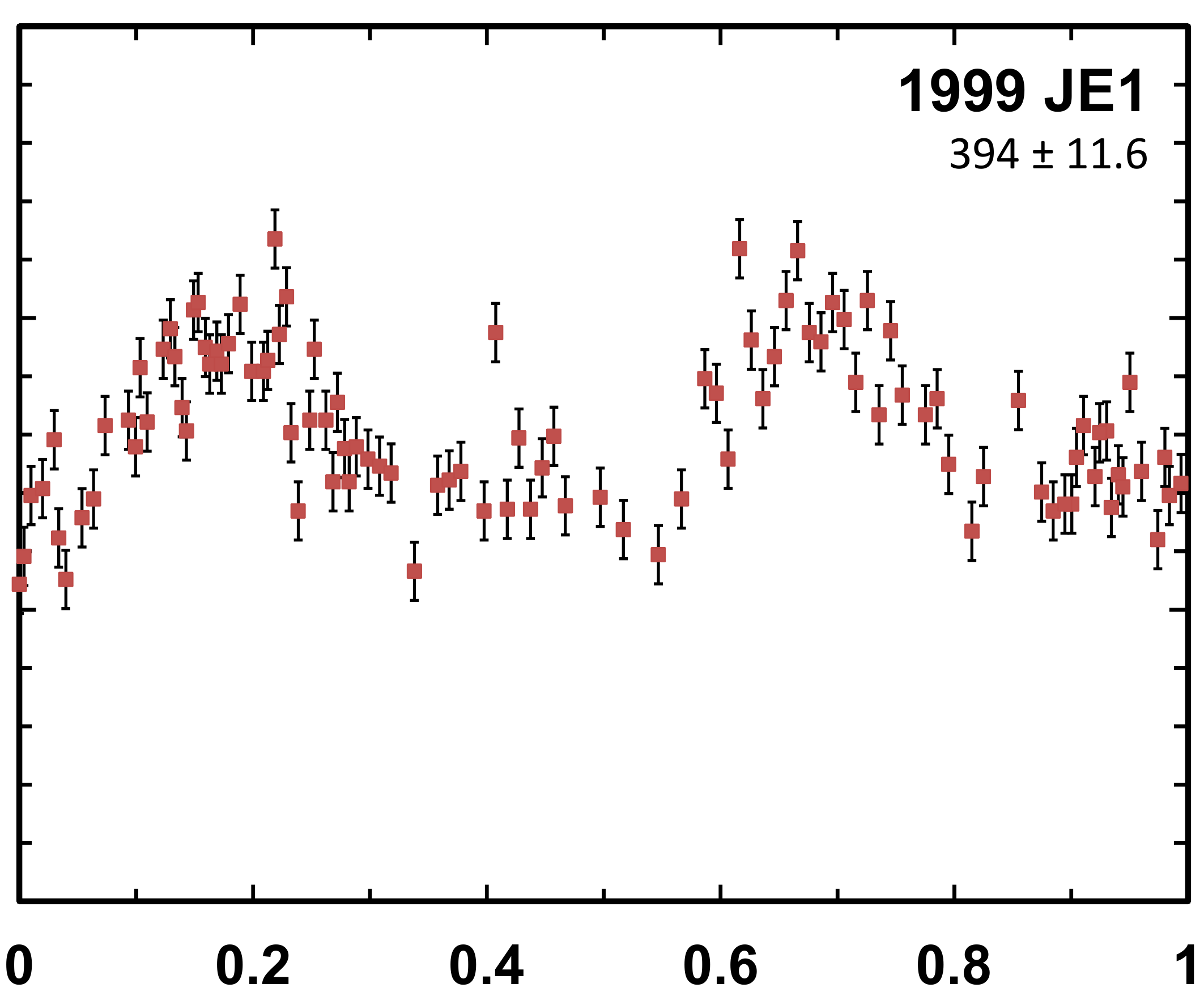}\\ \vskip 5pt
	\includegraphics[height=0.243\textwidth]{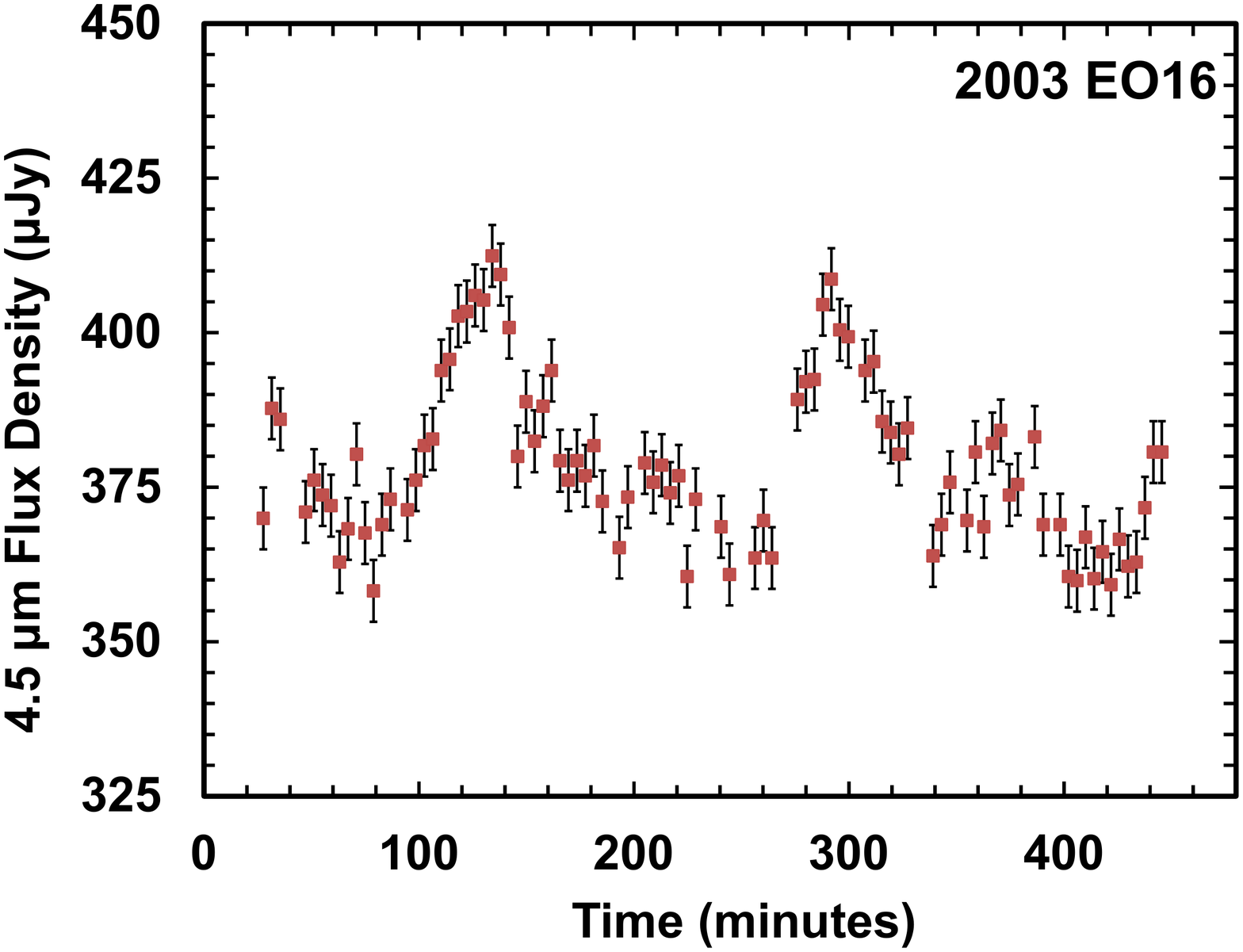}
	\includegraphics[height=0.240\textwidth]{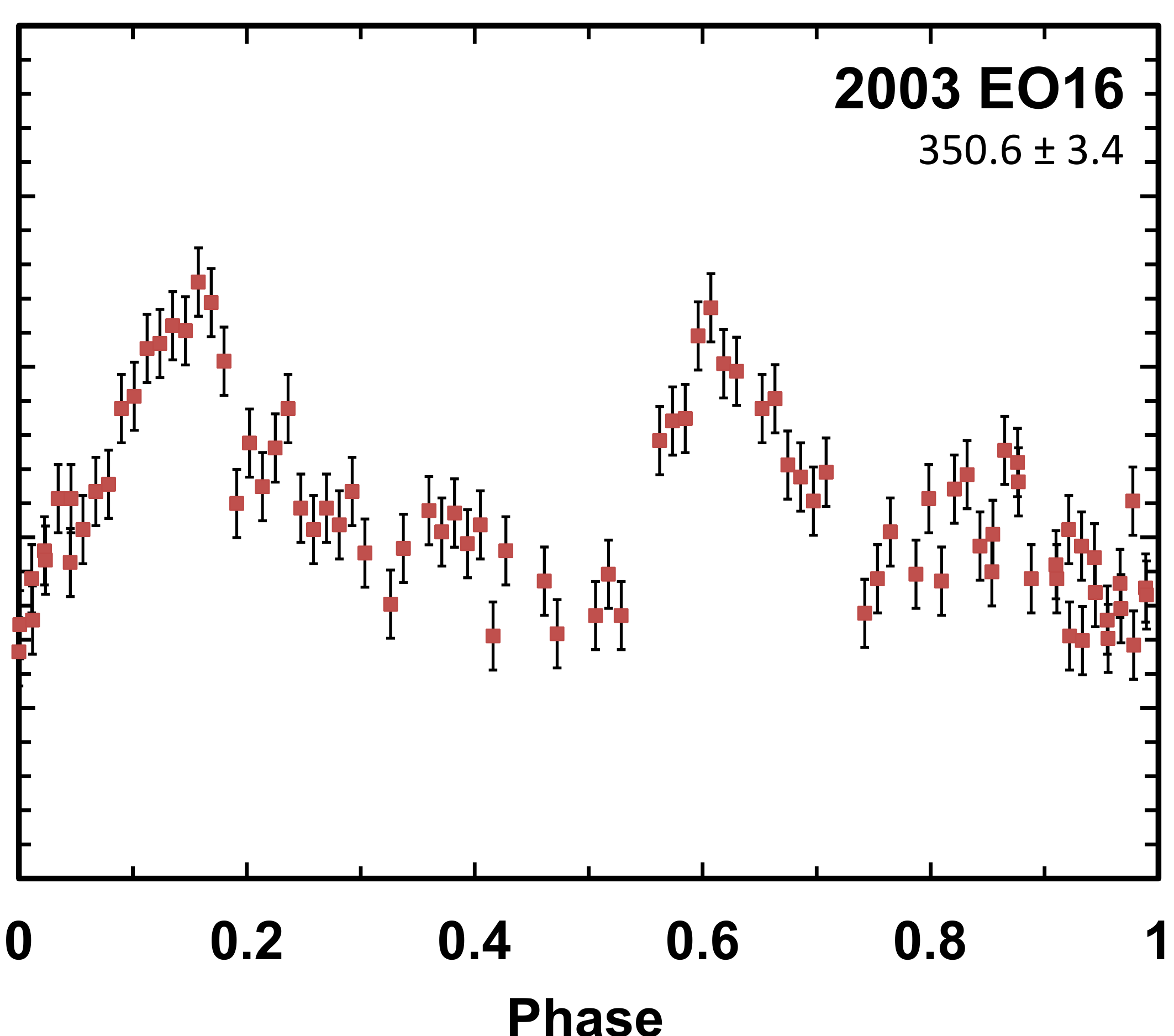}

\caption{Lightcurves and phase plots from the reduced and cleaned \Sp\ lightcurves for the objects where the lightcurve duration is longer than the derived rotational period. On the left are shown the lightcurves for the duration of the observation. The horizontal axis gives the time in minutes relative to the first point in the lightcurve. The plots on the right show the folded lightcurves, assuming the periods listed in Table \ref{LSfits}. The derived period (in minutes) is shown below the object name.}
\end{figure*}

\renewcommand{\thefigure}{\arabic{figure} (Cont.)}
\addtocounter{figure}{-1}

\begin{figure*}
\centering
\includegraphics[height=0.222\textwidth]{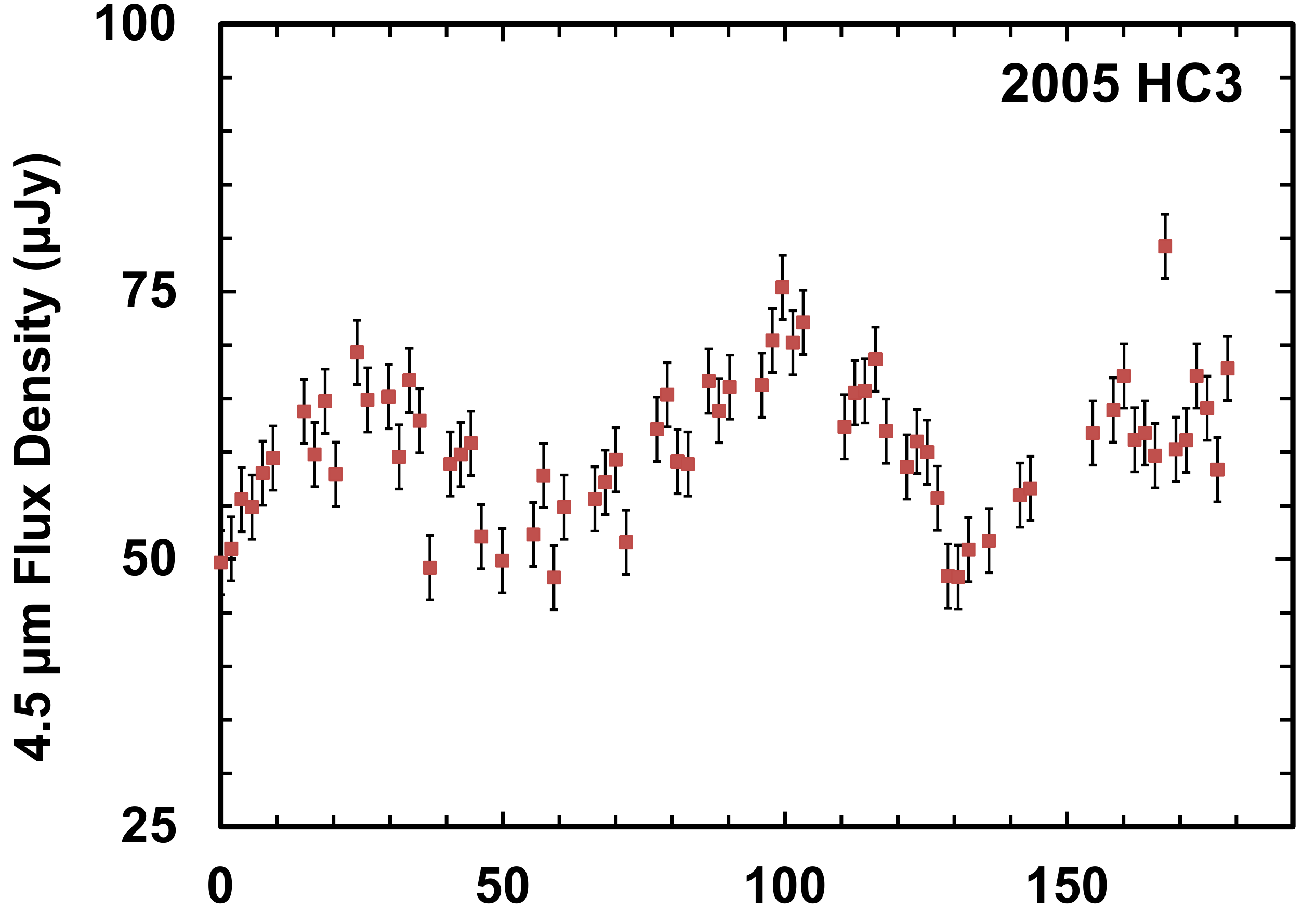}
\includegraphics[height=0.222\textwidth]{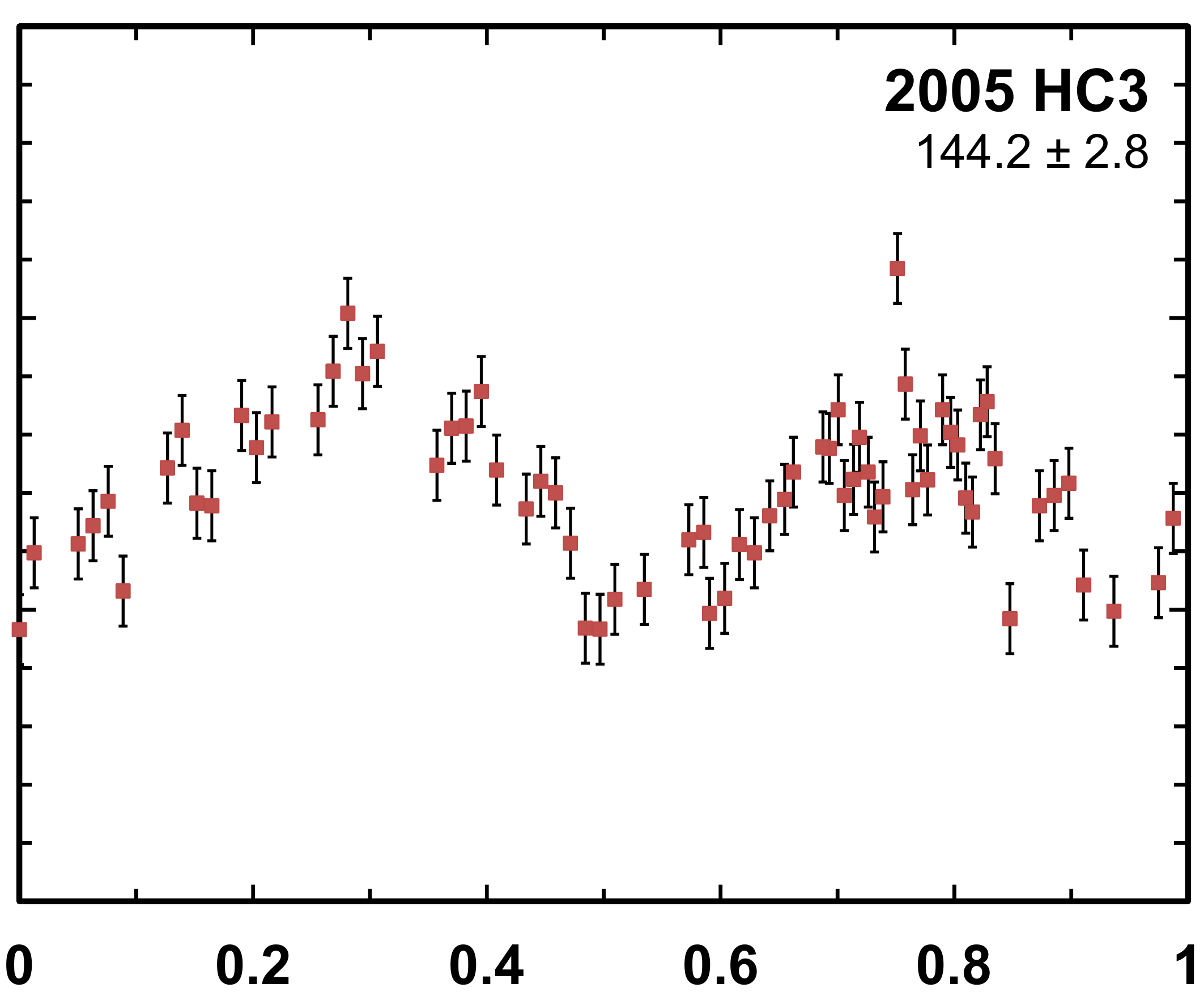}      \\ \vskip 5pt
\includegraphics[height=0.225\textwidth]{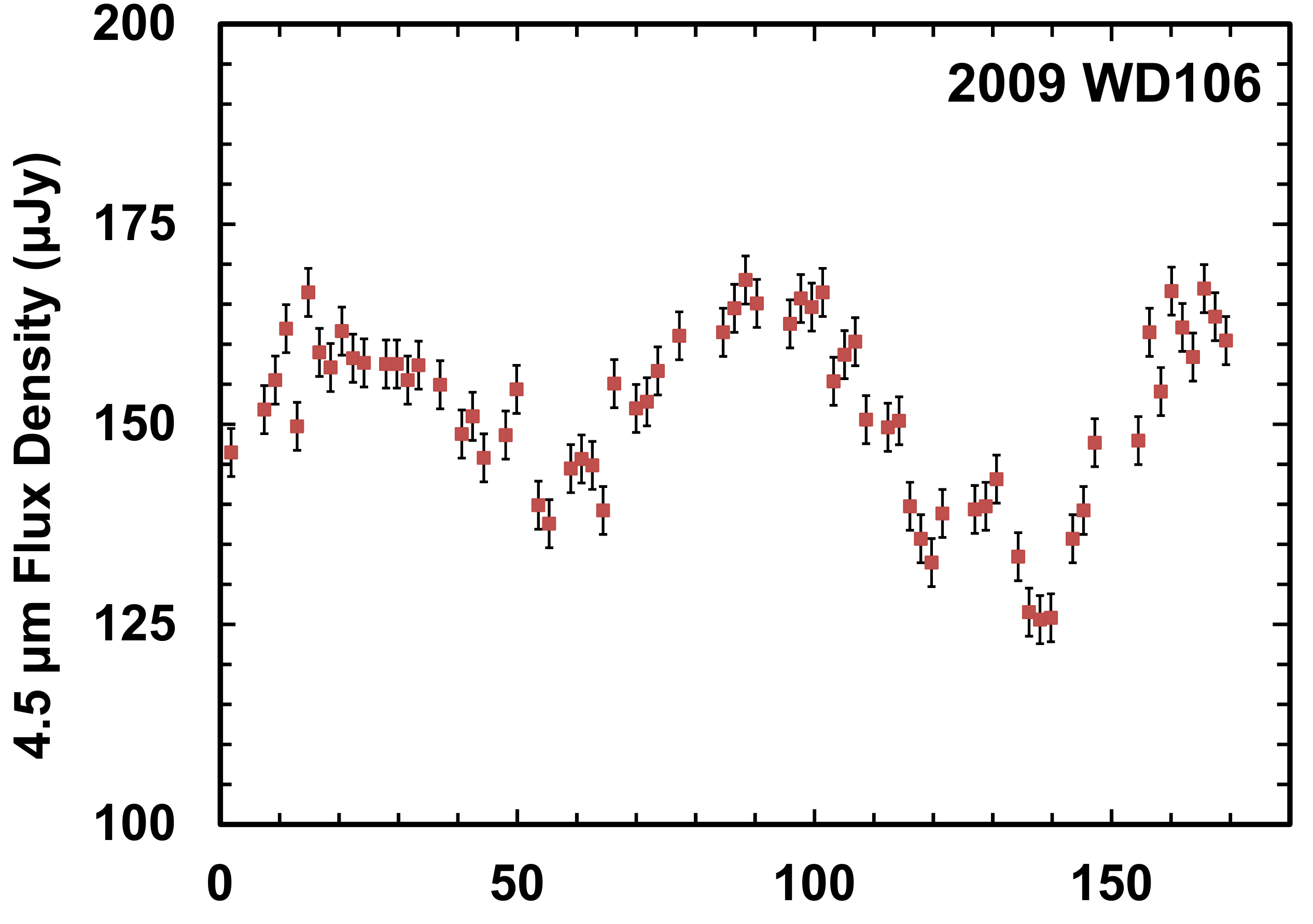}
\includegraphics[height=0.225\textwidth]{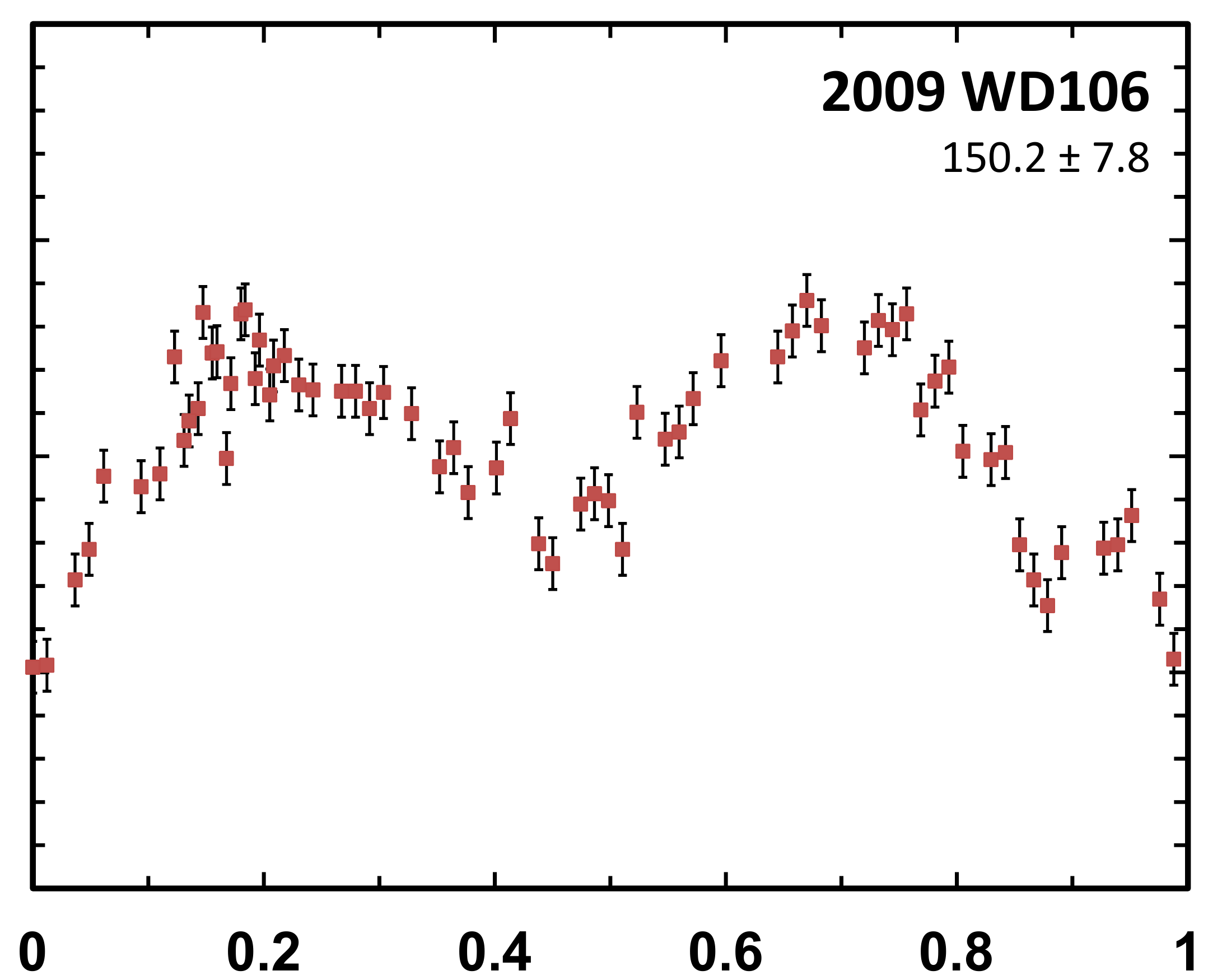}	\\ \vskip 5pt
\includegraphics[height=0.224\textwidth]{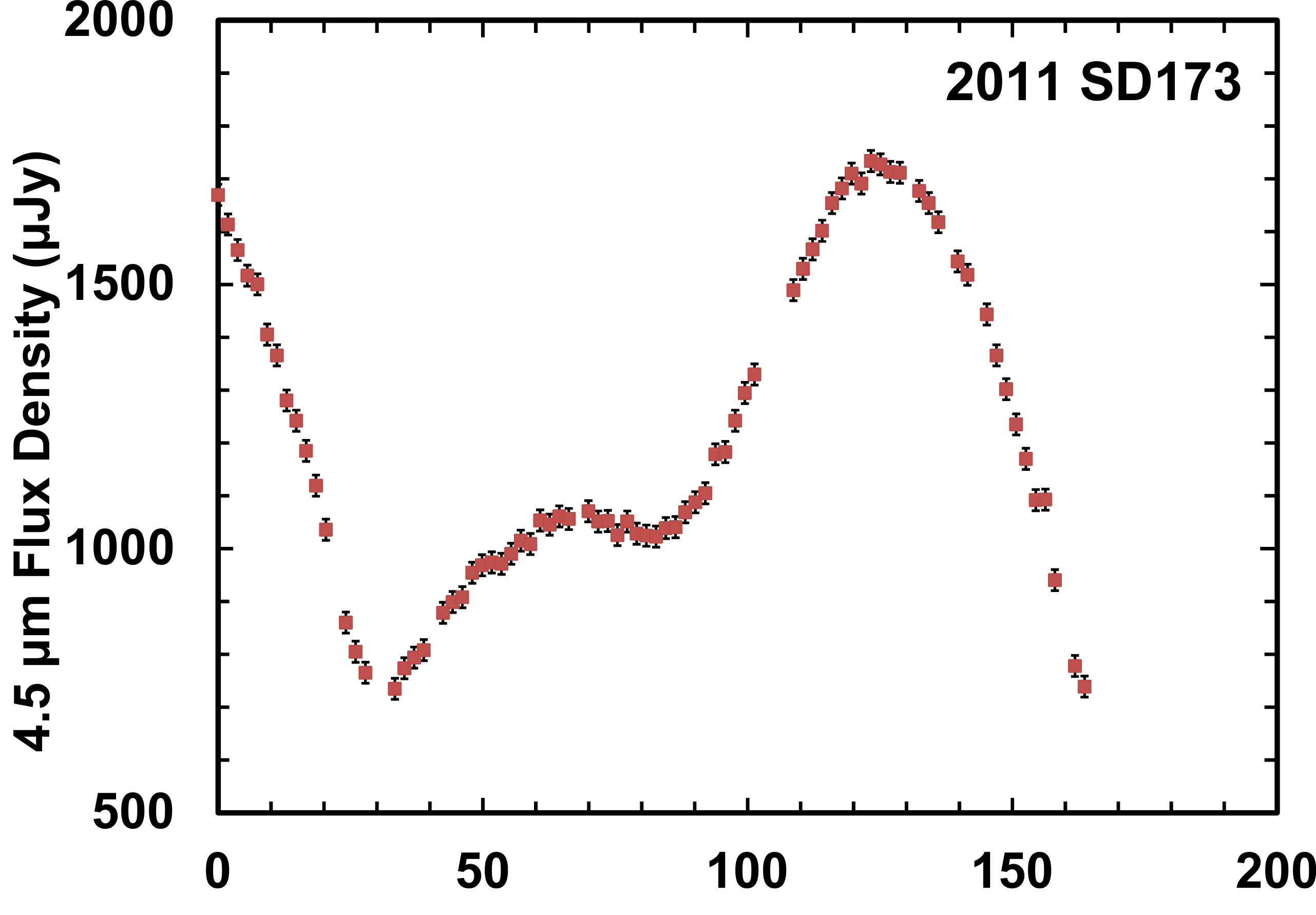}
\includegraphics[height=0.224\textwidth]{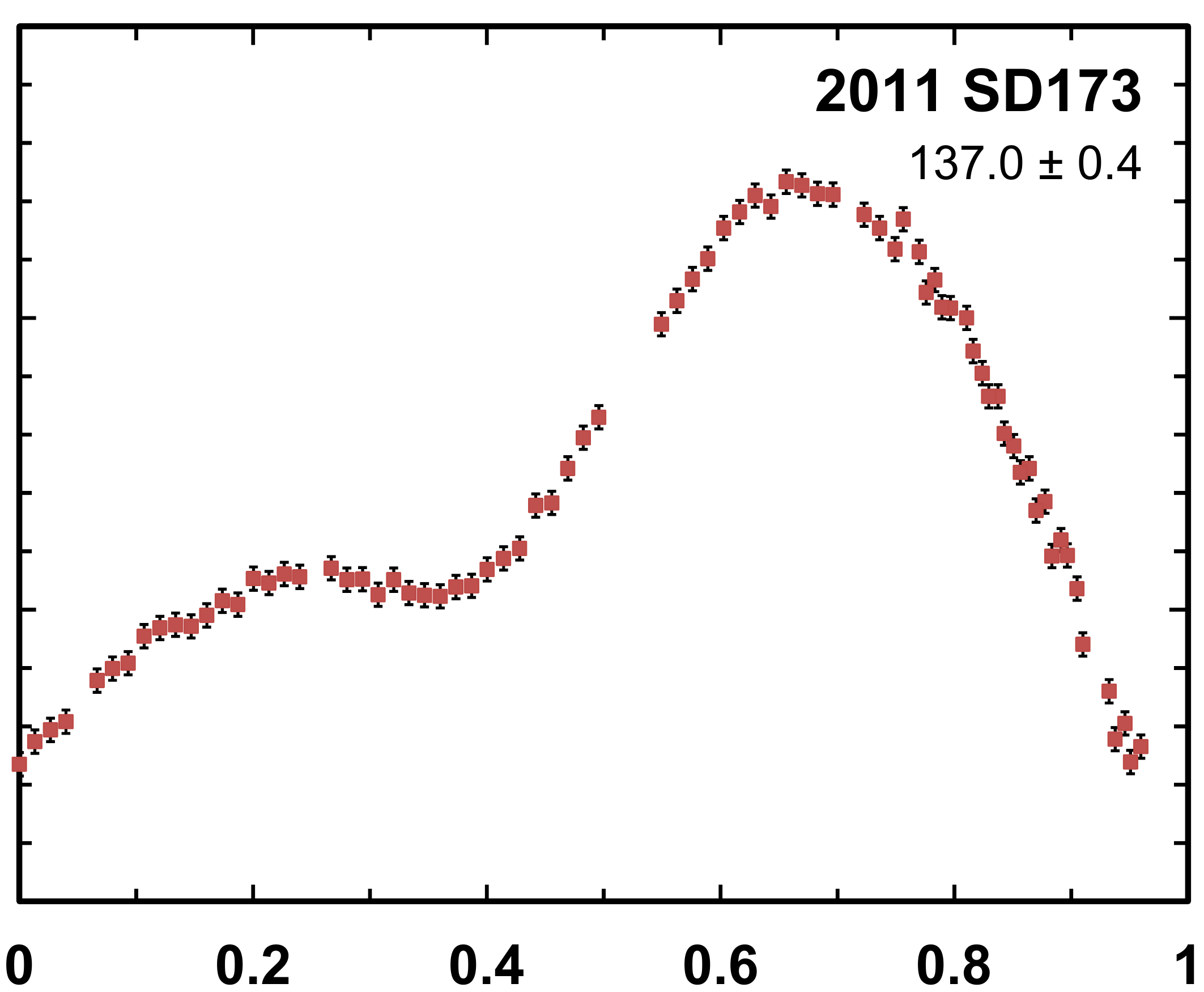}\\ \vskip 5pt
\includegraphics[height=0.225\textwidth]{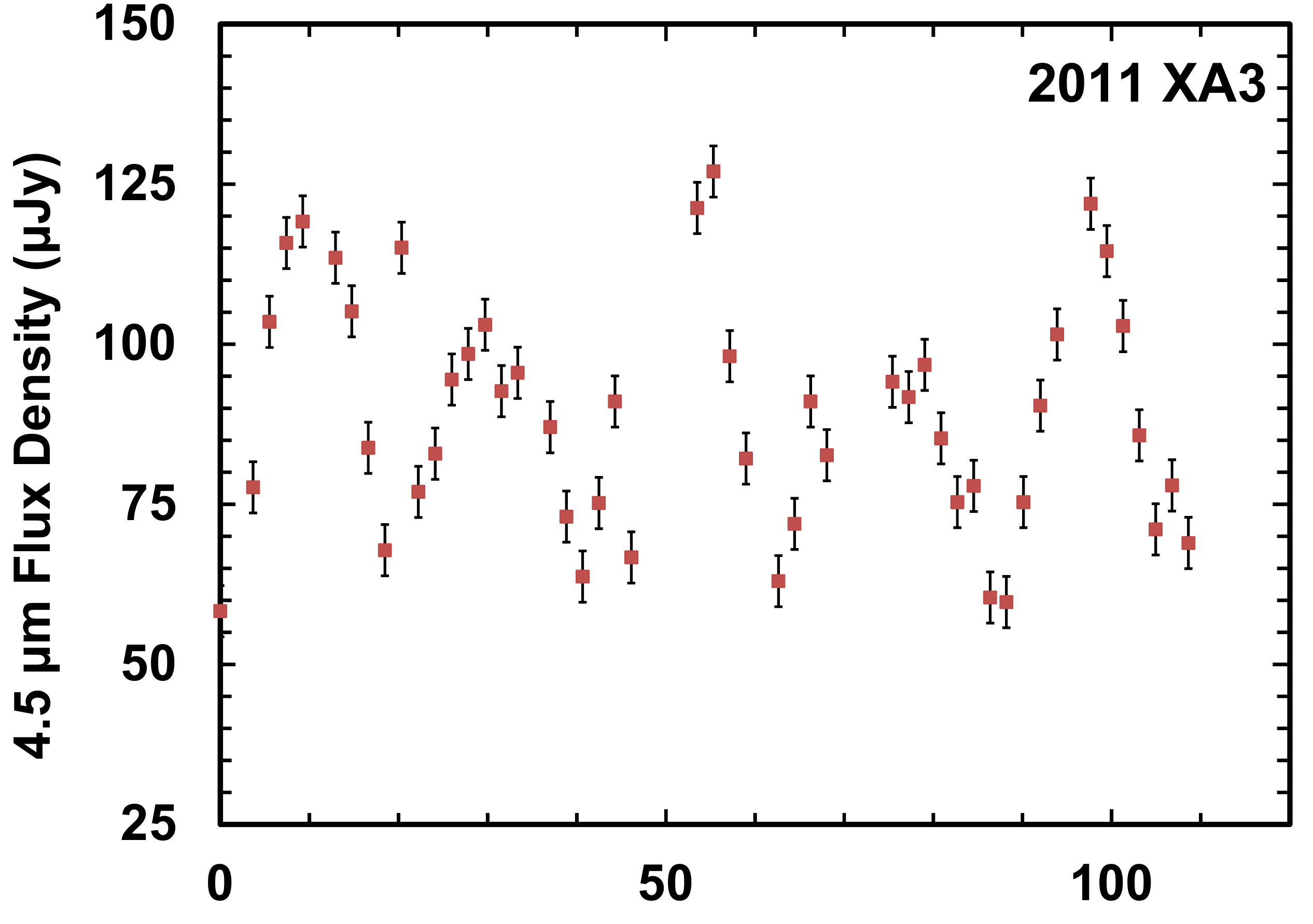}
\includegraphics[height=0.225\textwidth]{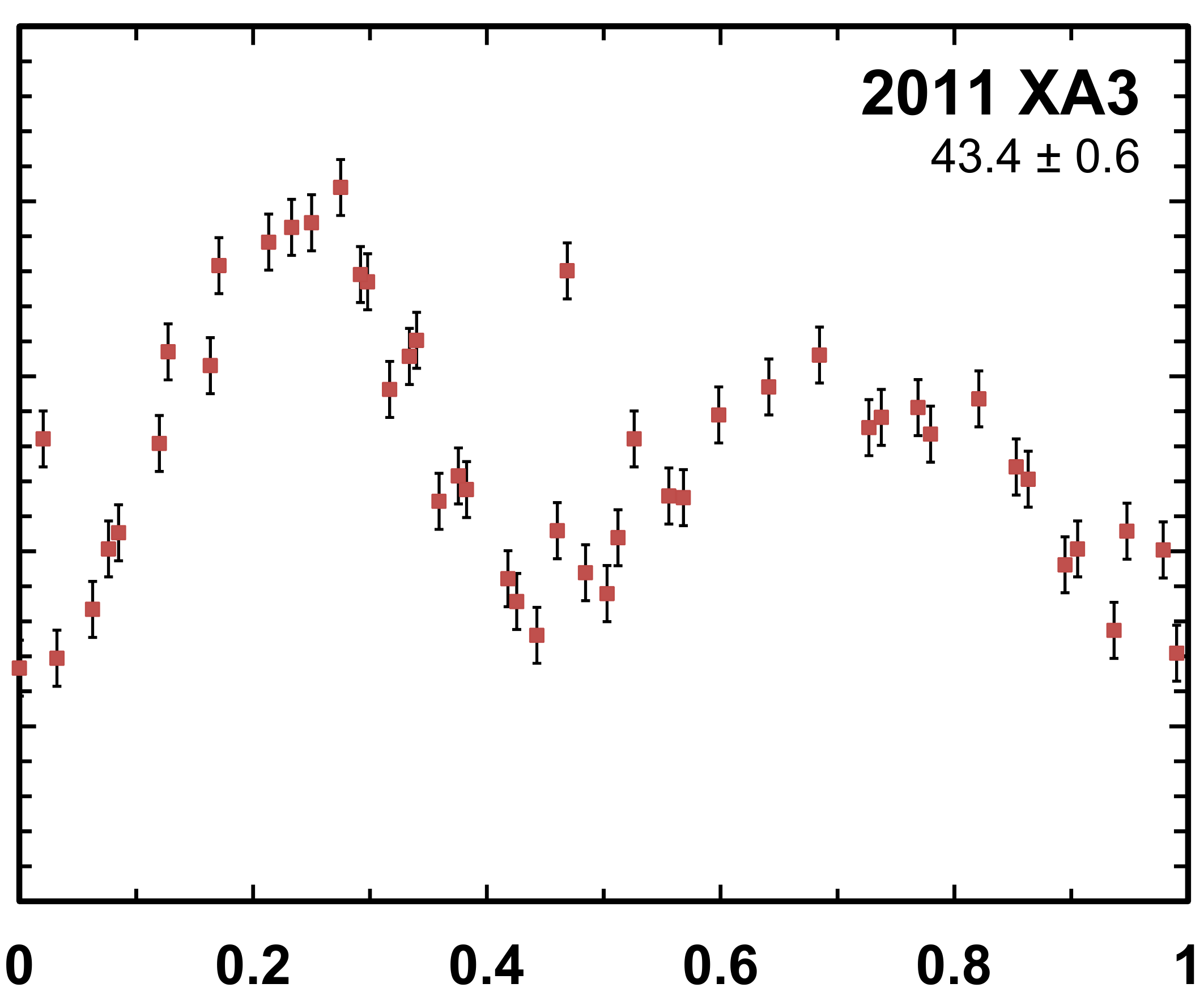}\\ \vskip 5pt
\includegraphics[height=0.242\textwidth]{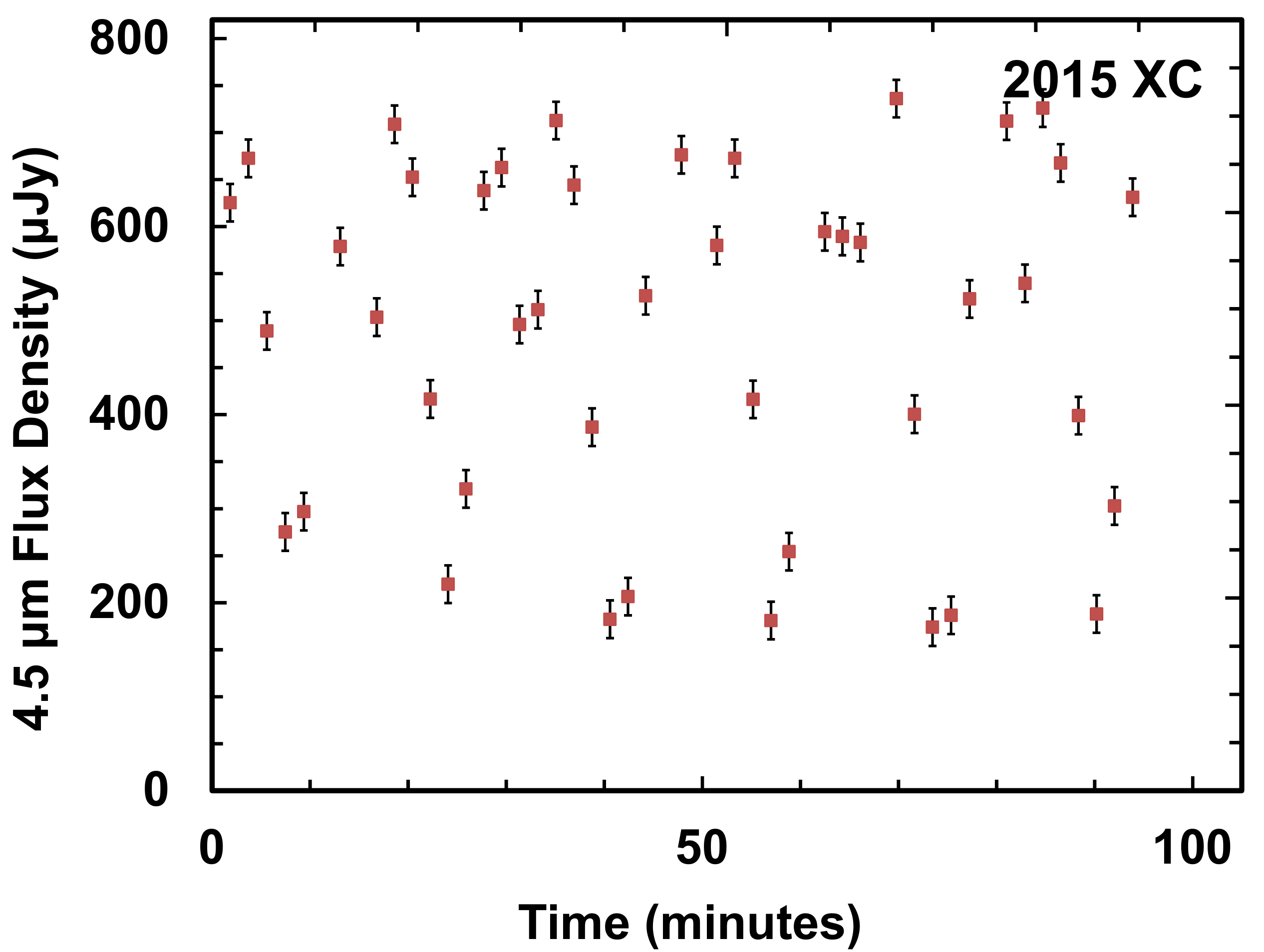}
\includegraphics[height=0.242\textwidth]{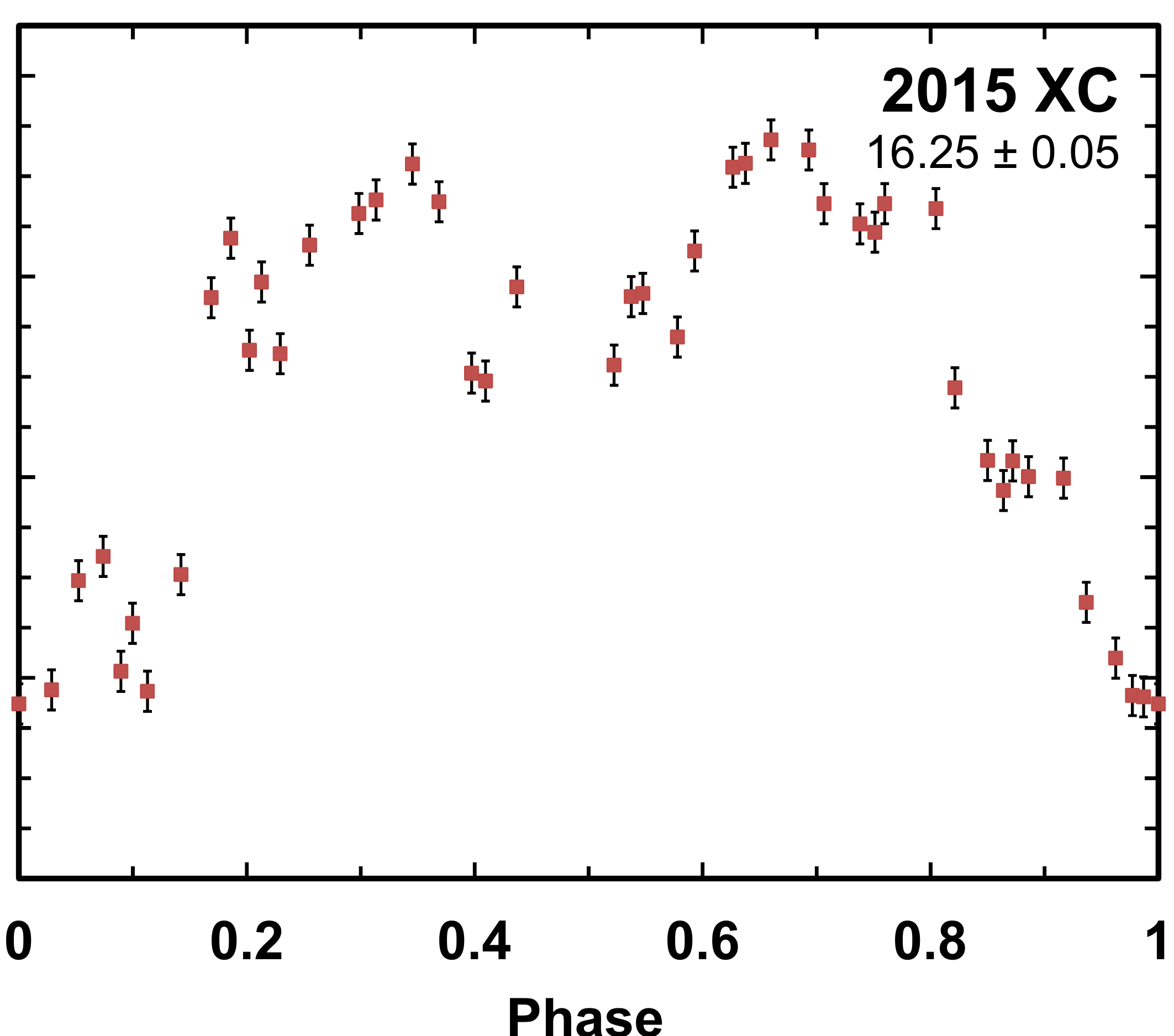} \\
\caption{}
\end{figure*}

\begin{deluxetable*}{lrccrccc}
\tablecaption{Periodogram Fits of Full NEO Lightcurves \label{LSfits}}
\tablecolumns{8}
\tablewidth{0pt}
\tablehead{
& & & HMJD &Lightcurve & \\
\colhead{Object} & \colhead{AORID} &
\colhead{UTC start time\tablenotemark{a}} &
\colhead{start time\tablenotemark{a}} &
\colhead{Duration} &
\colhead{Rotation Period} & \colhead{Amplitude} \\
\colhead{} & \colhead{} & \colhead{(YYYY-MM-DD hh:mm:ss)} & \colhead{(d)} & \colhead{(minutes)} &
\colhead{(minutes)} & \colhead{(mag)}
}
\startdata
1990 MF & 52514560 & 2016-02-27\enspace07{:}39{:}32 & 57445.3197069 & 105.0 & 54.4 $\pm$5.2 & 0.069 $\pm$ 0.012  \\
1990 UA & 42169088 & 2011-07-08\enspace17{:}20{:}53 & 55750.7229074 & 315.8 & 180.1$\pm1.9$\tablenotemark{b} & 0.216  $\pm$ 0.008 & \\
1998 FF14 & 61788672 & 2016-12-27\enspace16{:}27{:}21 & 57749.6862440 & 139.8 & 111.8$\pm$4.2 & 0.271 $\pm$ 0.047  \\
1999 JE1 & 42163456 & 2011-07-23\enspace02{:}34{:}22 & 55765.1077800 & 544.3 & 394$\pm$11.6 & 0.136 $\pm$ 0.015  \\
2003 EO16 & 42164480 & 2011-06-16\enspace09{:}11{:}35 & 55728.3836200 & 417.9 & 350.6$\pm$3.4  & 0.135 $\pm$ 0.010  \\
2005 HC3 & 52392192 & 2016-04-08\enspace16{:}22{:}38 & 57486.6829712 & 178.5 & 144.2$\pm$2.8  & 0.422 $\pm$ 0.046 \\
2009 WD106 & 52496128 & 2015-04-10\enspace16{:}43{:}14 & 57122.6972784 & 167.4 & 150.2$\pm$7.8  & 0.299 $\pm$ 0.020 \\
2011 SD173 & 52501760 & 2015-05-26\enspace08{:}34{:}56 & 57168.3581743 & 163.6 & 137.0$\pm$0.4 & 0.892 $\pm$ 0.022  \\
2011 XA3 & 61855488 & 2017-04-17\enspace09{:}50{:}19 & 57860.4105209 & 108.6 & 43.4$\pm$0.6 & 0.648 $\pm$ 0.133 \\
2015 XC & 61809152 & 2017-03-14\enspace18{:}31{:}52 & 57826.7727121 & 93.9 & 16.25$\pm$0.05 & 1.566 $\pm$ 0.066 \\
\enddata
\tablenotetext{a}{Time at the midpoint of the first frame of the observation.}
\tablenotetext{b}{The Plavchan algorithm was used to calculate this period, see Section \ref{objectdiscuss}}
\tablecomments{Columns: asteroid designations, \textit{Spitzer} Astronomical Observation Request identifier, observation start time in UT and heliocentric MJD, respectively, the observation duration, derived rotation period, and light-curve amplitude (mag).}
\end{deluxetable*}

\renewcommand{\thefigure}{\arabic{figure}}

\begin{figure*}\label{LSplotsPart}
\centering
\includegraphics[height=0.221\textwidth]{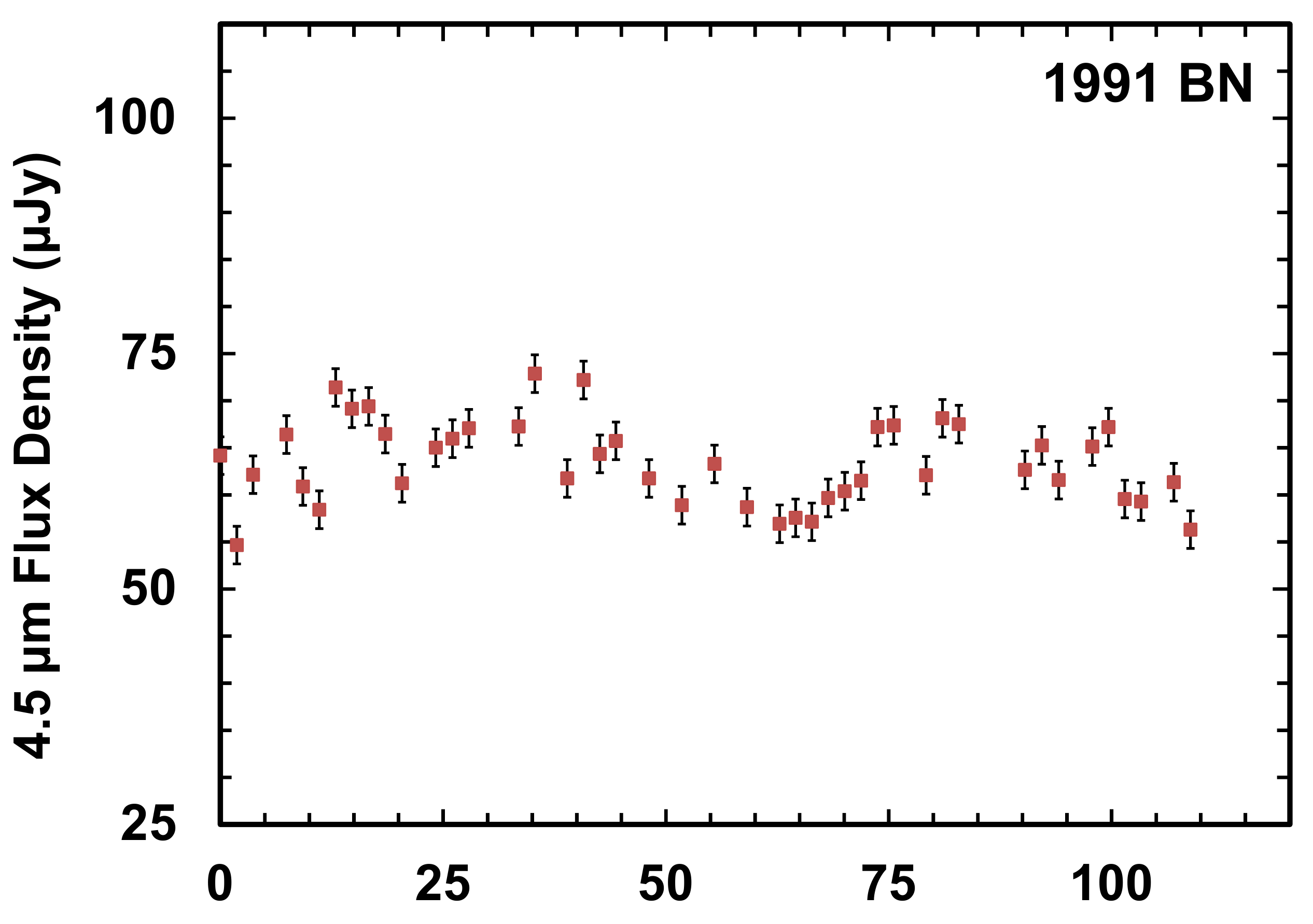}
\includegraphics[height=0.221\textwidth]{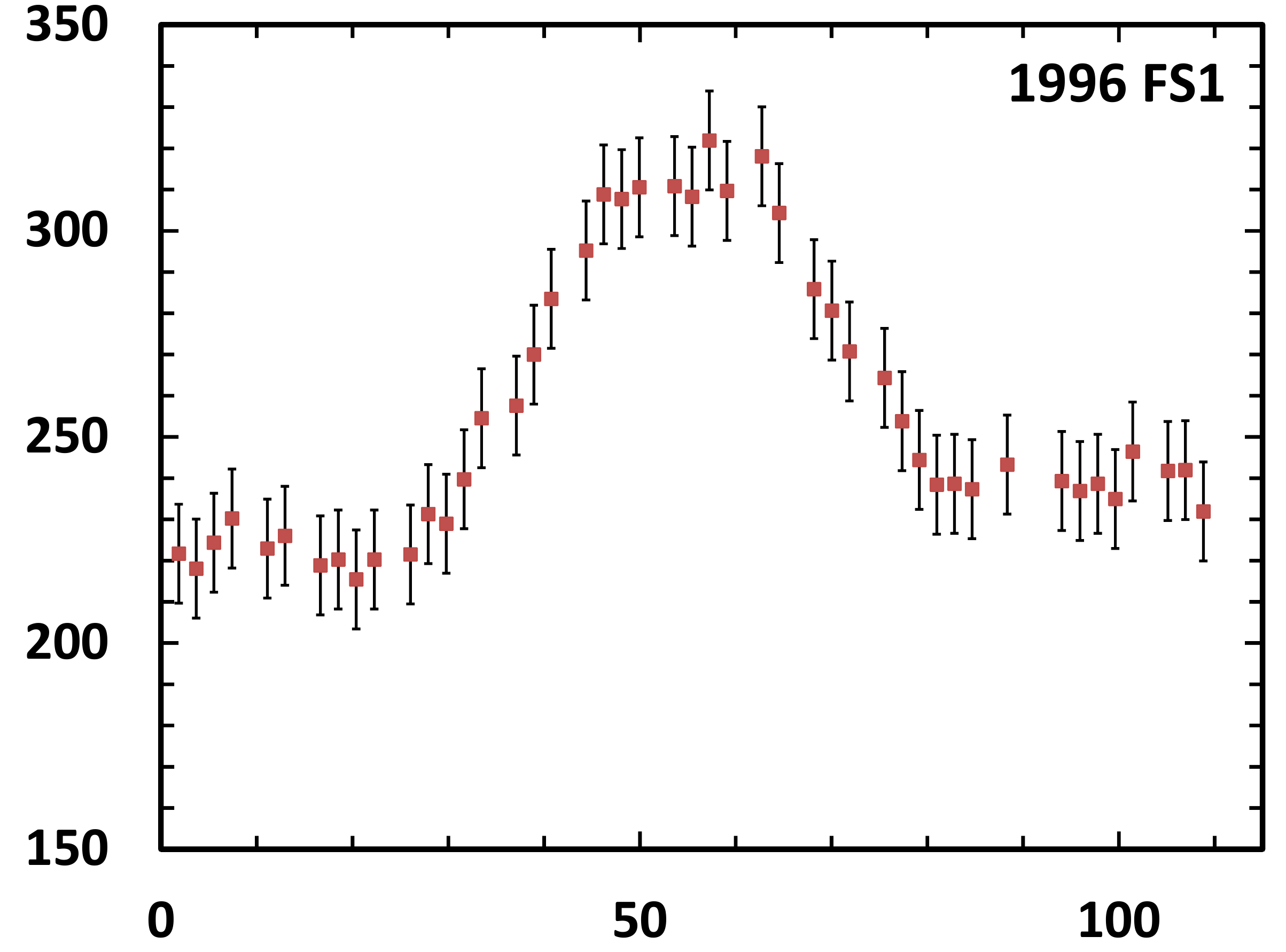}
\includegraphics[height=0.221\textwidth]{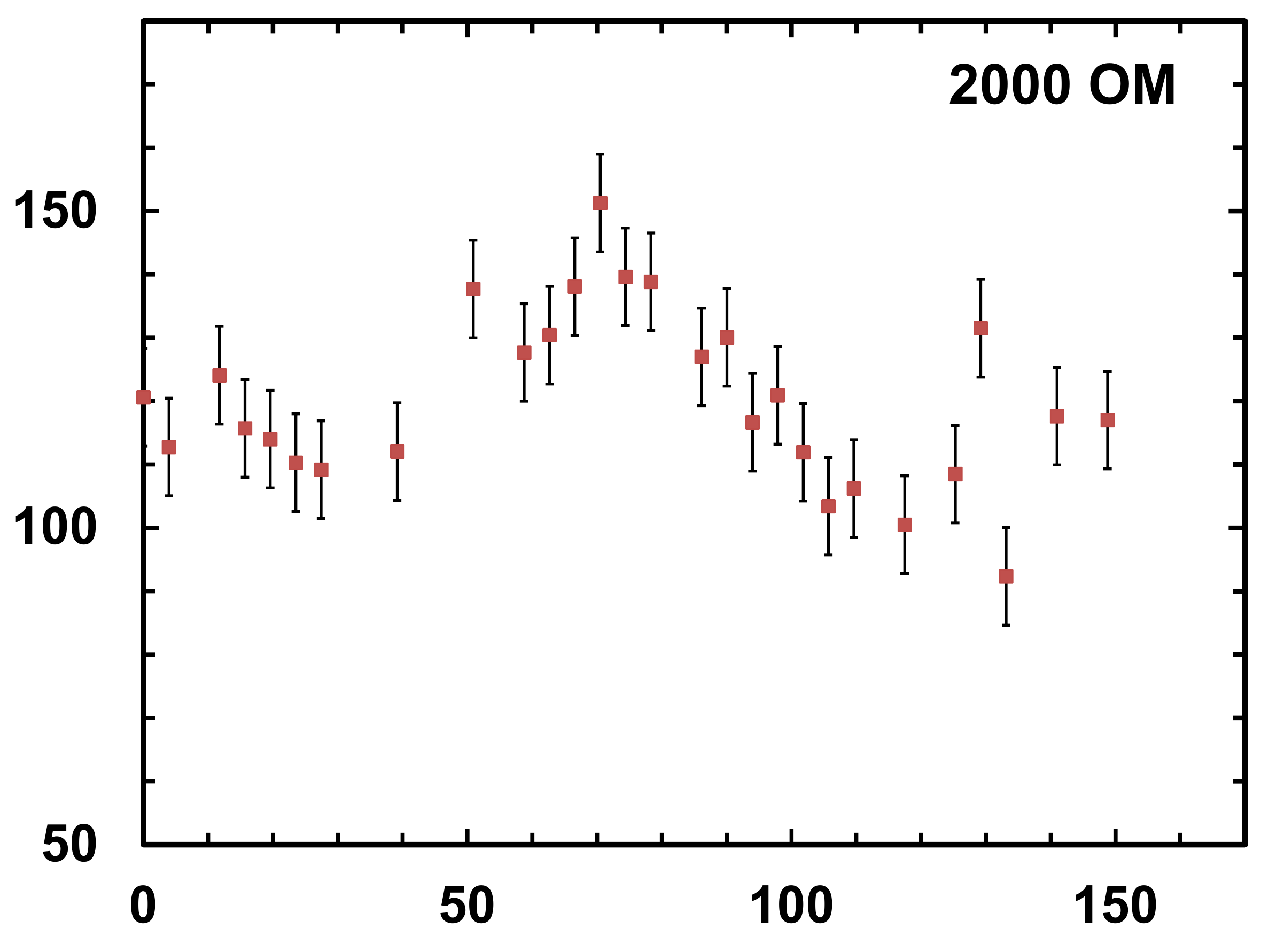}\\ \vskip 5pt
\includegraphics[height=0.221\textwidth]{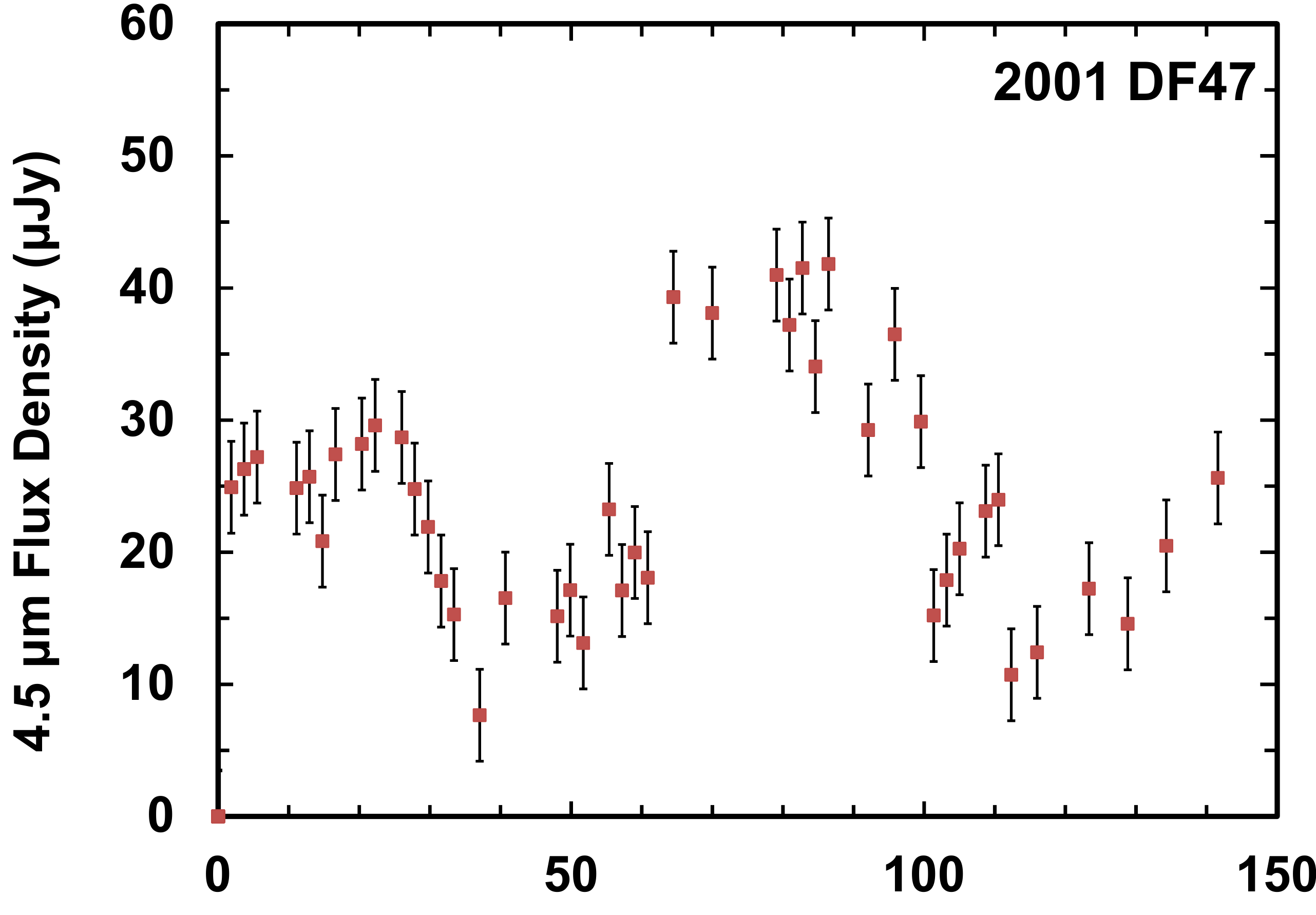}
\includegraphics[height=0.221\textwidth]{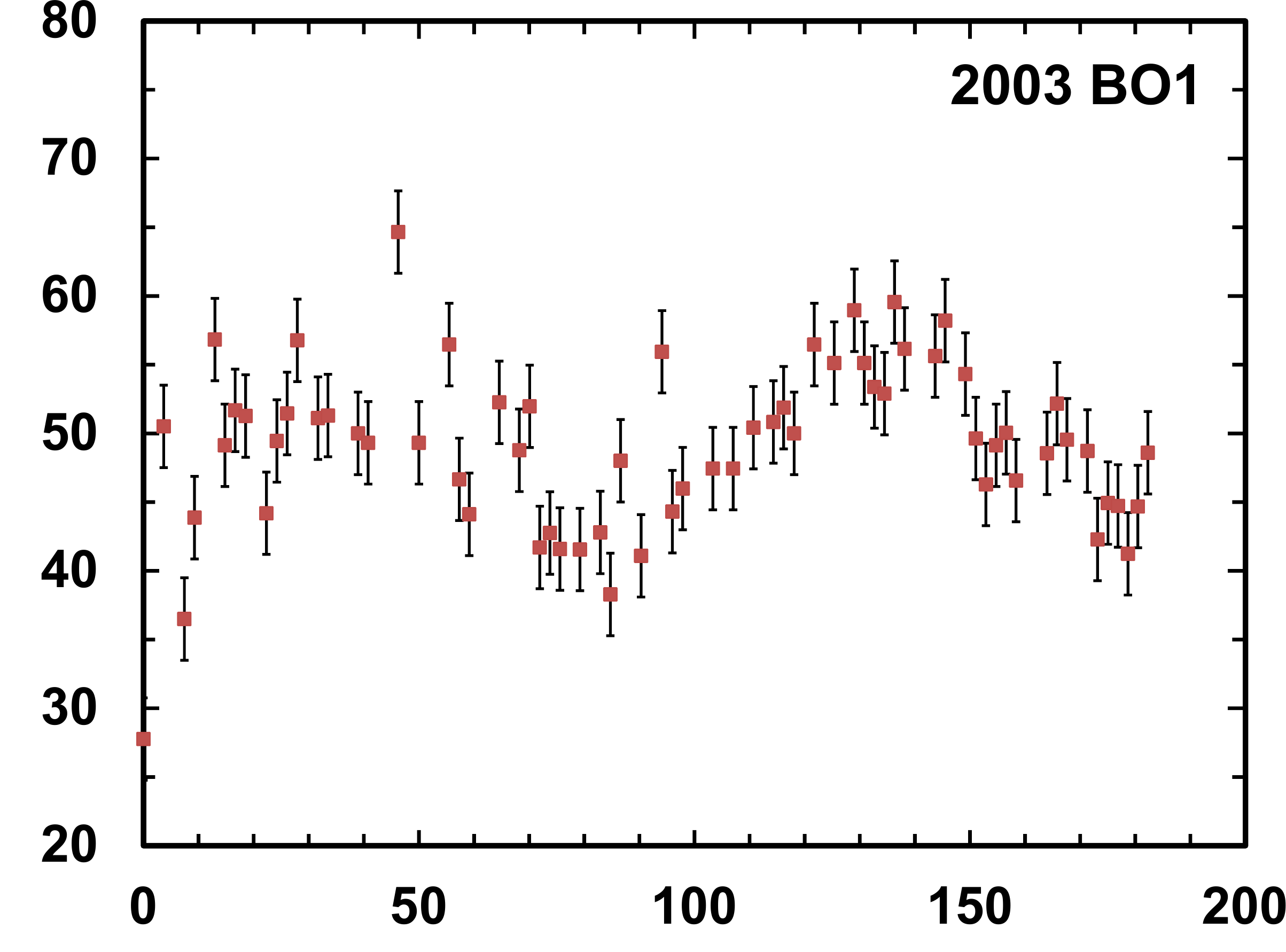}
\includegraphics[height=0.221\textwidth]{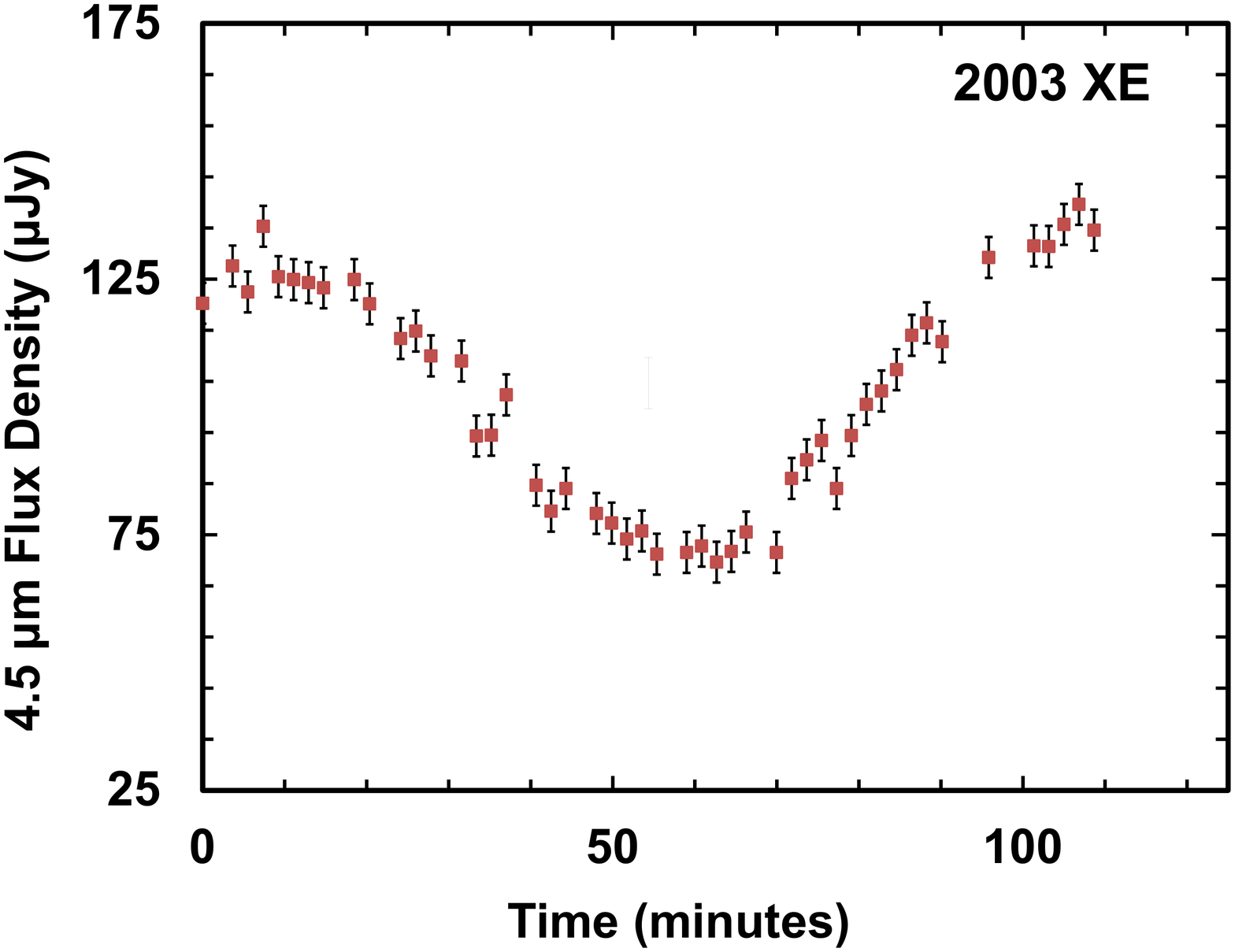}\\ \vskip 5pt
\includegraphics[height=0.221\textwidth]{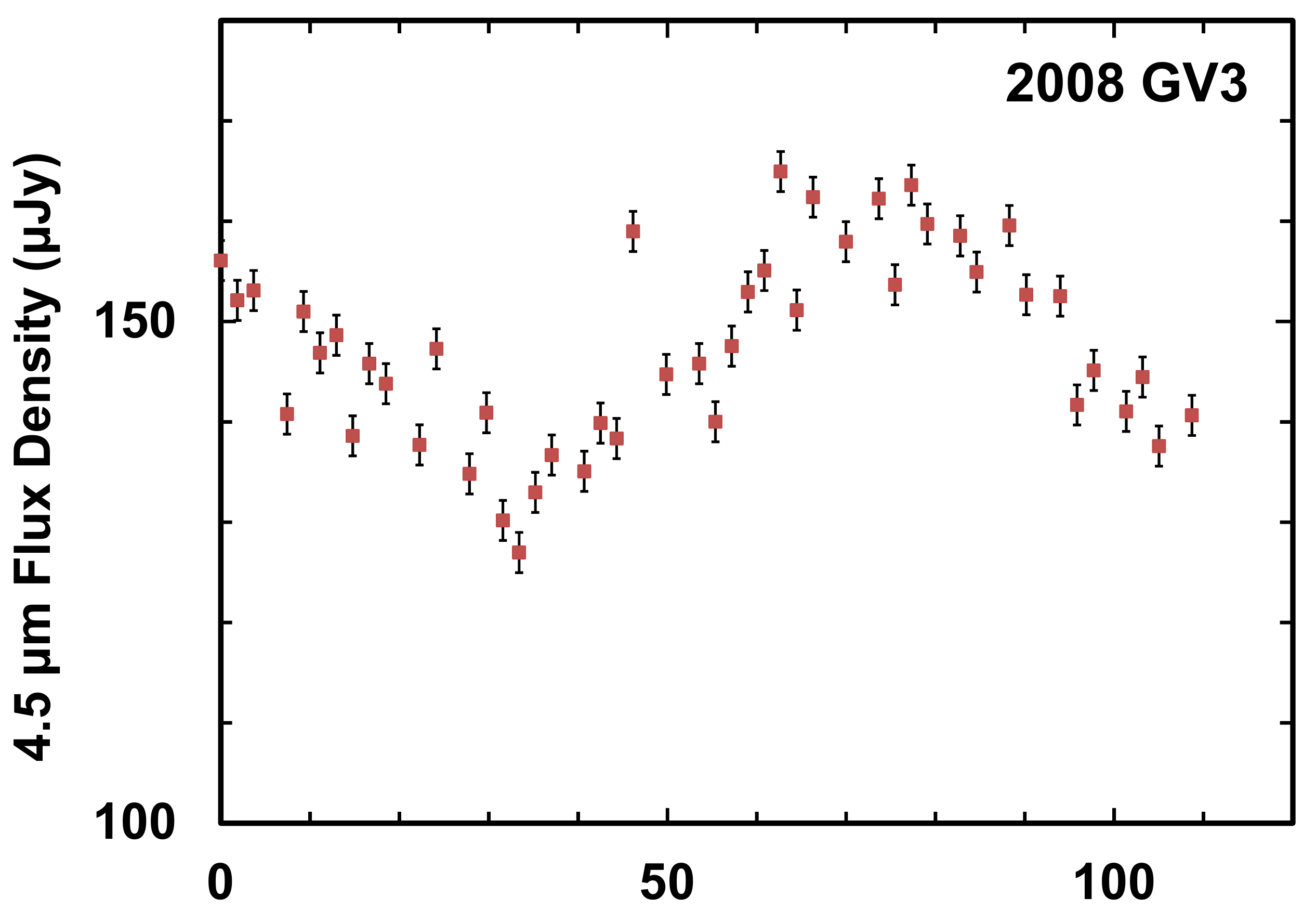}
\includegraphics[height=0.221\textwidth]{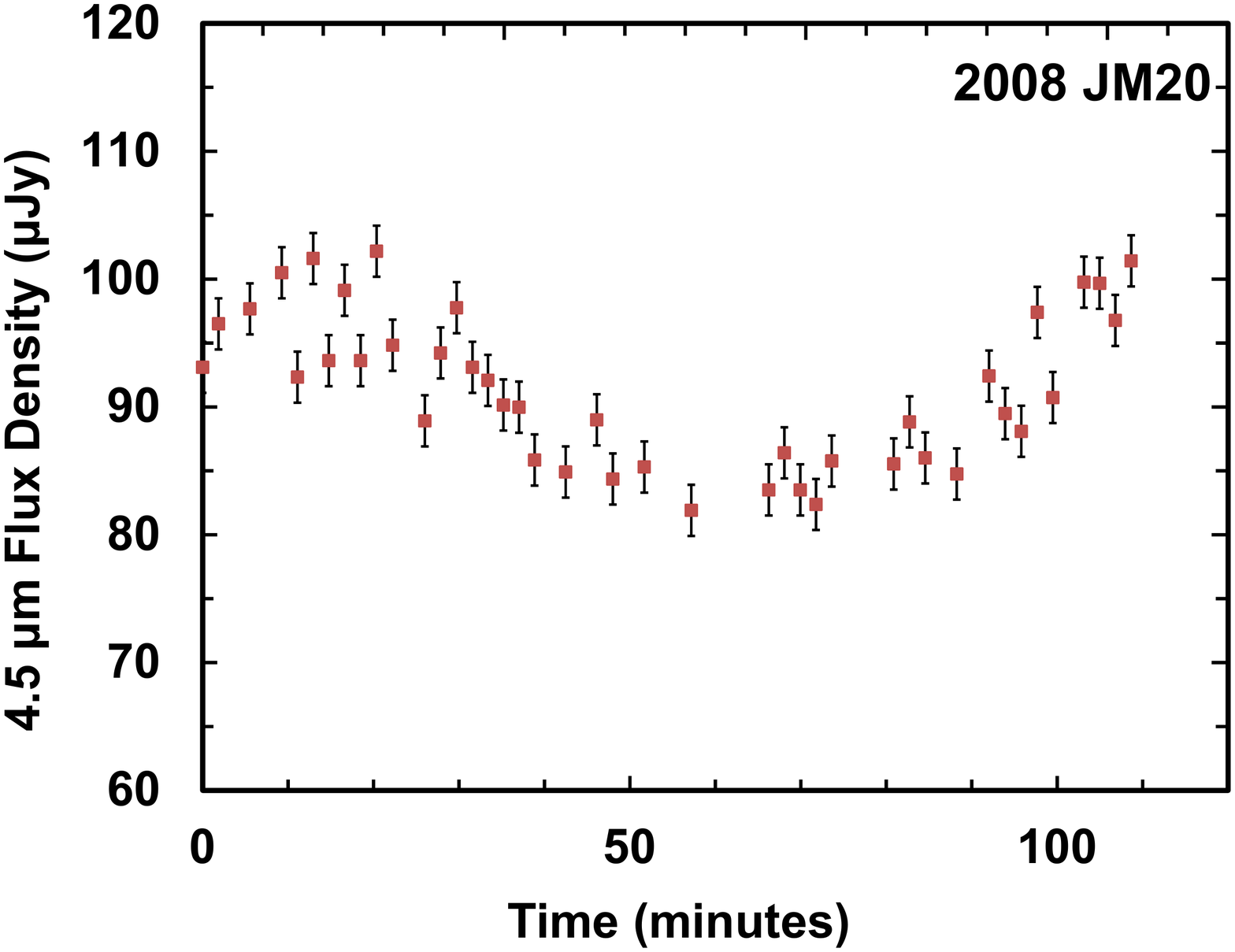}
\includegraphics[height=0.221\textwidth]{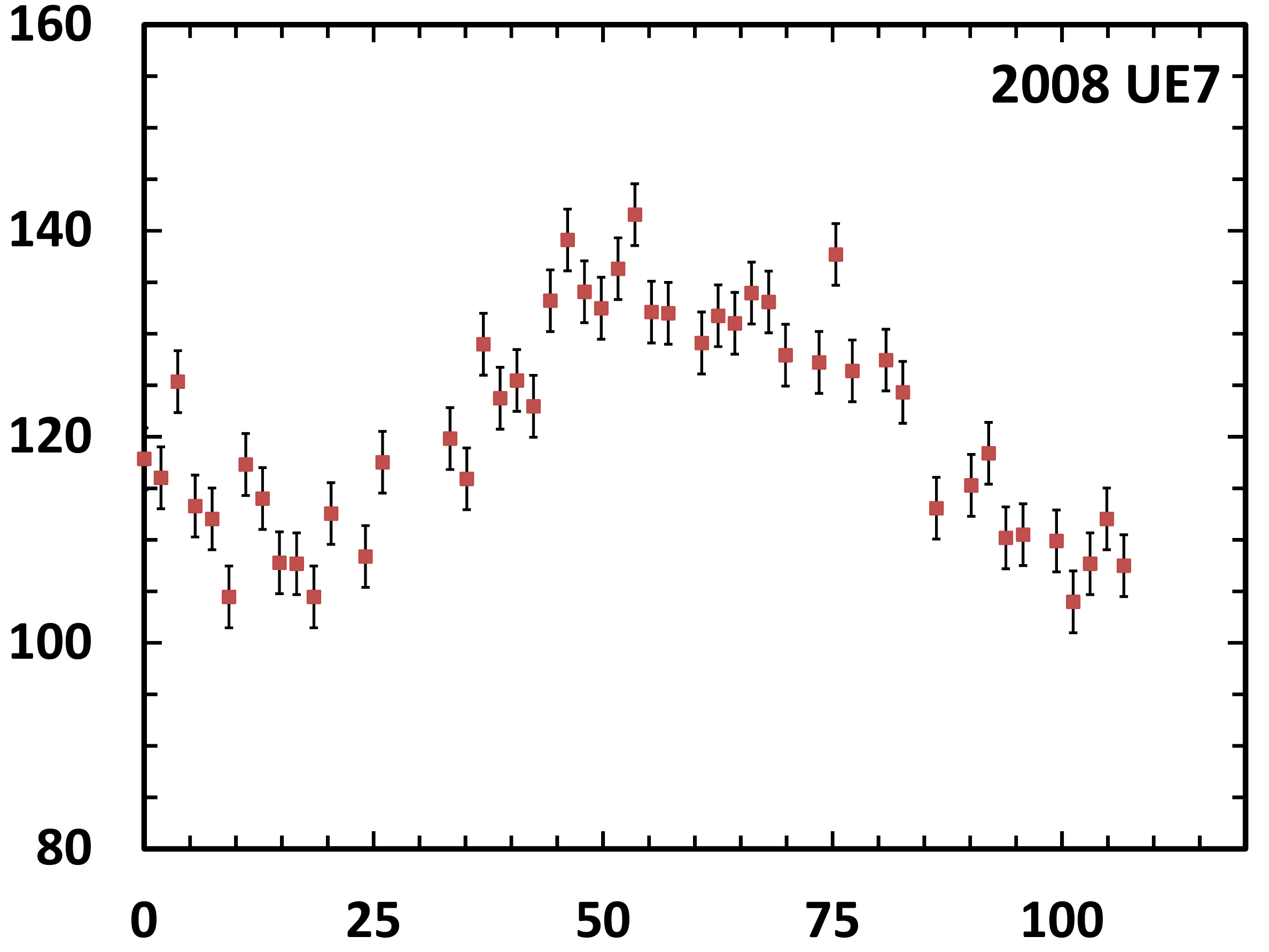}\\ \vskip 5pt
\includegraphics[height=0.238\textwidth]{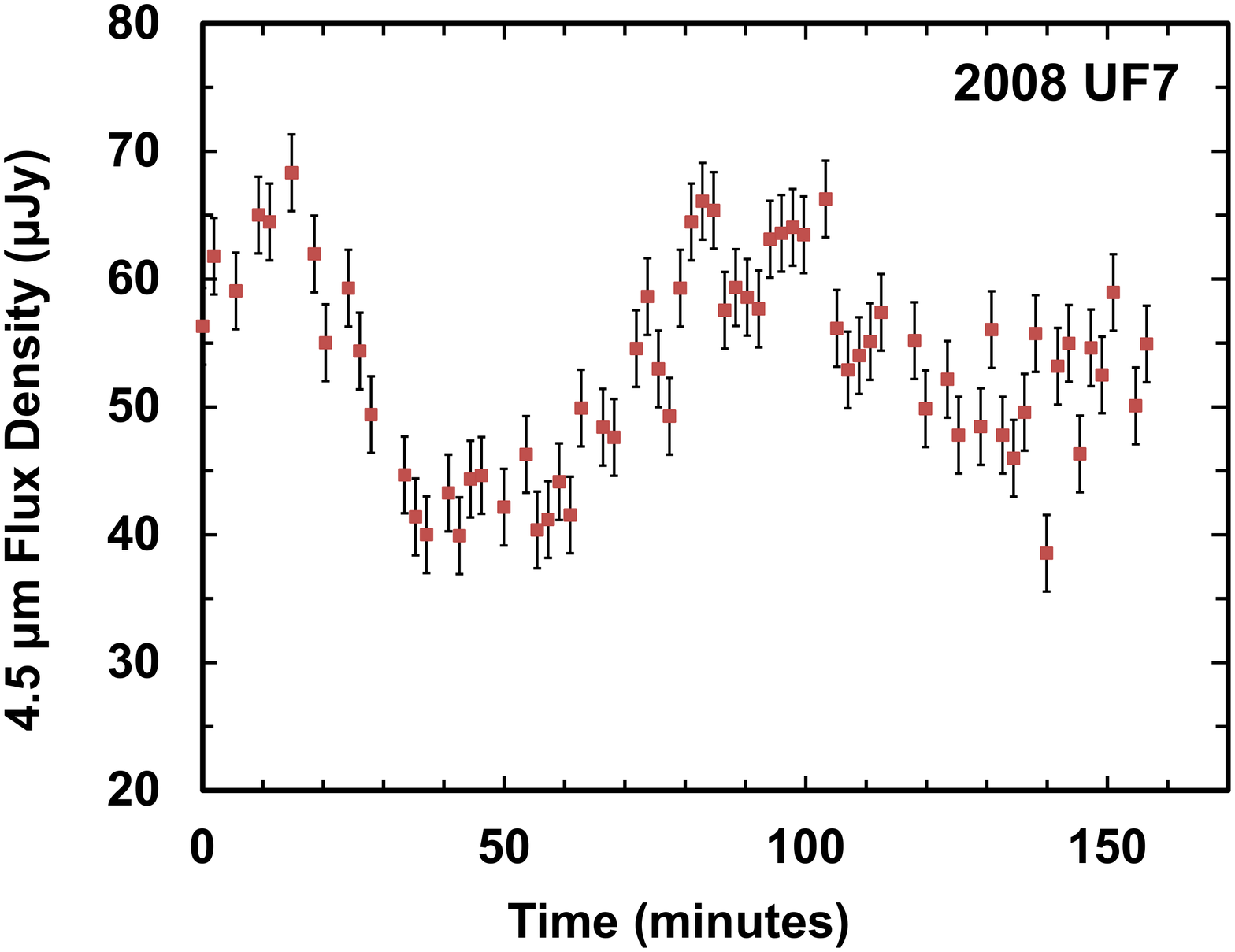}
\includegraphics[height=0.238\textwidth]{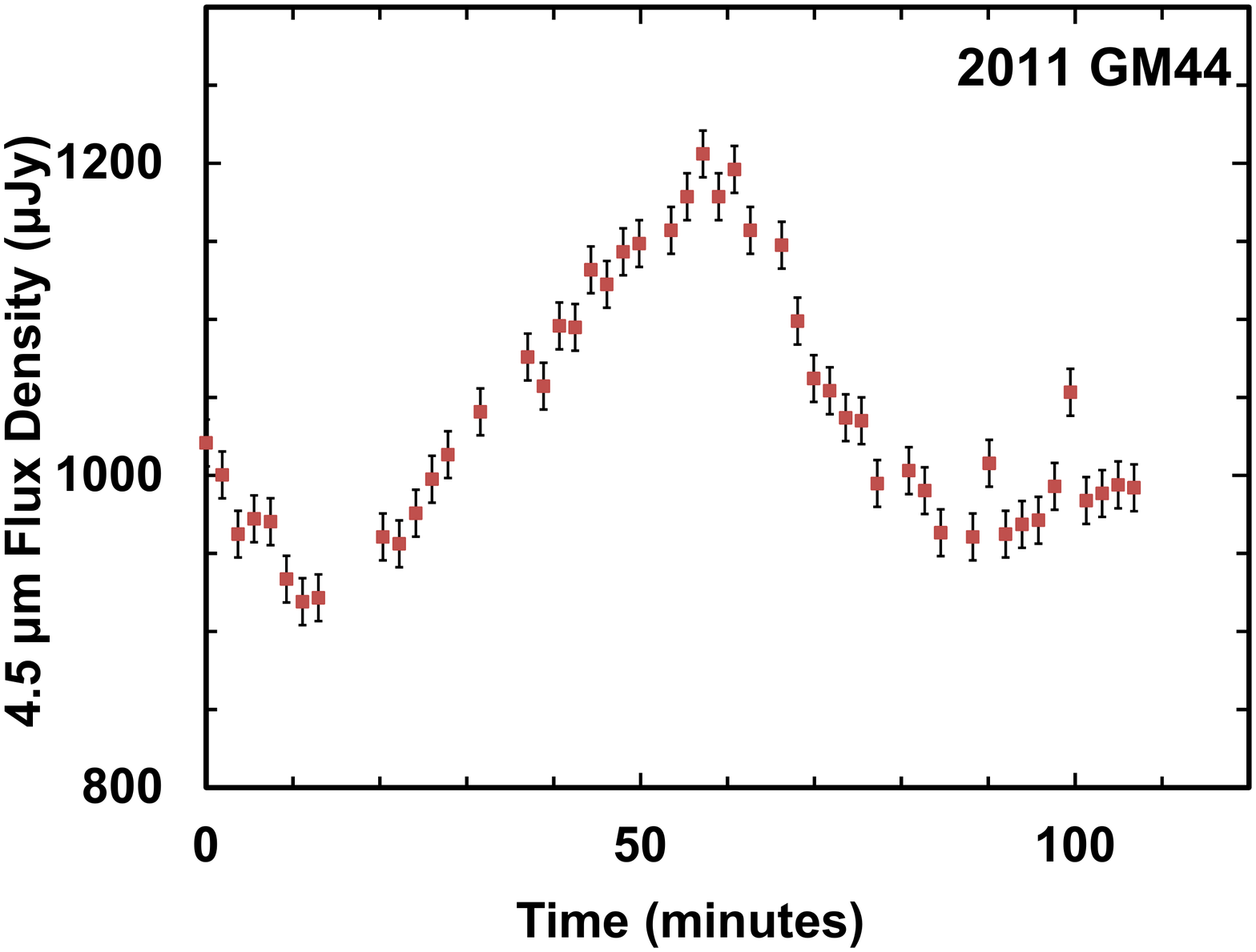} 

\caption{\Sp\ lightcurves that cover less than one rotation period. The horizontal axis gives the time in minutes relative to the first point in the lightcurve.The extrapolated lower limits to the rotation periods are given in Table \ref{LSfits2}. }\end{figure*}

\begin{deluxetable*}{lrccrccc}
\tablecaption{Periodogram Fits of Partial NEO Lightcurves \label{LSfits2}}
\tablecolumns{8}
\tablewidth{0pt}
\tablehead{
& & & HMJD &Lightcurve & Lower Limit to\\
\colhead{Object} & \colhead{AORID} &
\colhead{UTC start time\tablenotemark{a}} &
\colhead{start time\tablenotemark{a}} &
\colhead{Duration} &
\colhead{Rotation Period} & \colhead{Amplitude} \\
\colhead{} & \colhead{} & \colhead{(YYYY-MM-DD hh:mm:ss)} & \colhead{(d)} & \colhead{(minutes)} &
\colhead{(minutes)} & \colhead{(mag)}
}
\startdata
1991 BN & 52457216 & 2015-02-28\enspace01{:}55{:}51 & 57081.0810333 & 108.9 & 114.6$\pm$8.4  & 0.218 $\pm$ 0.040  \\
1996 FS1 & 52366592 & 2015-05-19\enspace06{:}24{:}06 & 57161.2621636 & 108.8 & 163.2$\pm$1.7  & 0.404 $\pm$  0.015 \\
2000 OM & 44166656 & 2011-08-28\enspace20{:}20{:}08 & 55801.8479200 & 152.7 & 190.2$\pm$7.4 & 0.354 $\pm$ 0.049 \\
2001 DF47 & 61870080 & 2016-10-25\enspace21{:}58{:}09 & 57686.9159586 & 139.7 & 158.8$\pm$4.8 & 1.208 $\pm$ 0.261  \\
2003 BO1 & 52507648 & 2015-06-23\enspace19{:}09{:}23 & 57196.7987658 & 182.3 & 187.4$\pm$6.4 & 0.363 $\pm$ 0.063 \\
2003 XE & 52498944 & 2015-04-26\enspace03{:}01{:}09 & 57138.1263838 & 108.7 & 203.8$\pm$5.4 & 0.709 $\pm$ 0.047  \\
2008 GV3 & 52410112 & 2016-01-16\enspace12{:}07{:}41 & 57403.5059145 & 108.7 & 172.2$\pm$7.2  & 0.231 $\pm$ 0.022 \\
2008 JM20 & 58815488 & 2016-08-03\enspace03{:}53{:}54 & 57603.1630151 & 108.7 &  208.8$\pm$ 18.8\tablenotemark{b} & 0.187 $\pm$ 0.020 \\
2008 UE7 & 52413184 & 2015-07-16\enspace18{:}52{:}07 & 57219.7867809 & 106.7 & 174.4$\pm$10.6 & 0.270 $\pm$ 0.024  \\
2008 UF7 & 52413440 & 2015-02-01\enspace06{:}39{:}26 & 57054.2779595 & 156.5 & 166.1$\pm$1.8 & 0.514 $\pm$ 0.053  \\
2011 GM44 & 52424704 & 2015-07-02\enspace06{:}21{:}57 & 57205.2658287 & 106.8 & 155.8$\pm$4.2  & 0.277 $\pm$ 0.015  \\
2013 CW32 & 61845504 & 2017-05-13\enspace05{:}50{:}42 & 57886.2441315 & 42.6 & 79.2$\pm$2.6 & 0.224 $\pm$ 0.030  \\
\enddata
\tablenotetext{a}{Time at the midpoint of the first frame of the observation.}
\tablenotetext{b}{The Plavchan algorithm was used to calculate this period, see Section \ref{objectdiscuss}}
\tablecomments{Columns: asteroid designations, \textit{Spitzer} Astronomical Observation Request identifier, observation start time in UT and heliocentric MJD, respectively, the observation duration, derived rotation period, and light-curve amplitude (mag). The periods and amplitudes should be treated as lower limits.}
\end{deluxetable*}
\subsection{Lightcurve extraction}
Extracting lightcurves from the \Sp\ NEO Survey was performed in several steps. Mosaics of the \Sp\ data for each object were constructed using the \texttt{IRACproc} software \citep{schuster06} which is based on the \texttt{mopex} mosaicking software \citep{makovoz06} distributed by the \Sp\ Science Center (SSC)\footnote{http://ssc.spitzer.caltech.edu/}. We downloaded the Basic Calibrated Data (BCD) frames for each observation from the \Sp\ Heritage Archive\footnote{http://sha.ipac.caltech.edu/applications/Spitzer/SHA/} which has data from the latest pipeline version for all IRAC observations. For each object, first a mosaic was made of the background field by masking the NEO from each BCD and making a mosaic in the non-moving frame. Since the observations were obtained by tracking at the non-sidereal NEO rate, the background objects are more or less trailed in the image, depending on the NEO's apparent rate of motion and the frame time being used. The mosaicking process removes any array artifacts and cosmic ray and other transient effects and creates a clean image of the field that the NEO was moving through. This background mosaic was then subtracted from each individual BCD image. This process removes most of the flux from the field objects, but the cosmic rays, hot or dead pixels, and other array artifacts remain in the BCD image. Also, the subtraction is incomplete near the core of bright stars and often artifacts are present in those locations. However, usually the fields are not very crowded with bright stars and the NEO falls on regions free from these effects for most of the observation. Aperture photometry on each BCD is then performed with the \texttt{phot} task in IRAF. An aperture radius of 6 pixels (7\farcs32) was used, with a sky background annulus of 6 pixels separated from the aperture by 6 pixels. The zero point magnitude for the photometry was determined from IRAC observations of calibration stars that we downloaded, reduced, and extracted in the same way (except without the background field subtraction). We also construct a mosaic from the BCDs in the moving reference frame of the NEO and perform photometry on that image, and we get excellent agreement between the fluxes derived from the BCD and the mosaic photometry. For some BCDs, the photometry process fails to generate valid results. For example, if the source happened to fall on a group of dead pixels, or there was a cosmic ray event that affected the region near the NEO, the \texttt{phot} task would fail to produce photometry, or give invalid results. If the source is faint and close to the sensitivity limit of the IRAC frames, there can be photometry dropouts when the source becomes too faint to photometer during certain parts of the lightcurve. However, in most cases, 95-100\% of the BCDs yield valid photometry in this step.

After collecting the BCD photometry, two additional steps are performed to clean the lightcurve data. First, a check of the source positions is made in the extracted data. During these relatively short observations, the path of the NEO on the plane of the sky can be approximated by a linear or in some cases a quadratic function. The position of the source as a function of time is fit with a linear function in both RA and Dec, and the deviation from the fit is calculated for each data point. For a few cases where a long lightcurve was obtained, this was switched to a quadratic function when it was apparent the linear fit was not sufficient. We then compare each point to the position predicted by the fitted function, and reject those data points with deviations greater than about one pixel (1\farcs2). This rejects points that were affected by cosmic rays or other array artifacts that caused the source position and photometry to be affected.

The second step is to determine the noise level in the lightcurve and exclude photometry that exceeds a cutoff value, in order to reject photometry affected by cosmic rays or other effects like incomplete background subtraction. Since the source is likely variable, we must separate out the measurement noise from the source variation. The noise depends not only on the instrumental parameters such as integration time but also on the details of the background field and subtraction process. We therefore estimate the noise in the photometry by calculating for each data point the standard deviation including the two points immediately preceding and following it (5 points in total). This is determined for points 3 through N-2 in the lightcurve, and the median of these values is taken to be the estimate of the measurement noise. We also determine the median value of the nearest 5 lightcurve points, and calculate the difference between the data point and this local median value. If it differs by more than 3$\times$ the noise estimate, then it is rejected from the lightcurve. This process is fairly robust and works well in most cases, but it assumes that the source is slowly changing during the course of 5 frames. Also, in some cases there are larger gaps in the lightcurve which can cause issues with this method. In these cases, we adjusted the noise estimate value slightly to allow more points to be declared valid. In most cases, >90\% of the lightcurve points pass all of these checks and appear in the lightcurve. Plots of our sample of lightcurves showing periodicity are shown in Figures \ref{LSplotsFull} -- \ref{sineplots}.

\section{Results and Discussion}\label{periods}

\subsection{Period and Amplitude Derivation}

We visually analyzed the set of lightcurves we reduced to search for apparent periodicities. We identified 38 NEOs that had obvious periodicity with a significant amplitude, and where the \Sp\ data apparently covered a large fraction of the rotational period, or the lightcurve appeared close to sinusoidal but the observation time did not fully cover one period. These are shown in Figures \ref{LSplotsFull} -- \ref{sineplots}, and analyzed in the sections below. 
\subsubsection{Lomb-Scargle (LS) Periodograms}
We analyzed the NEOs with lightcurves that appeared to sample more than half of a rotational period using the Lomb-Scargle (LS) algorithm \citep{1976Ap&SS..39..447L, 1982ApJ...263..835S}, as implemented by the NASA Exoplanet Science Institute Periodogram service\footnote{https://exoplanetarchive.ipac.caltech.edu/cgi-bin/Pgram/nph-pgram}. The rotational period of the NEO was then assumed to be double the period value of the highest peak of the periodogram for the 4.5 \micron\ flux density data of each lightcurve is reported in Table \ref{LSfits}. 

We estimated the 1-$\sigma$ uncertainty for the rotational period reported by generating simulated data for each lightcurve and running the LS analysis to derive periods for them. The simulated data was constructed in the following way: for each measured lightcurve, a smoothed curve was calculated using a running average of 5 data points. Then noise was added to the smoothed curve using a normal distribution with the same standard deviation estimated for the measurement. We then performed the same LS analysis for these simulated lightcurves and derived periods. We calculated the standard deviation of the period estimates for the simulated data, which we report as the error estimate of the period in Table \ref{LSfits}. The 1-$\sigma$ uncertainty value is contingent on the assumption that the highest peak in the LS analysis indeed represents the true period. A low value for the 1-$\sigma$ uncertainty signals high confidence in the precision of the period reported for the highest peak, but if the LS peak does not represent the true period, then the error would be much higher. 

For each of these objects, the 4.5 \micron\ flux density data were processed through a simple moving mean algorithm with a sample width of 6 observations. We then used the maximum and minimum values of the processed flux density data to calculate the amplitude, in magnitudes, for each object$'$s rotation. We used the associated photometric uncertainties to calculate the 1-$\sigma$ uncertainty for each amplitude. These values are reported in Table \ref{LSfits}.

\subsubsection{Sine Fits}\label{sinefitsection}
\begin{figure*}\label{sineplots}
\centering
\includegraphics[height=0.221\textwidth]{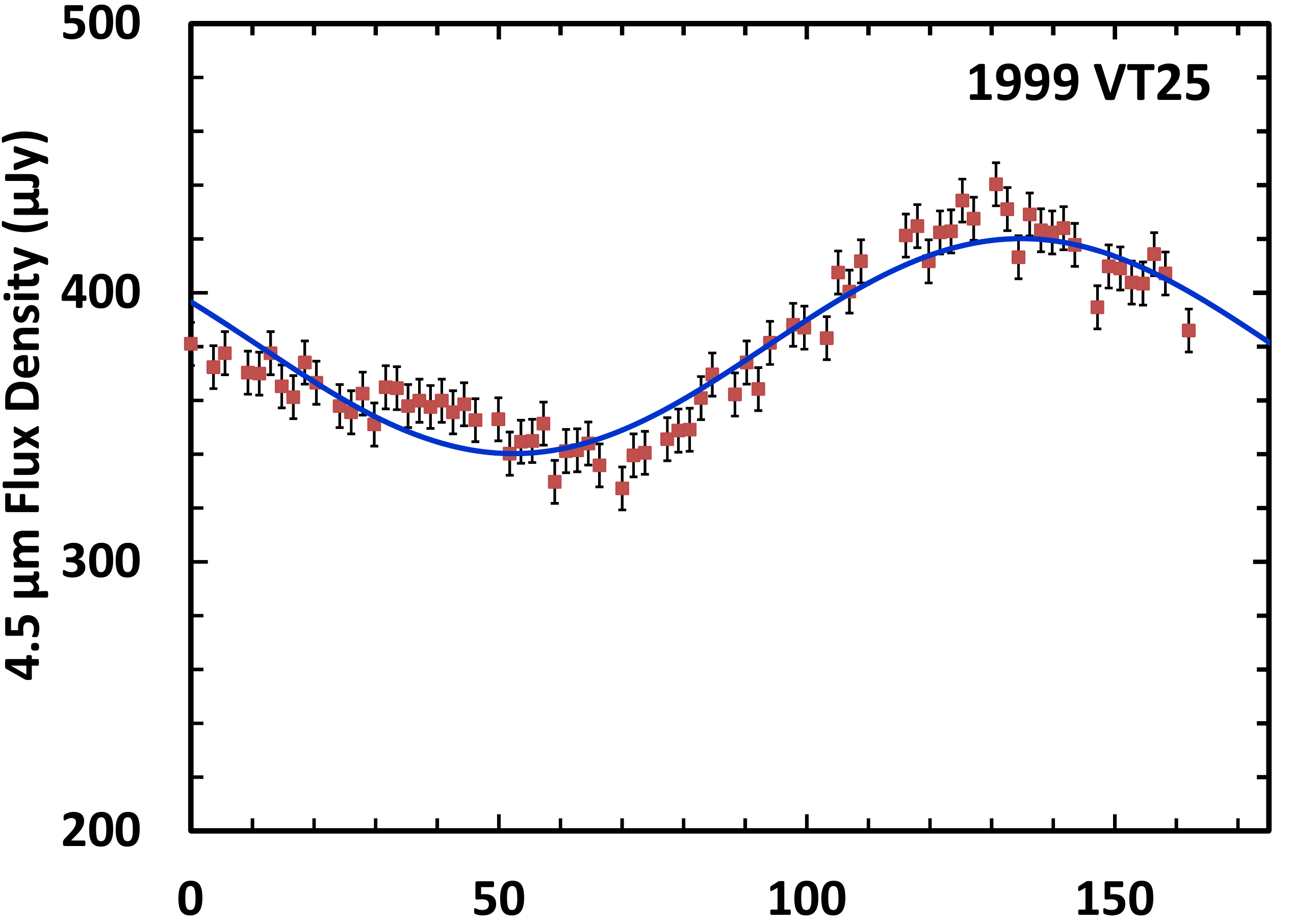}
\includegraphics[height=0.221\textwidth]{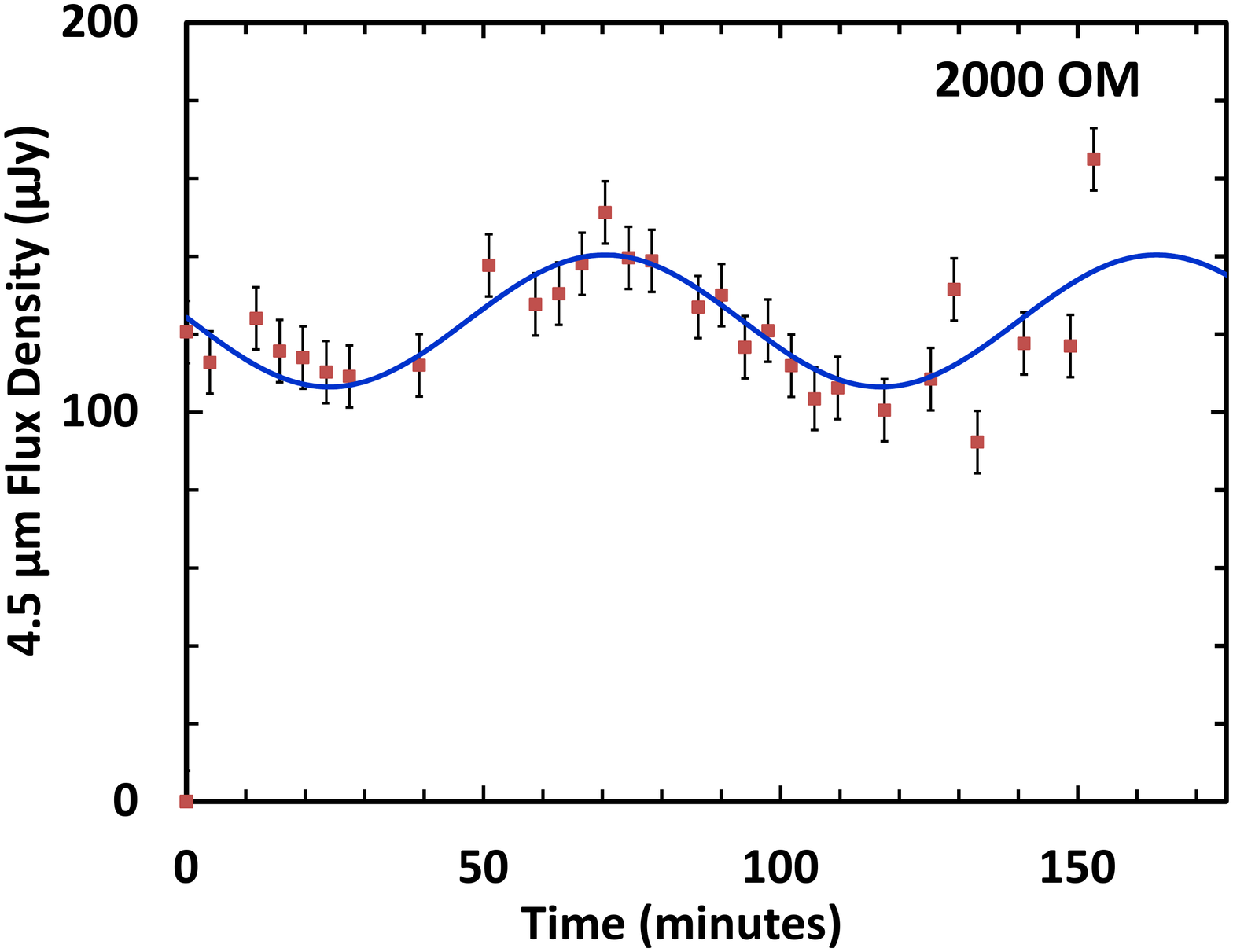}
\includegraphics[height=0.221\textwidth]{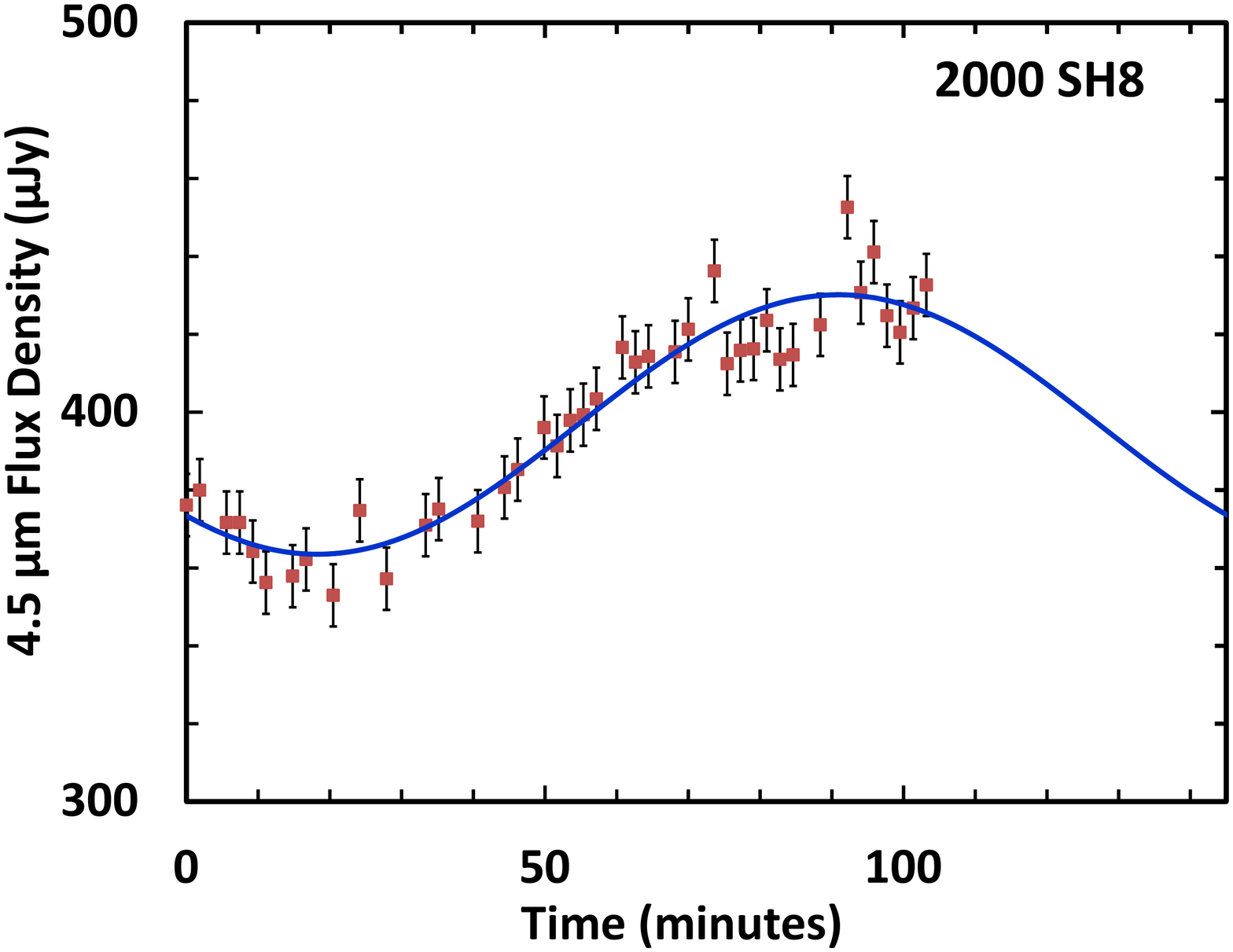}\\ \vskip 5pt
\includegraphics[height=0.221\textwidth]{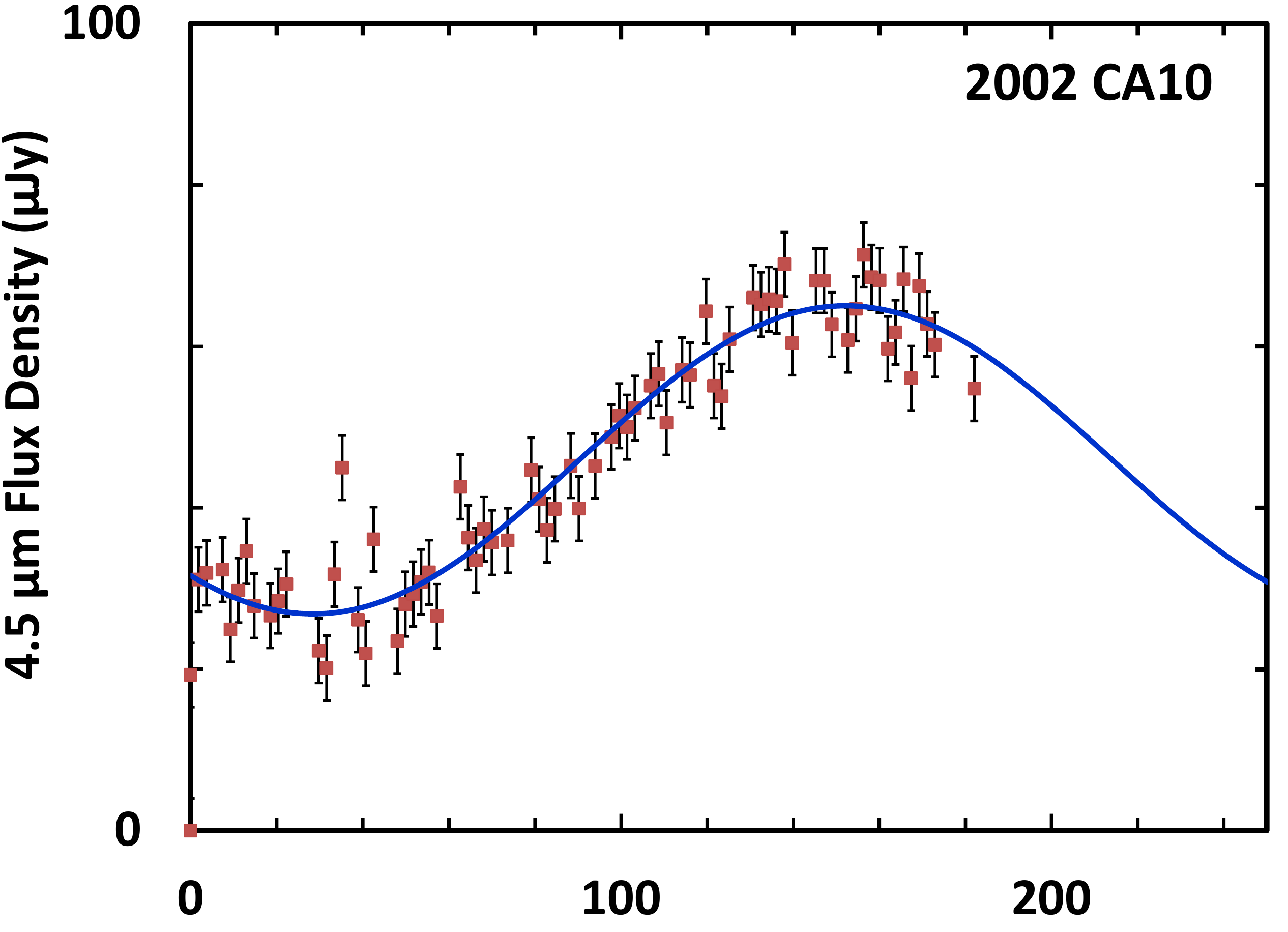}
\includegraphics[height=0.221\textwidth]{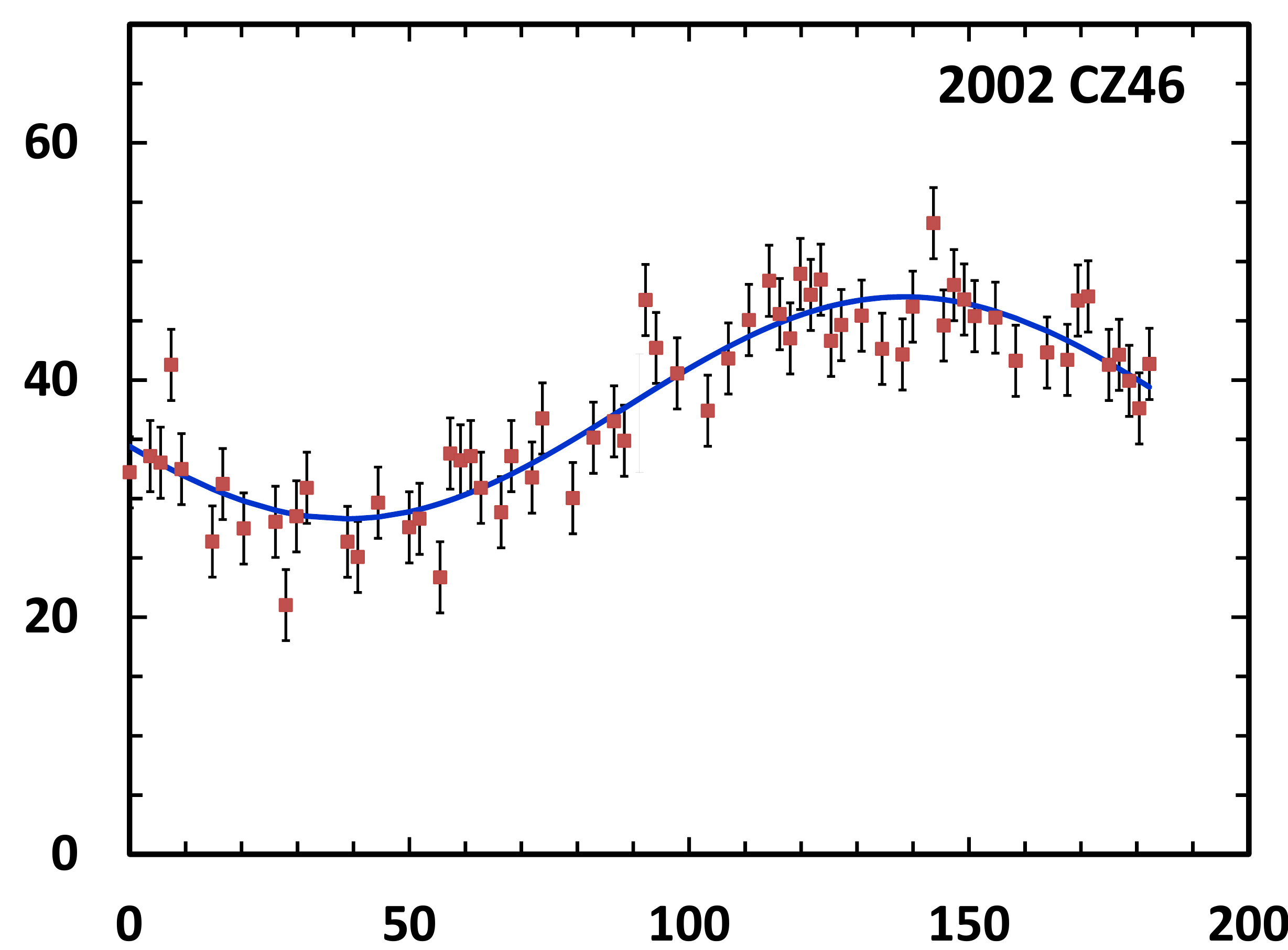}
\includegraphics[height=0.221\textwidth]{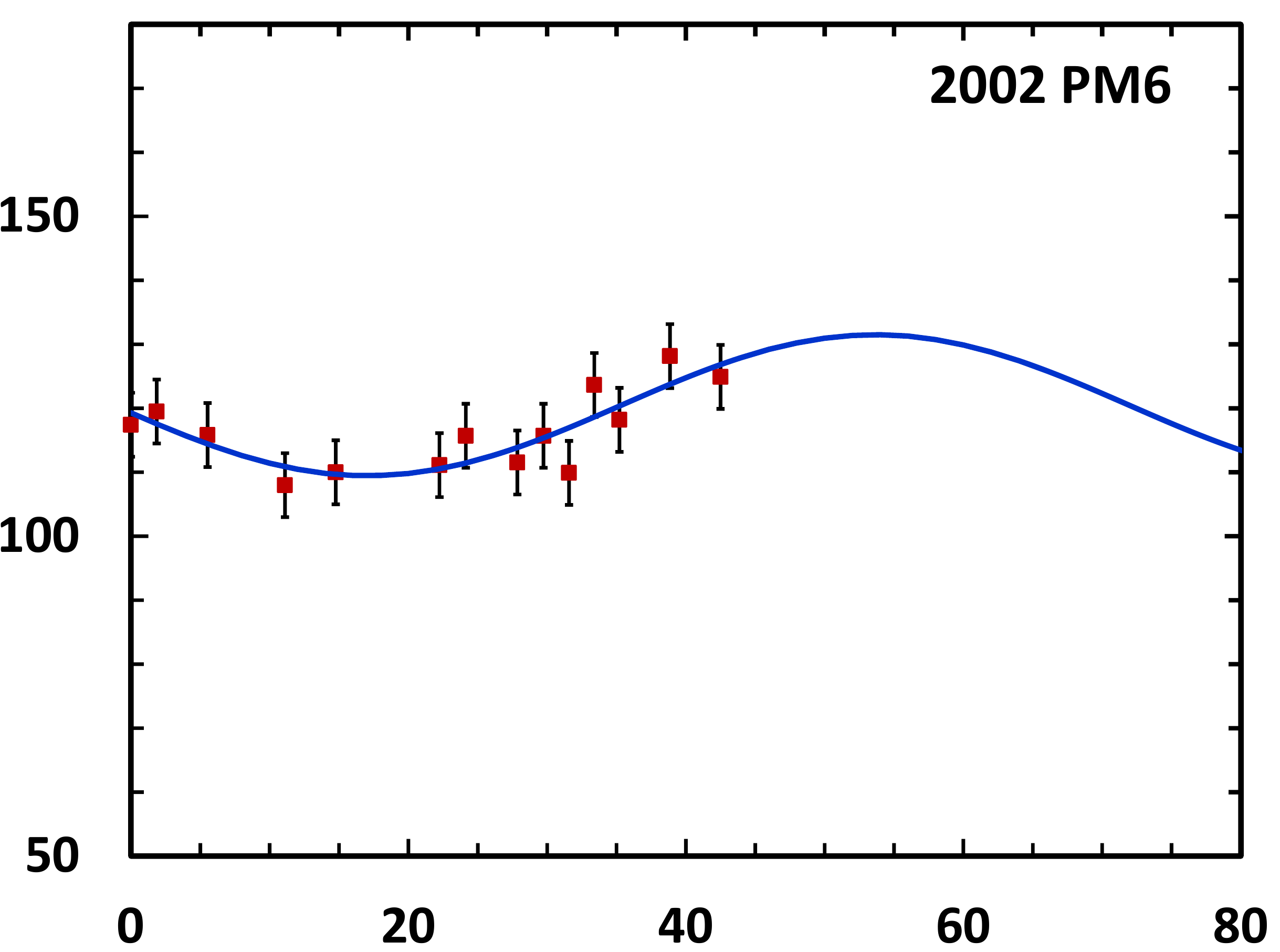}\\ \vskip 5pt
\includegraphics[height=0.221\textwidth]{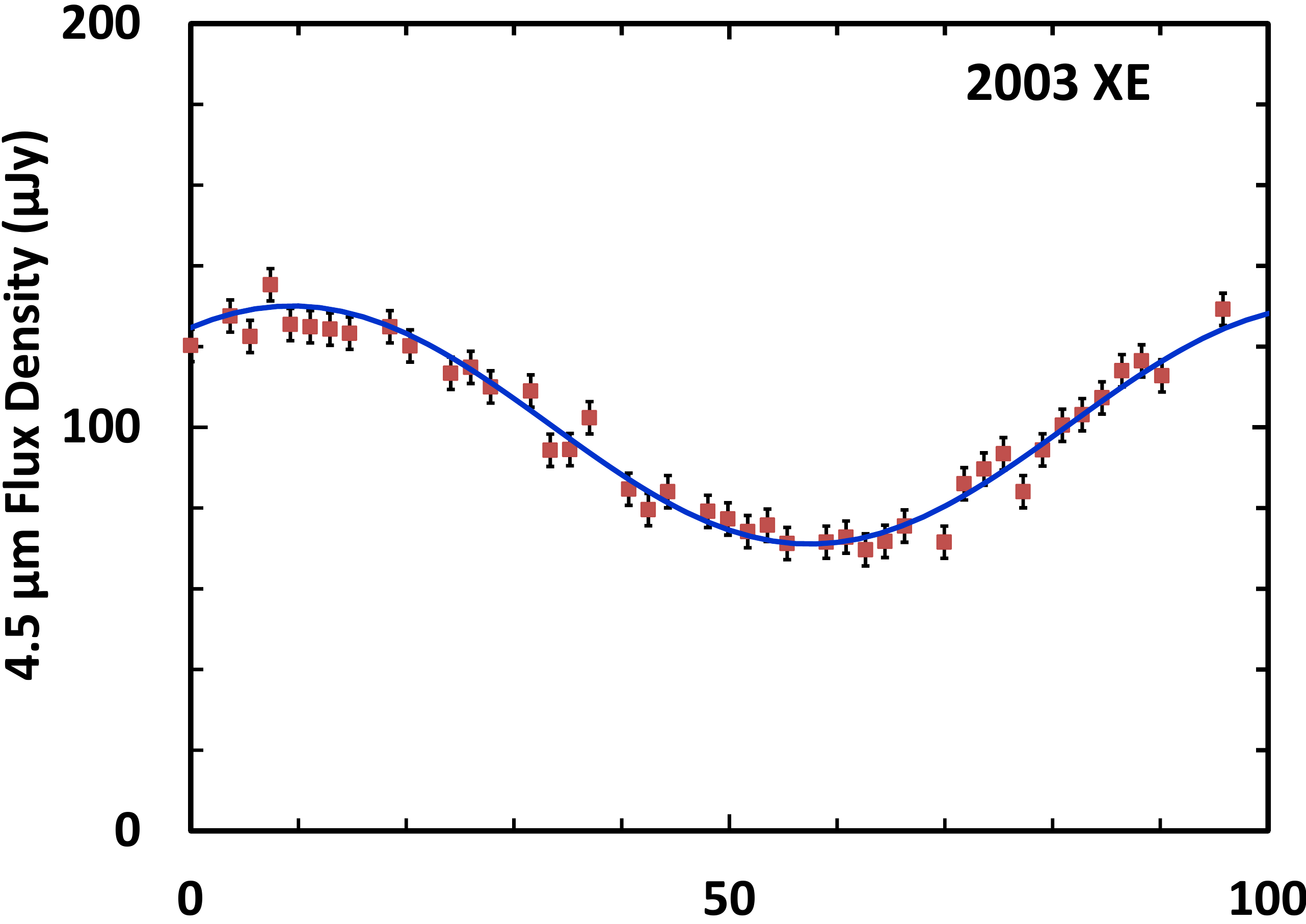}
\includegraphics[height=0.221\textwidth]{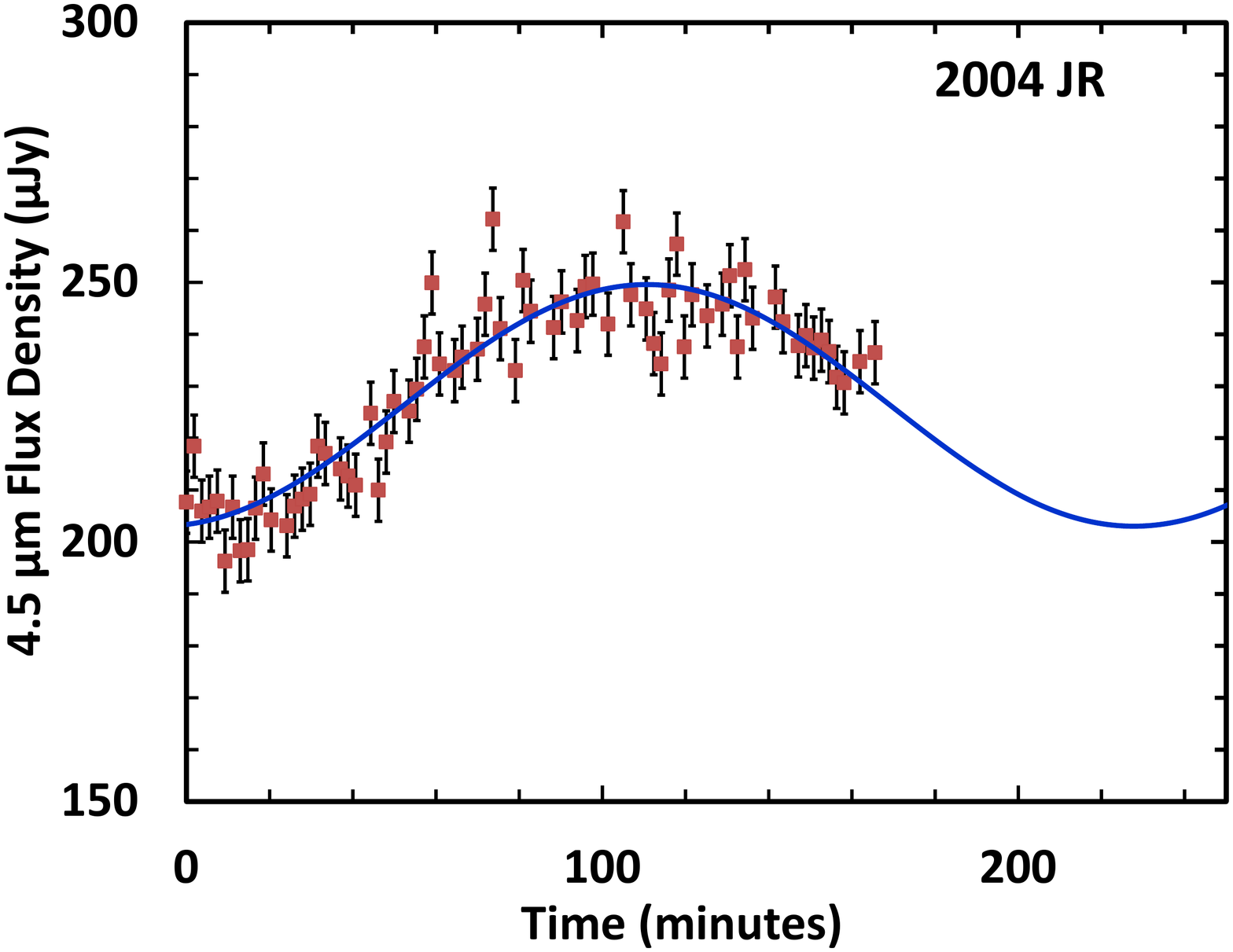}
\includegraphics[height=0.221\textwidth]{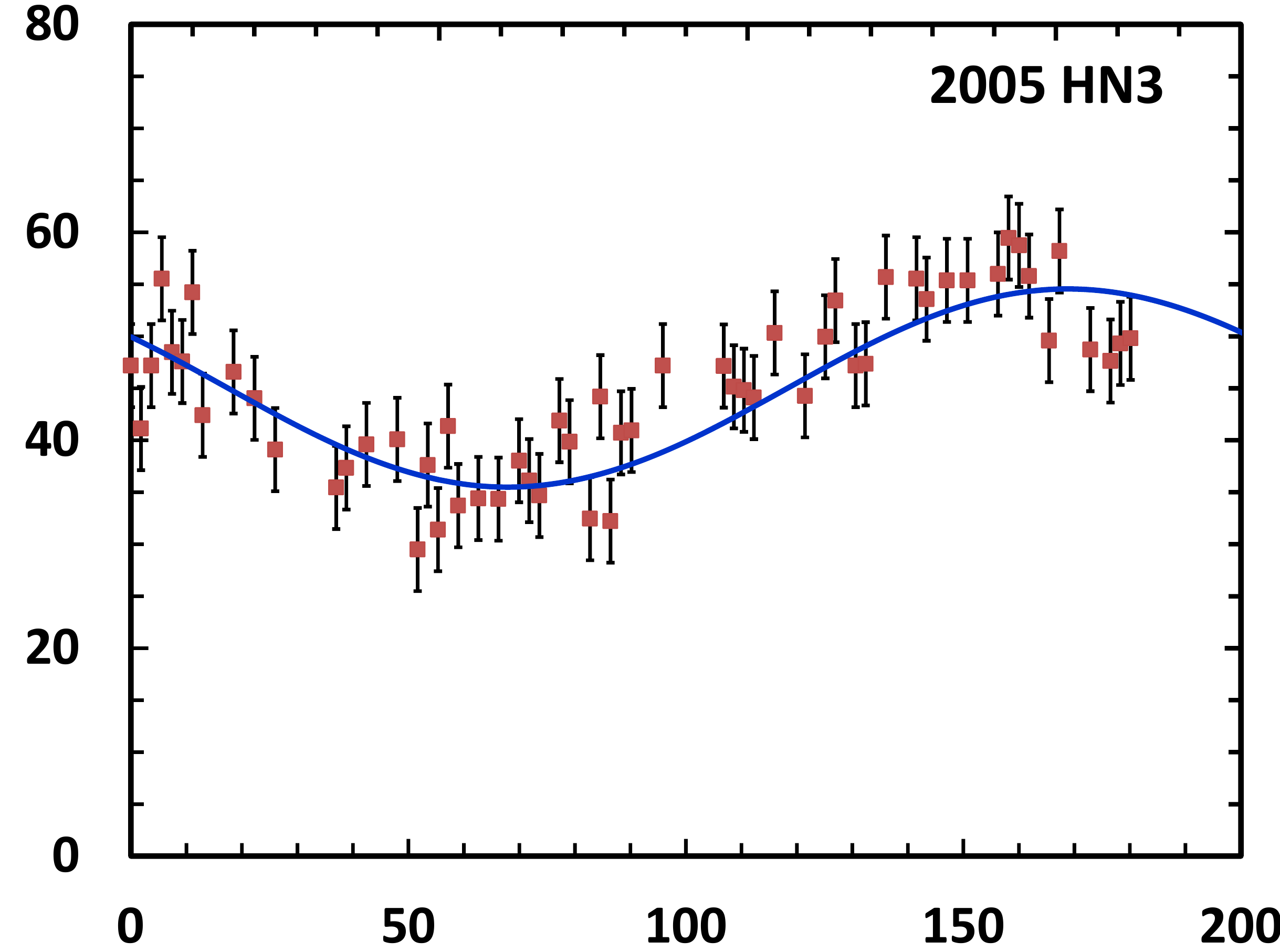}\\ \vskip 5pt
\includegraphics[height=0.221\textwidth]{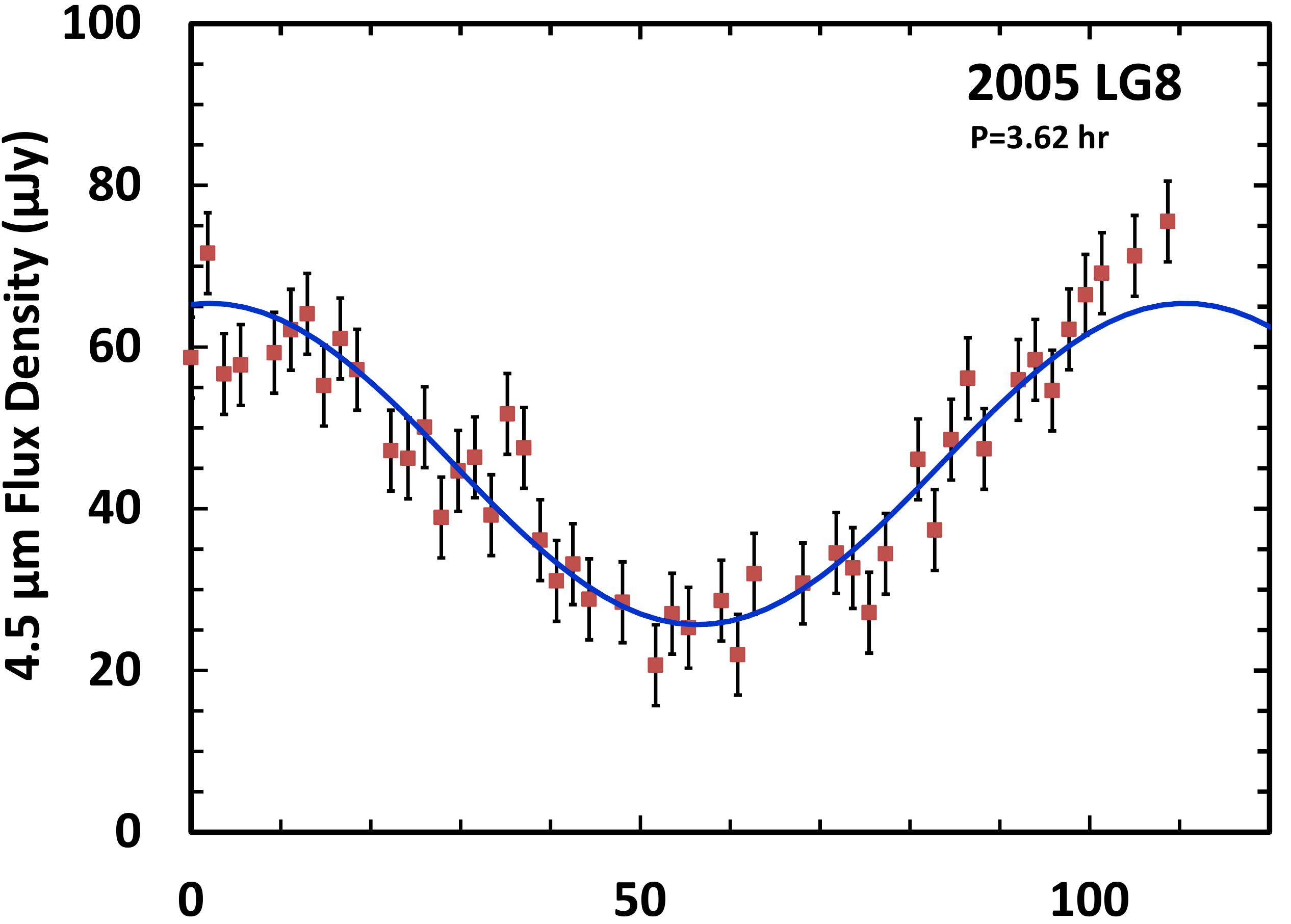}
\includegraphics[height=0.221\textwidth]{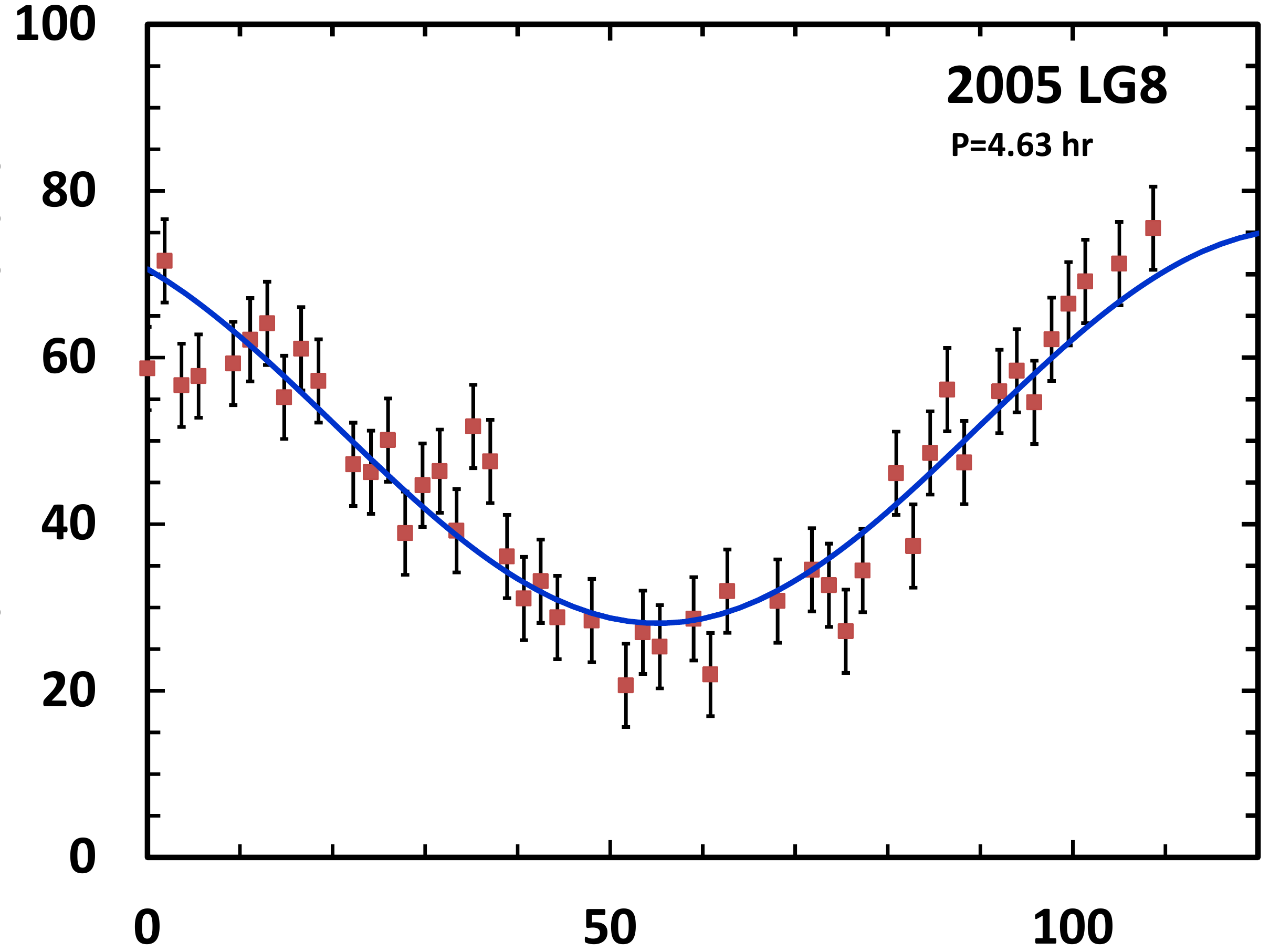}
\includegraphics[height=0.221\textwidth]{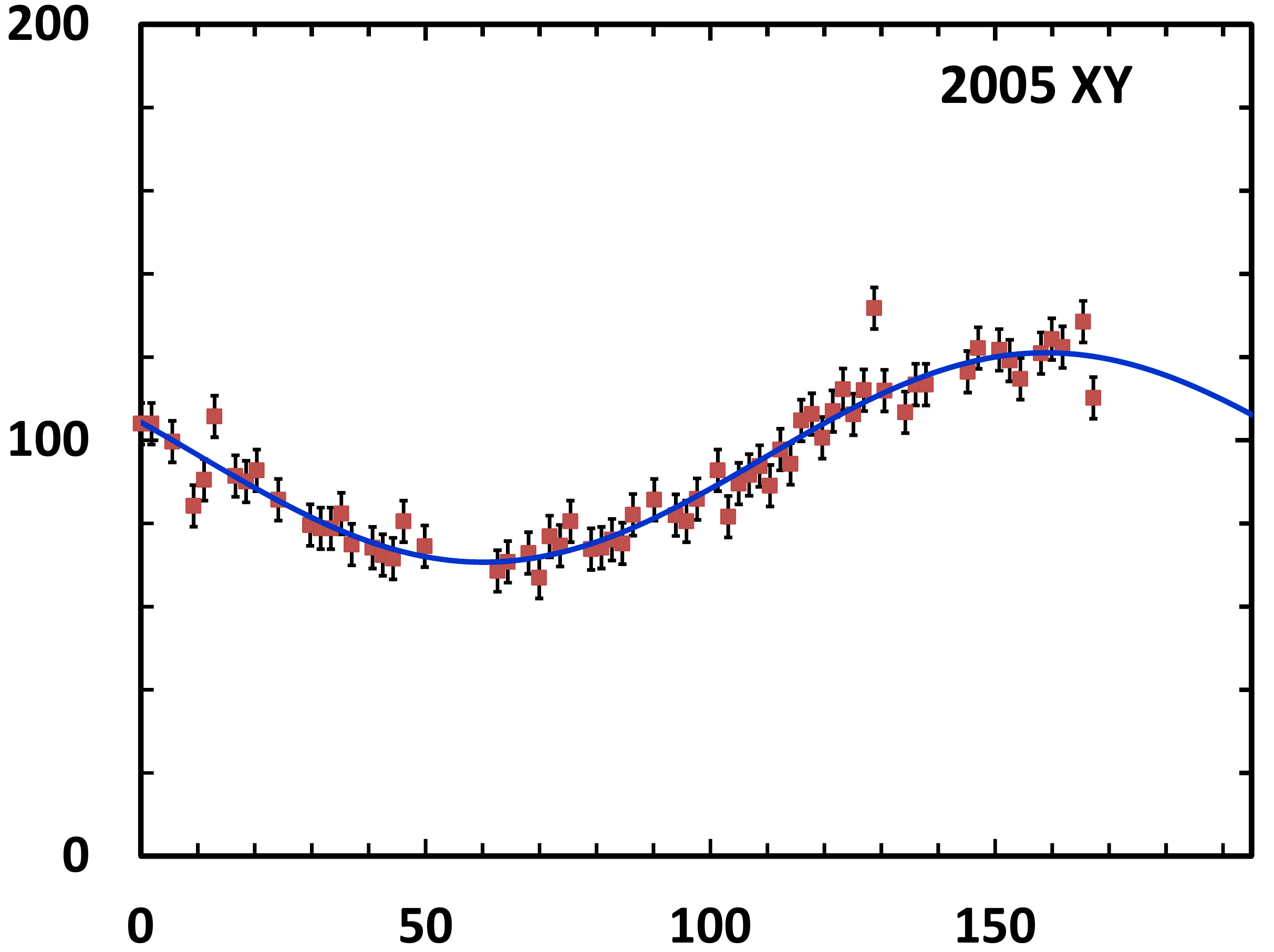}\\ \vskip 5pt
\includegraphics[height=0.237\textwidth]{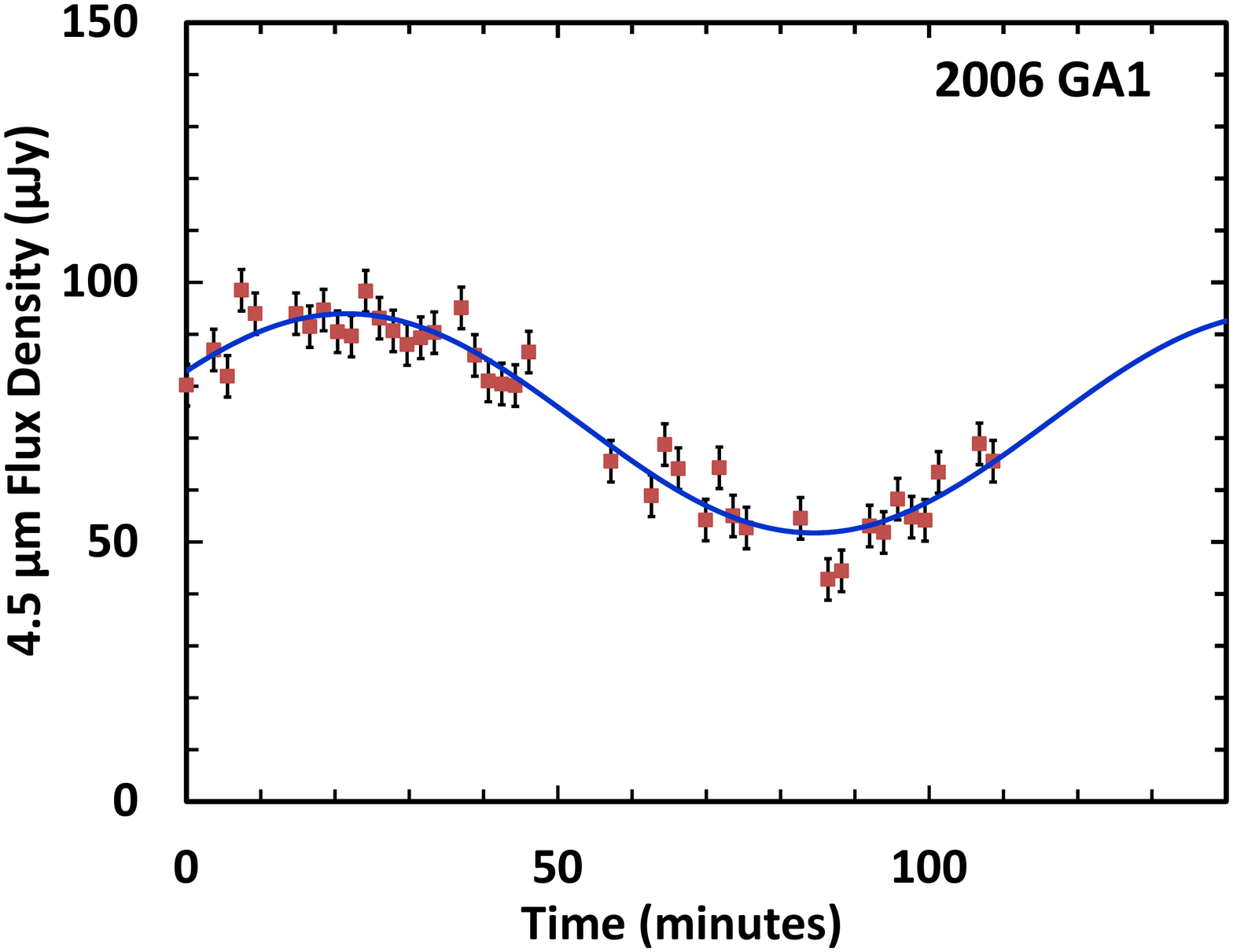}
\includegraphics[height=0.237\textwidth]{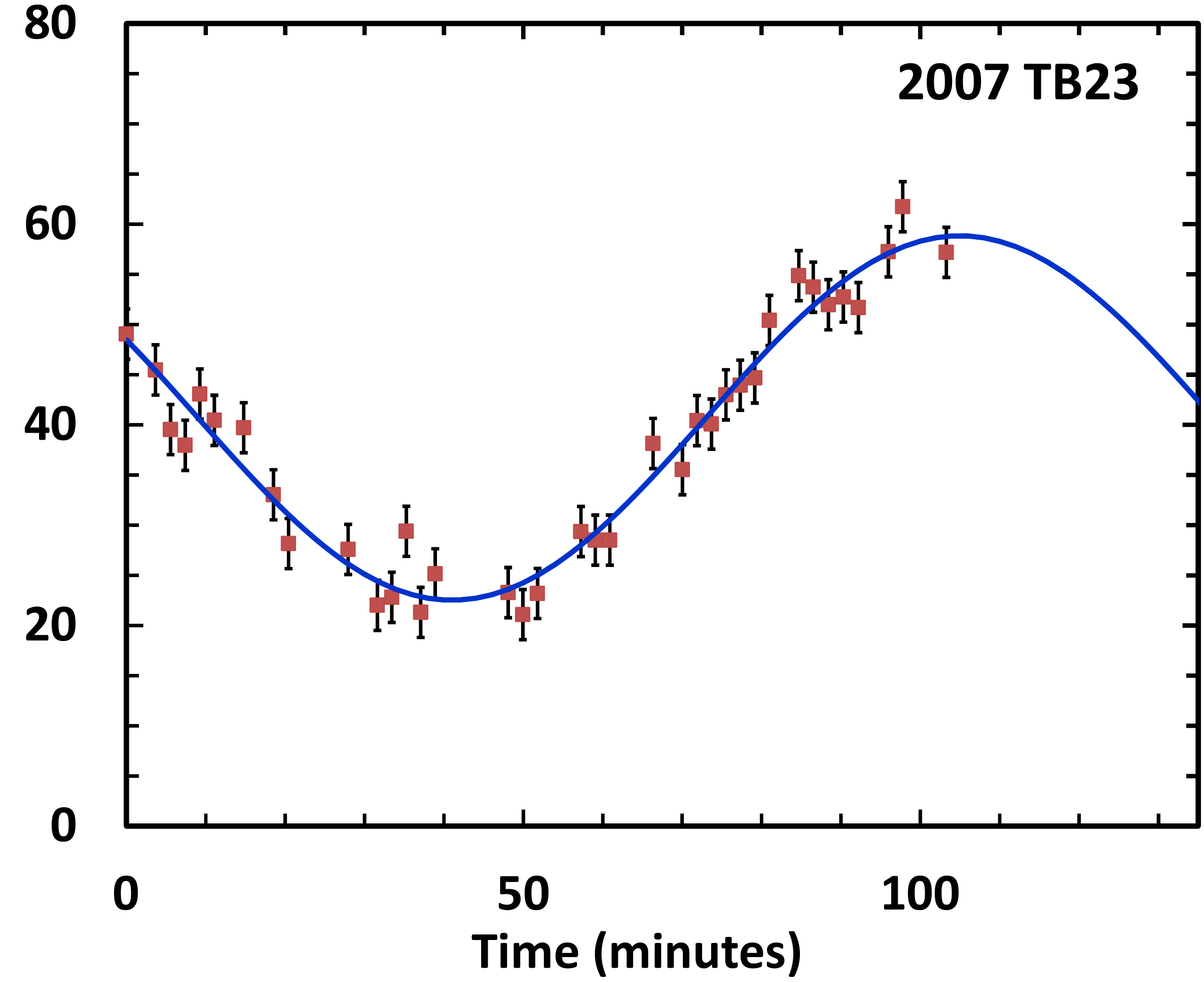}
\includegraphics[height=0.237\textwidth]{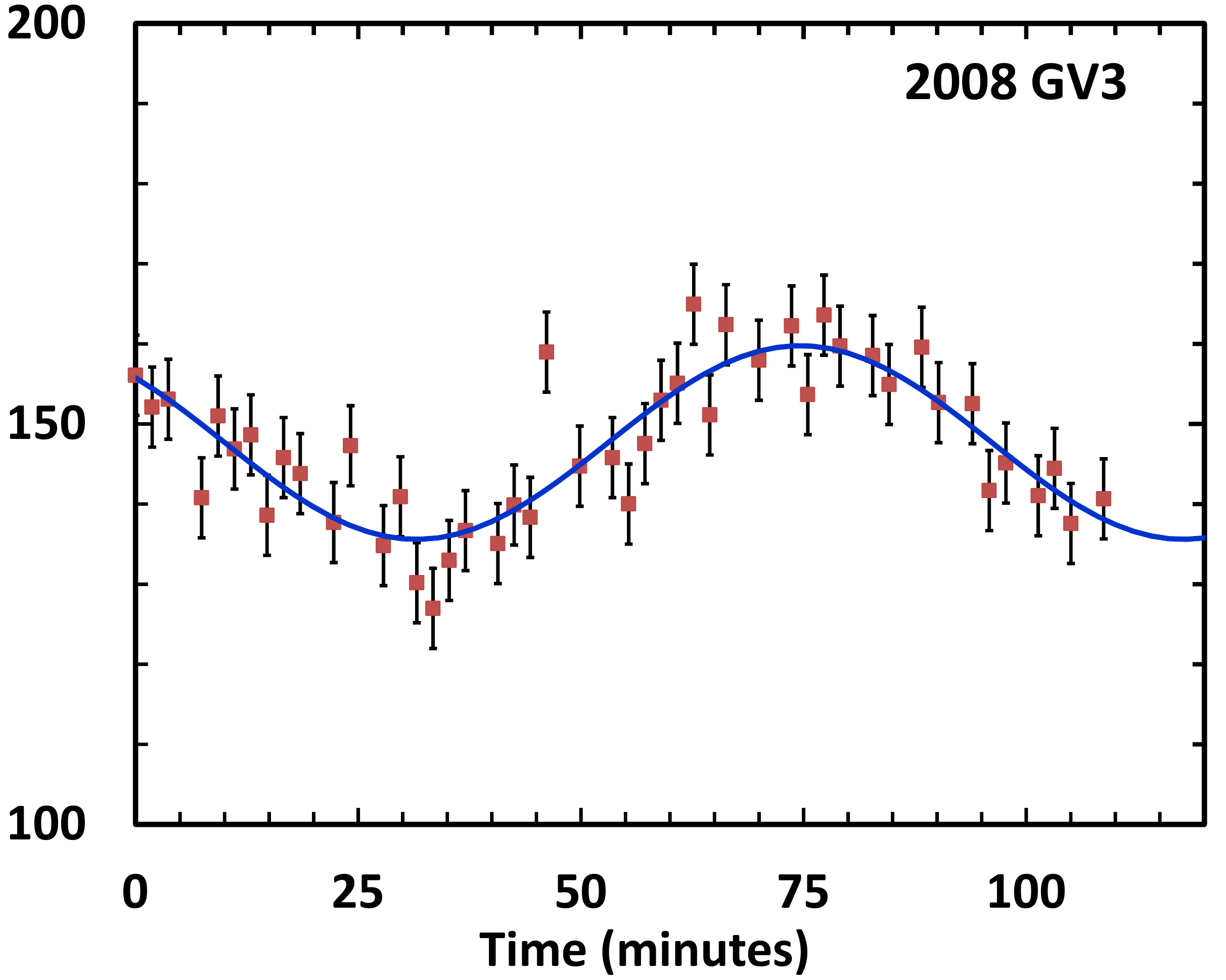}
\caption{Plots of \Sp\ lightcurves where less than one rotational period was observed, fit with sine functions.  The blue line is the sine fit to the data, with the parameters given in Table \ref{sinefits}. }
\end{figure*}

To obtain a lower limit on the rotation period for the NEOs with lightcurves that indicated a long sinusoidal rotation period relative to the observation window, as well as to corroborate the period estimates for 6 of the NEOs analyzed with the periodogram that exhibited sinusoidal variation, we fit a sinusoidal function to the 4.5 \micron\ flux density data for each object using a nonlinear least squares method, with all data points equally weighted. The periods reported in the table are the extrapolated rotational periods of the NEO, assuming that the full rotational light curve is a bimodal sine function with the fitted period. The uncertainties were  estimated by simulating datasets with the sine function determined from the observations of each object, sampled at the same time intervals as the observation but with simulated flux data with random errors having the same $\sigma$ as the observation. The estimated error was then taken to be the standard deviation of each parameter from the fits to the simulated data. Periods and amplitudes, and their respective 1-$\sigma$ uncertainties, are reported in Table \ref{sinefits}. The fitted sine curves are plotted with the data in Figure \ref{sineplots}.

\renewcommand{\thefigure}{\arabic{figure} (Cont.)}
\addtocounter{figure}{-1}
\begin{figure*}
\centering
\includegraphics[height=0.221\textwidth]{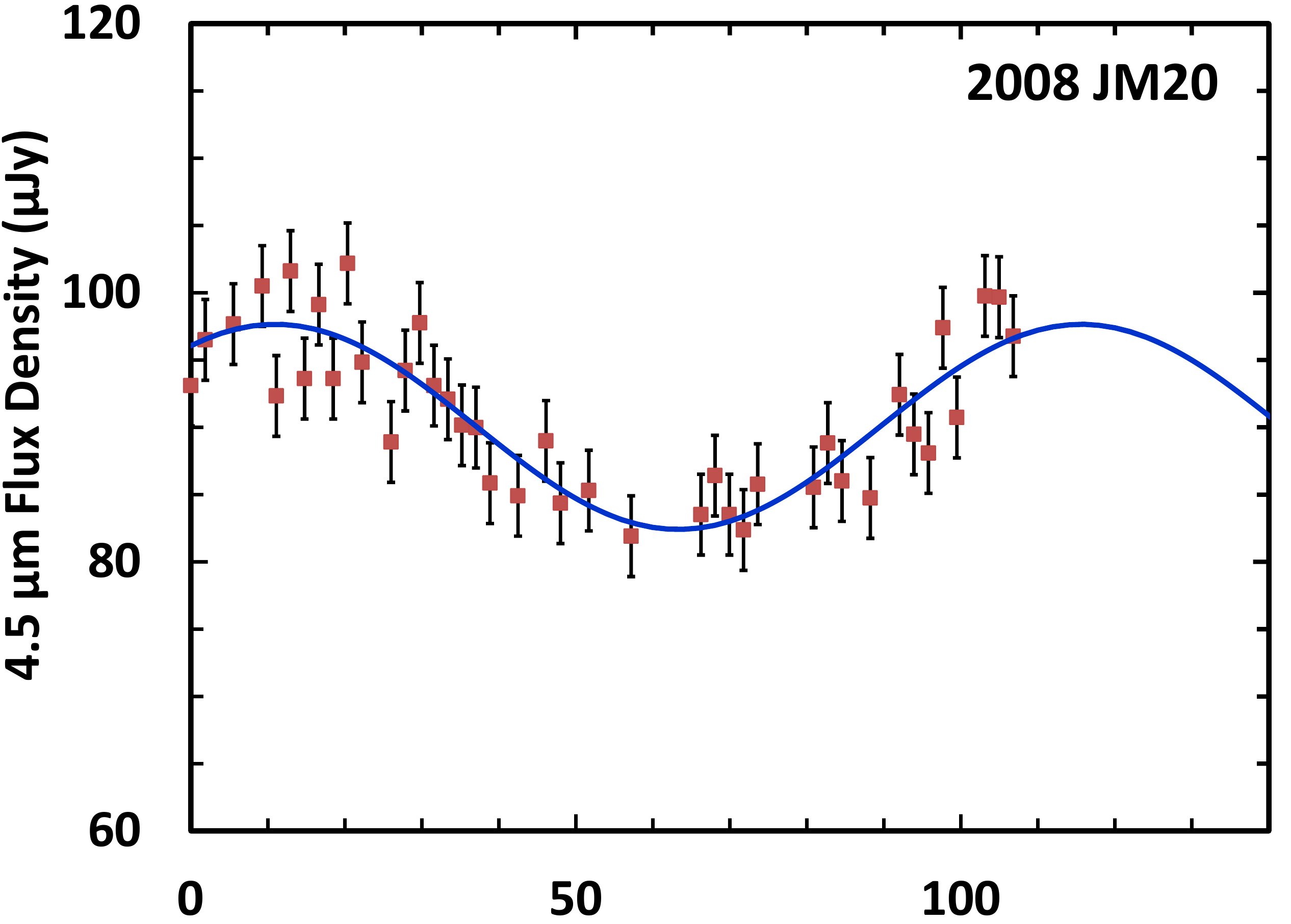}
\includegraphics[height=0.221\textwidth]{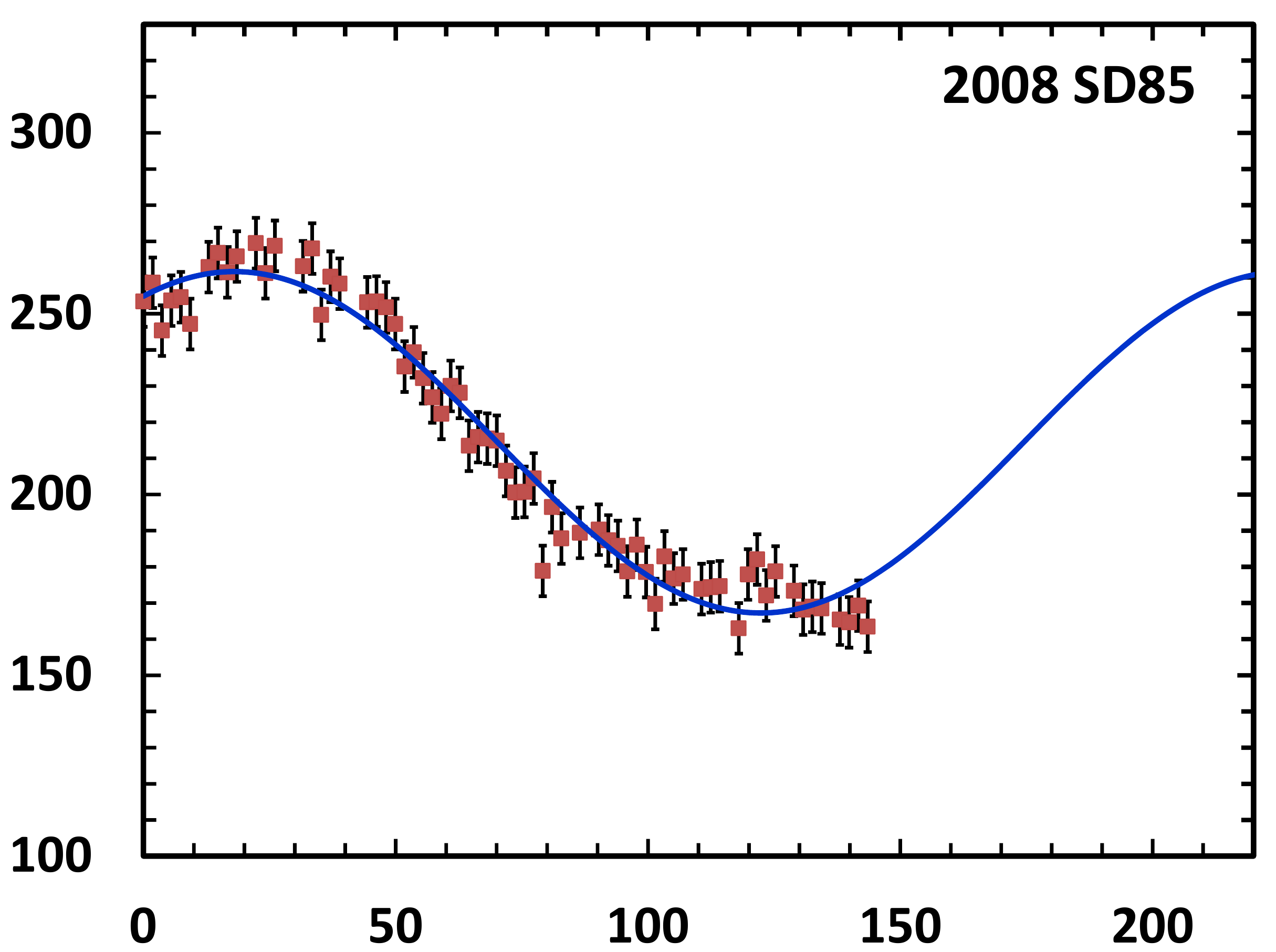}
\includegraphics[height=0.221\textwidth]{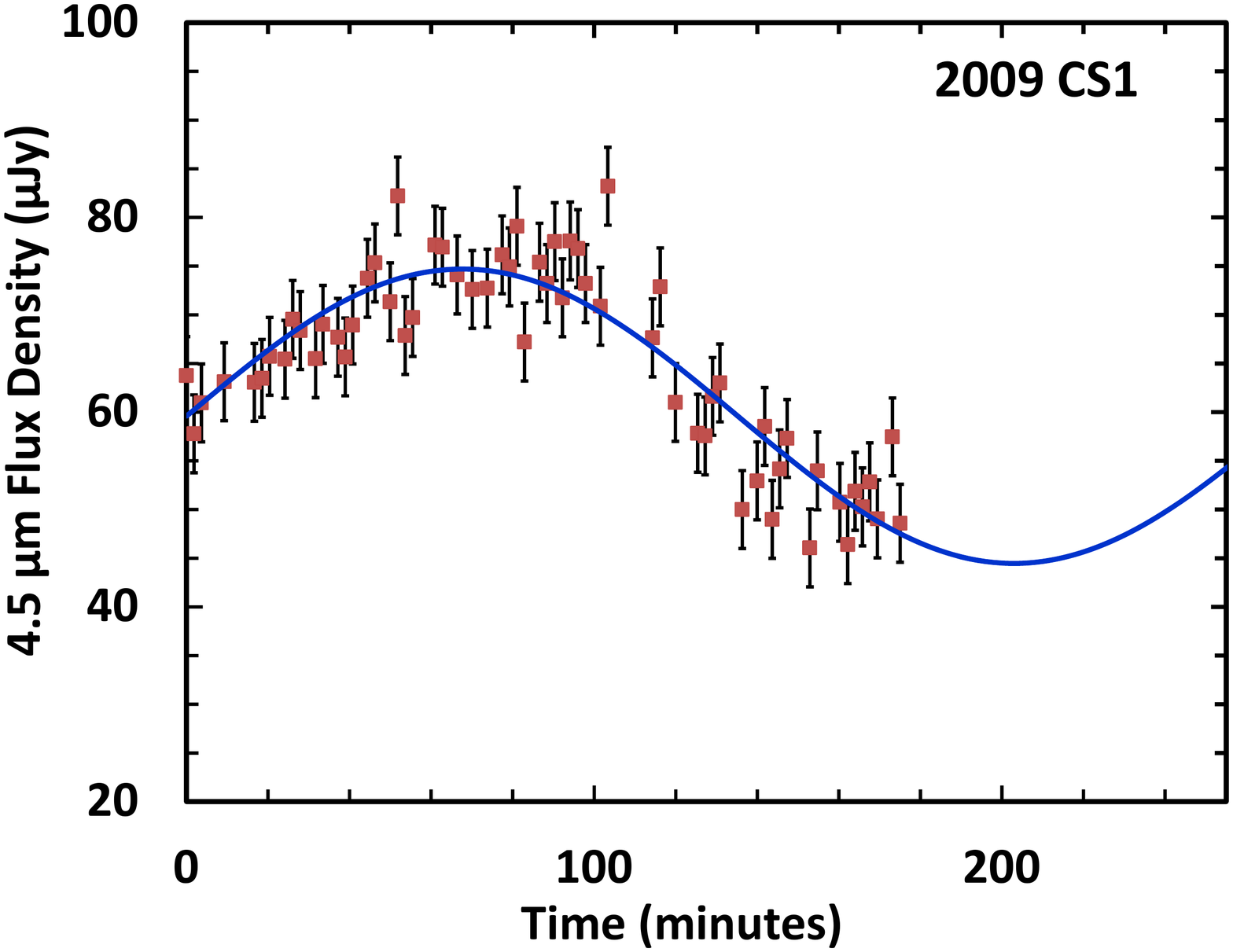}\\ \vskip 5pt
\includegraphics[height=0.2212\textwidth]{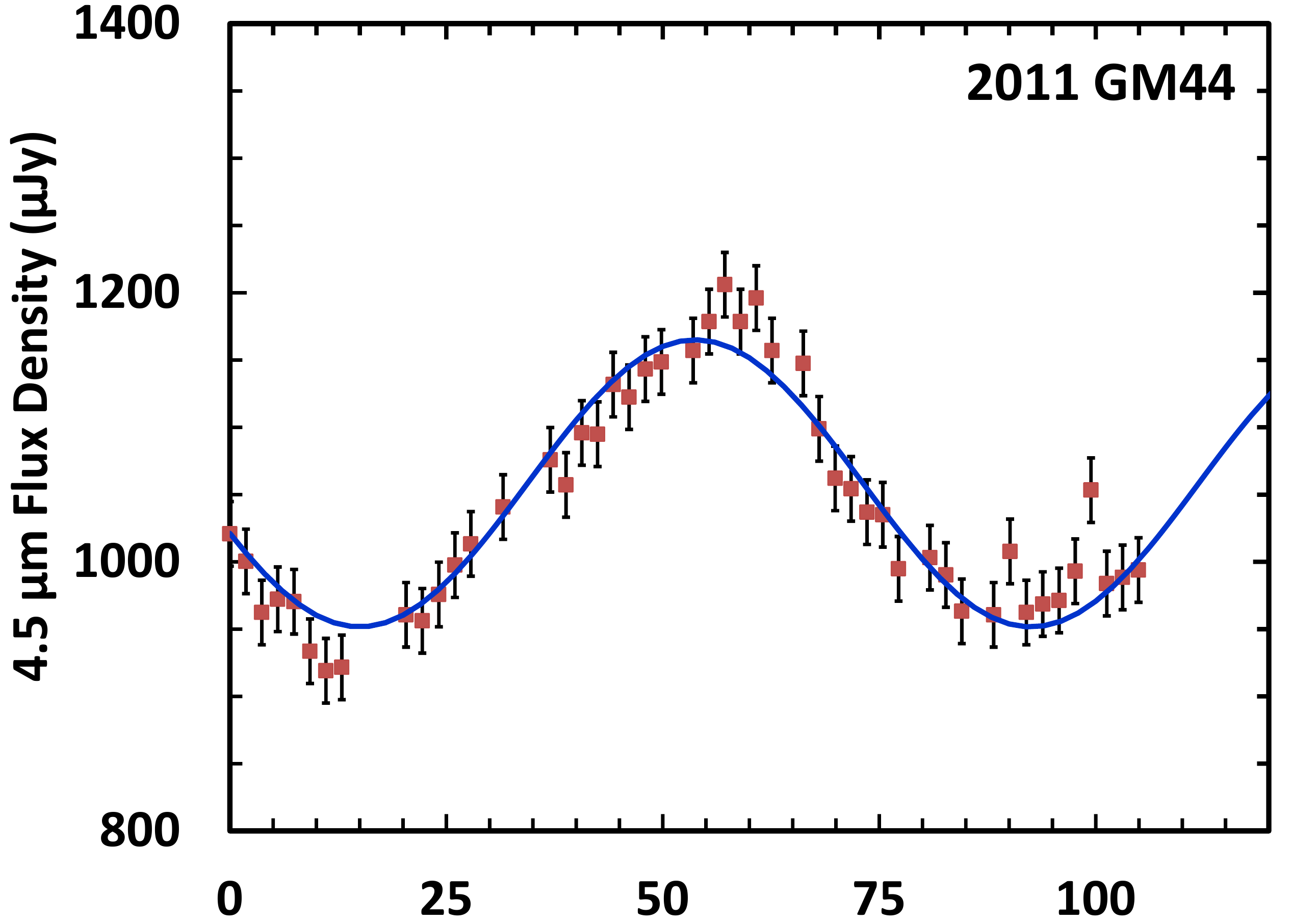}
\includegraphics[height=0.221\textwidth]{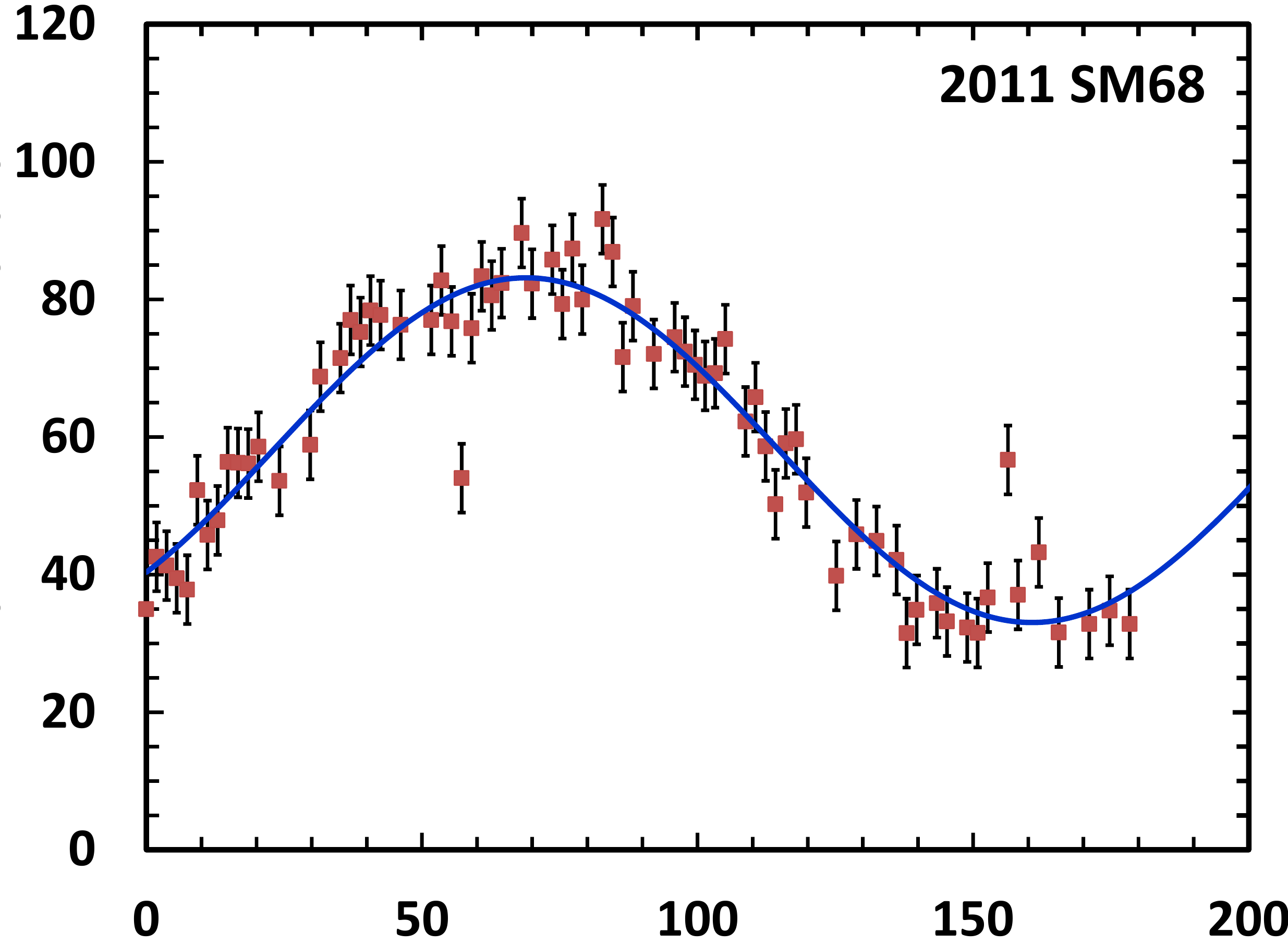}
\includegraphics[height=0.221\textwidth]{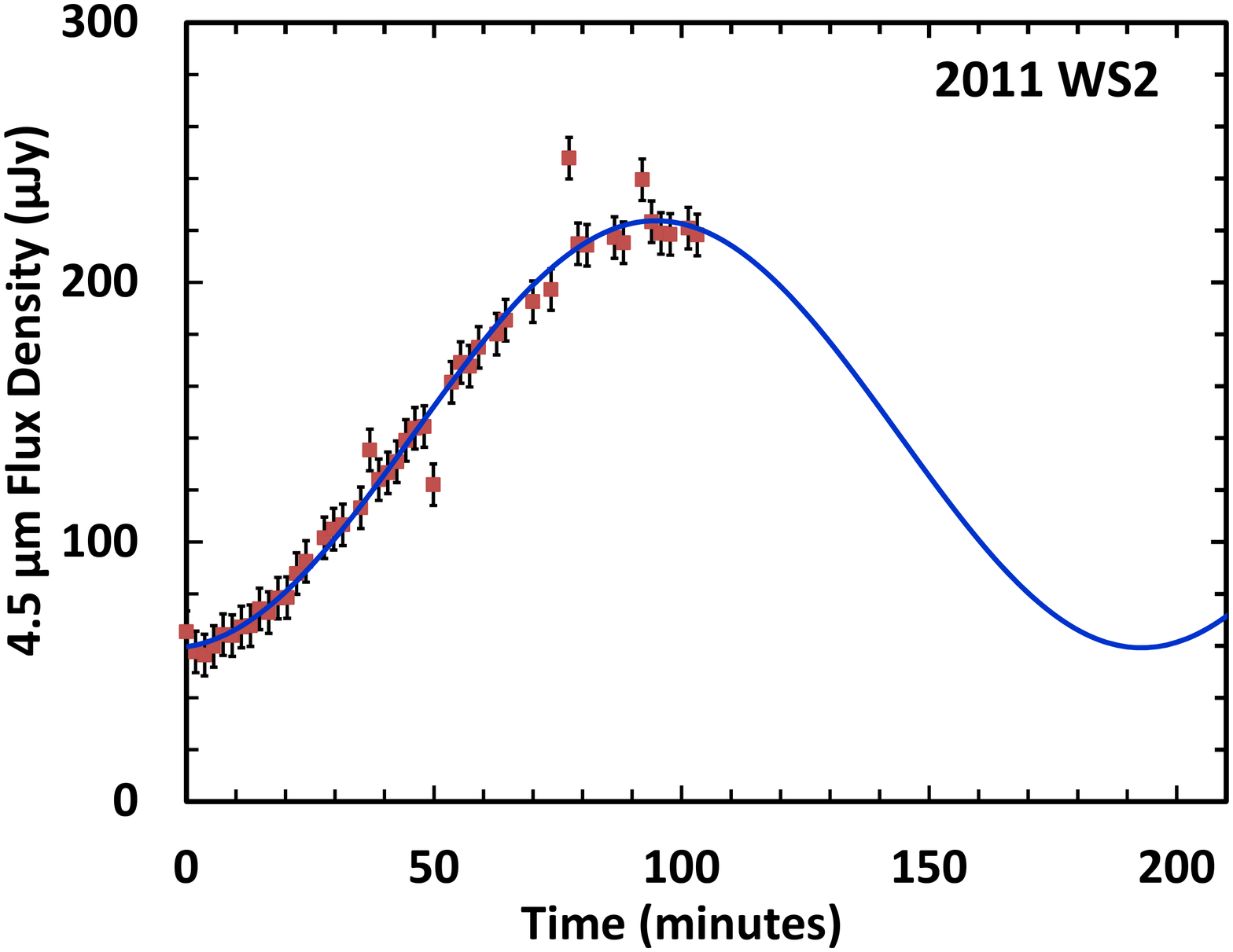}\\ \vskip 5pt
\includegraphics[height=0.24\textwidth]{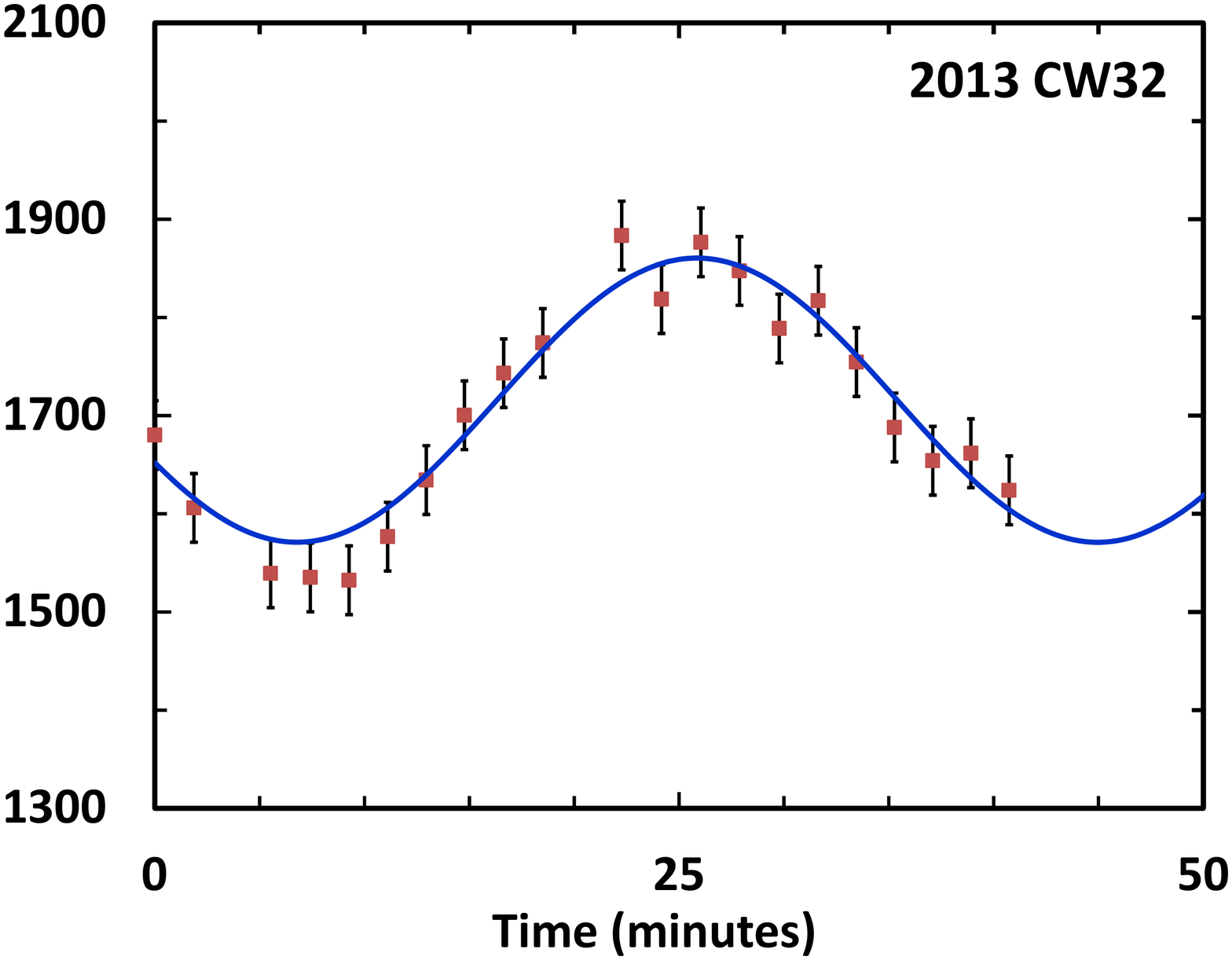}
\qquad\includegraphics[height=0.24\textwidth]{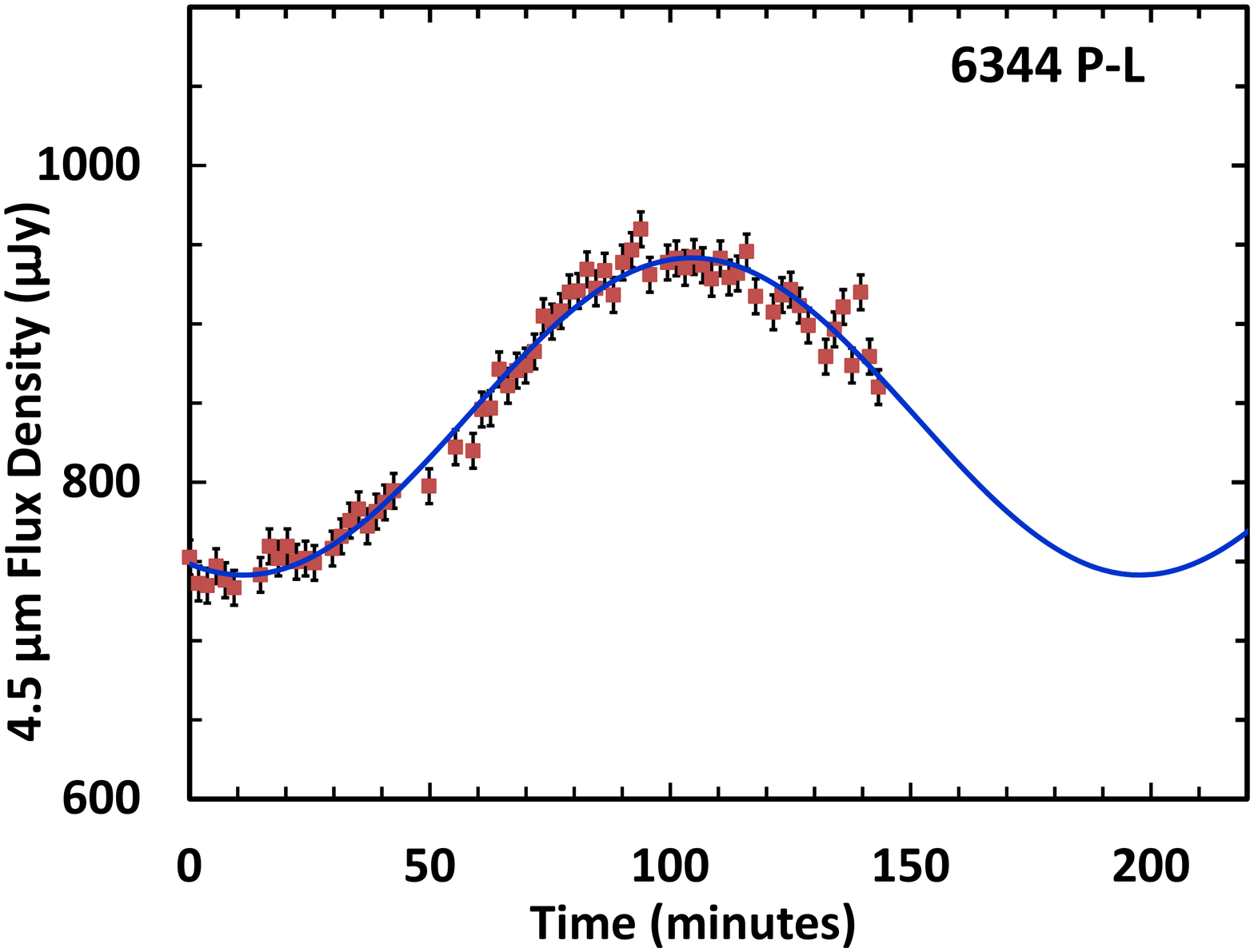}
\caption{}
\end{figure*}
\renewcommand{\thefigure}{\arabic{figure}}

\begin{deluxetable*}{lrccrcc}
\tablecaption{Sinusoidal Fits of Partial NEO Lightcurves \label{sinefits}}
\tablecolumns{7}
\tablewidth{0pt}
\tablehead{
& & & HMJD &Lightcurve &Lower Limit to \\
\colhead{Object} & \colhead{AORID} &
\colhead{UTC start time\tablenotemark{a}} &
\colhead{start time\tablenotemark{a}} &
\colhead{Duration} &
\colhead{Rotation Period} & \colhead{Amplitude} \\
\colhead{} & \colhead{} & \colhead{(YYYY-MM-DD hh:mm:ss)} & \colhead{(d)} & \colhead{(minutes)} &
\colhead{(minutes)} & \colhead{(mag)}
}
\startdata
1999 VT25 & 52457728 & 2016-03-10\enspace00{:}33{:}20 & 57457.0237318 & 162.0 & 329.6 $\pm$ 0.64 & 0.228 $\pm$ 0.002 \\
2000 OM & 44166656 & 2011-08-28\enspace20{:}20{:}08 & 55801.8479200 & 152.7 & 185.6 $\pm$ 0.30 & 0.299 $\pm$ 0.001 \\
2000 SH8 & 58817280 & 2016-06-08\enspace12{:}37{:}52 & 57547.5268862 & 103.2 & 290.6 $\pm$ 0.12 & 0.182 $\pm$ 0.001 \\
2002 CA10 &52459008 & 2015-05-15\enspace15{:}10{:}54 & 57157.6331529 & 182.1 & 495.0 $\pm$ 3.0 & 0.956 $\pm$ 0.021\\
2002 CZ46 & 52462336 & 2015-07-01\enspace09{:}04{:}46 & 57204.3788919 & 182.2 & 399.0 $\pm$ 1.4 & 0.552 $\pm$ 0.005\\
2002 PM6 & 52454656 & 2015-09-26\enspace22{:}30{:}04 & 57291.9381325 & 42.5 & 146.4 $\pm$ 1.4 & 0.200 $\pm$ 0.011 \\
2003 XE & 52498944 & 2015-04-26\enspace03{:}01{:}09 & 57138.1263838 & 108.7 & 192.2 $\pm$ 4.6 & 0.323 $\pm$ 0.010 \\
2004 JR & 52387584 & 2015-02-27\enspace08{:}23{:}03 & 57080.3499218 & 165.6 & 468.2 $\pm$ 4.6 & 0.221 $\pm$ 0.006 \\
2005 HN3 & 52392448 & 2015-07-27\enspace14{:}28{:}50 & 57230.6039391 & 180.1 & 363.2 $\pm$ 6.8 & 0.480 $\pm$ 0.033 \\
2005 LG8 & 52481280 & 2015-05-11\enspace09{:}25{:}07 & 57153.3930231 & 108.7  & 217.0\tablenotemark{b} $\pm$ 3.0 & 1.027 $\pm$ 0.054\\
2005 XY & 52396544 & 2015-02-07\enspace11{:}18{:}06 & 57060.4714836 & 167.2 & 395.0 $\pm$ 7.6 & 0.588 $\pm$ 0.037 \\
2006 GA1 & 52460800 & 2015-05-04\enspace16{:}48{:}26 & 57146.7008871 & 108.6 & 251.8 $\pm$ 4.8 & 0.639 $\pm$ 0.065 \\
2007 TB23 & 52406272 & 2015-02-25\enspace04{:}27{:}32 & 57078.1863634 & 103.3 & 255.8 $\pm$ 1.1 & 1.046 $\pm$ 0.024\\
2008 GV3 & 52410112 & 2016-01-16\enspace12{:}07{:}41 & 57403.5059145 & 108.7 & 172.2 $\pm$ 0.4 & 0.179 $\pm$ 0.002 \\
2008 JM20 & 58815488 & 2016-08-03\enspace03{:}53{:}54 & 57603.1630151 & 108.7 & 209.4 $\pm$ 0.08 & 0.184 $\pm$ 0.001 \\
2008 SD85 & 61826048 & 2016-12-28\enspace11{:}34{:}16 & 57750.4827190 & 145.3 & 417.0 $\pm$ 0.88 & 0.486 $\pm$ 0.004 \\
2009 CS1 & 52414720 & 2015-10-25\enspace00{:}54{:}58 & 57320.0387609 & 180.4 & 541.0 $\pm$ 2.4 & 0.565 $\pm$ 0.009 \\
2011 GM44 & 52424704 & 2015-07-02\enspace06{:}21{:}57 & 57205.2658287 & 106.8 & 155.0 $\pm$ 0.6 & 0.220 $\pm$ 0.002 \\
2011 SM68 & 52428288 & 2015-02-01\enspace09{:}52{:}09 & 57054.4117981 & 182.0 & 367.0 $\pm$ 11 & 0.995 $\pm$ 0.084 \\
2011 WS2 & 52429568 & 2015-02-03\enspace02{:}31{:}17 & 57056.1056369 & 106.8 & 392.0 $\pm$ 1.8 & 1.453 $\pm$ 0.020 \\
2013 CW32 & 61845504 & 2017-05-13\enspace05{:}50{:}42 & 57886.2441315 & 42.6 & 76.4 $\pm$ 0.26 & 0.184 $\pm$ 0.002 \\
6344 P-L & 61869824 & 2017-05-27\enspace14{:}39{:}23 & 57900.6112606 & 145.1 & 373.0 $\pm$ 0.28&  0.259 $\pm$ 0.001\\
\enddata
\tablenotetext{a}{Time at the midpoint of the first frame of the observation.}
\tablenotetext{b}{Period is not consistent with prior measurements; see discussion in Section \ref{objectdiscuss}.}
\tablecomments{Columns: asteroid designations, \textit{Spitzer} Astronomical Observation Request identifier, observation start time in UT and heliocentric MJD, respectively, the observation duration, derived rotation period, and lightcurve amplitude (mag). The periods and amplitudes should be treated as lower limits.}
\end{deluxetable*}

\subsection{Discussion of Period-Fitting Results}\label{Presults}

The NEOs listed in Table \ref{LSfits} (1990~MF, 1990~UA, 1998~FF14, 1999~JE1, 2003~EO16, 2005~HC3, 2009~WD106, 2011~XA3, and 2005~XC) were found to have fully-sampled periods, assuming a bimodal lightcurve. While there is a high likelihood that the period estimates resulting from these well-sampled lightcurves are accurate, they should still be treated as lower bounds due to the possibility of multimodal distributions combined with photometric uncertainties that can make two independent lightcurve peaks indistinguishable from one another.

The NEOs 2002~PM6 and 2011~WS2 had, at most, 25 -- 30\% of their rotational period sampled. 2000~SH8, 2002~CA10, 2004~JR, 2005~XY, 2008~SD85, 2009~CS1, 1999~VT25, 2005~HN3, and 2011~SM68 had, at most, 35 --  50\% of their rotational period sampled. Their periods were estimated by fitting a sine function to the data as described in Section \ref{sinefitsection}, assuming they have symmetrical bimodal rotation curves. However, these period estimates should be treated strictly as lower bounds, as it is very possible that these objects have multimodal lightcurves which were not well-sampled.

Six of the NEOs with sinusoidal fits had lightcurves indicating total sampling near or greater than half a rotational period (2000~OM, 2003~XE, 2008~GV3, 2008~JM20, 2011~GM44, and 2013~CW32). These objects had, at most, a median of 0.6 sampled rotation periods. Thus, we analyzed these objects with the periodogram method in addition to the sinusoidal fitting one. The periodogram-based period estimates are all within 3-sigma of the sinusoidal fit estimates, indicating consistency between the two methods.

Four of the NEOs in our sample have previously measured rotational periods: 2005~LG8, 2008~UE7, 2011~XA3, and 2015~XC. We compare our measurements to the previous results for each object in Section \ref{objectdiscuss}.

\begin{figure}
\includegraphics[width=0.47\textwidth]{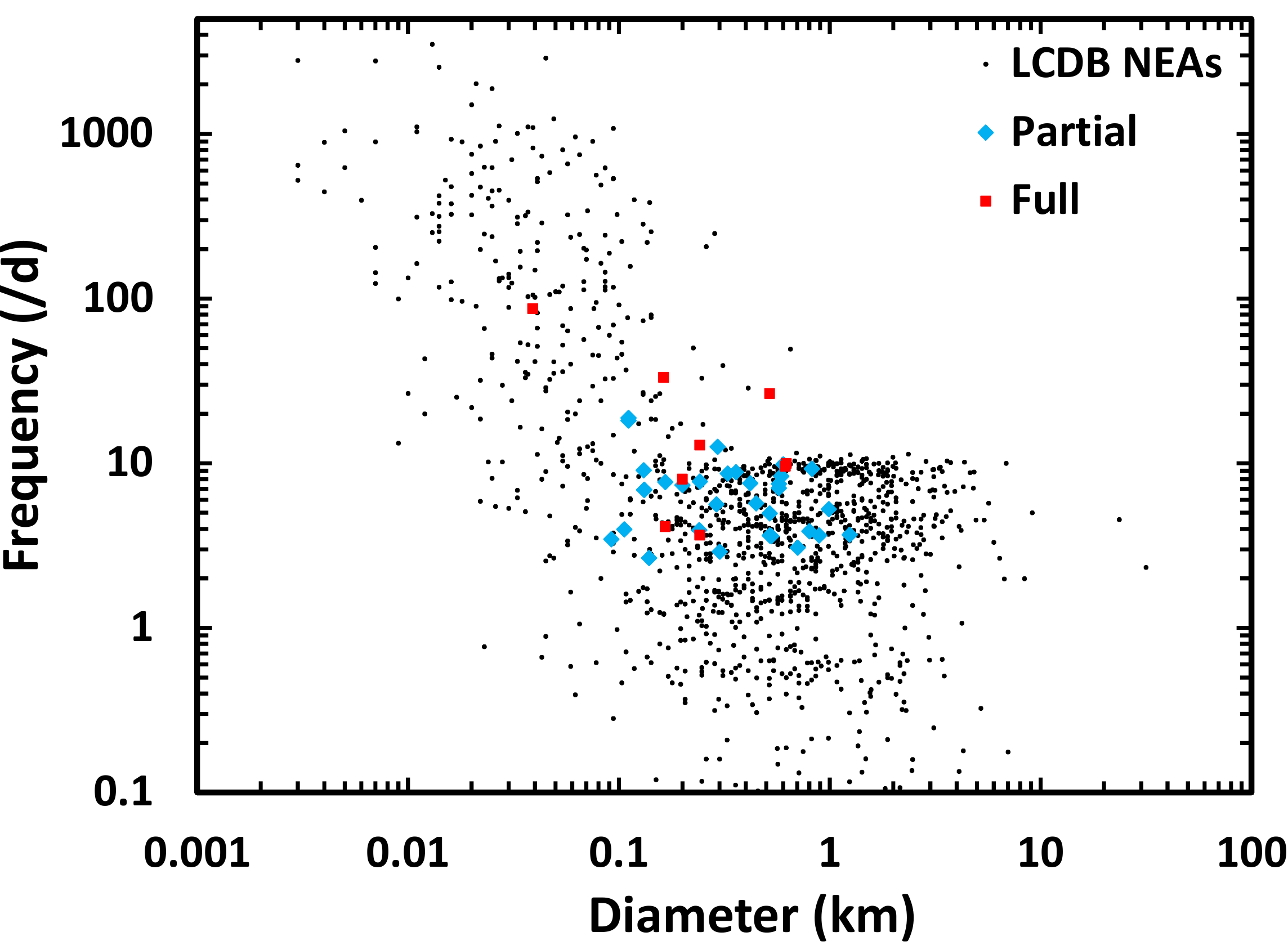}\label{fvsdiam}
\caption{A plot of the frequency versus diameter of the NEOs. The black dots are measurements of NEOs from the LCDB \citep{warner09}. The red points show the \Sp\ measurements for the cases where the lightcurve covered the full rotation period, and the blue dots show the extrapolated values for those where less than one rotational period was observed. The \Sp\ values fall in the range of previously observed NEOs. The red point at (0.517, 26.47) that is above the ``spin barrier'' line at a frequency of $\sim$ 10 d$^{-1}$ is 1990\ MF, which is discussed in Section \ref{cohesive}.  }
\end{figure}

We converted the rotational periods and estimates in Tables \ref{LSfits} -- \ref{sinefits} to spin frequencies and plotted of the frequency versus diameter of the new measurements compared to the NEOs listed in the lightcurve database 
\citep[LCDB;][updated 2018 March 7]{warner09} is shown in Figure \ref{fvsdiam}. The red points are for our measurements from Table \ref{LSfits} where the observations covered more than one rotational period, and the blue points are for the NEOs in Tables \ref{LSfits2} and \ref{sinefits} where we have derived lower limits. For the objects that had previous observations, we used those published rotational periods in this figure instead of the lower limits we derived. The \Sp\ measurements are within the same range as previous NEO spin frequencies and diameters. One point that is slightly discrepant is that of 1990~MF which lies at D=0.519~km, Freq=26.5~rev/day, above the ``spin barrier''  at $\sim$10~rev/day in this range of diameters. As seen in Figure \ref{LSplotsFull}, the amplitude of the \Sp\ lightcurve for this object is low compared to the noise, and possibly the full lightcurve was not sampled and the period is longer than that derived, which would move the point down in the diagram.

\subsection{Impact of Lightcurve Variations on the Thermal  Modeling}\label{NEATM}

We investigated the impact of the detected lightcurve variations during
our \Sp\ observations on thermal modeling results. The default
NEOSurvey thermal model \citep{2016AJ....152..172T} uses an adaption of the
Near-Earth Asteroid Thermal Model \citep[NEATM,][]{1998Icar..131..291H} to derive
diameter and geometric albedo estimates of the target in combination
with a Monte Carlo model to derive realistic uncertainties on these
parameters. The model uses \Sp-measured thermal flux densities and
combines them with optical data in the form of the target's absolute
magnitude to model the surface temperature distribution on a spherical
model asteroid. Target diameter and geometric albedo are found in a
least-squares fit of the modeled spectral energy distribution to the
observed one. The NEATM uses a variable ``beaming parameter'' $\eta$,
which accounts for surface roughness, thermal inertia, and other
effects in a zero-th order approximation. Being reliant on single-band
\Sp\ IRAC 4.5 \micron\ data, $\eta$ is drawn from a measured distribution of
such values \citep[see][for details]{2016AJ....152..172T}. 

In this analysis, we re-derived diameters and albedos for
the targets listed in Table \ref{LSfits} using a hypothetical nominal IRAC 4.5 \micron\
flux density for the target equal to the maximum and the minimum of
the measured lightcurves. This simulates the hypothetical case that
we observed the target exactly during the lightcurve maximum or
minimum and allows us to investigate the impact on the thermal
modeling results. We restricted ourselves to the targets listed in Table
1, which cover a wide range of lightcurve amplitudes. We compared the
thermal modeling diameters and geometric albedos with the previously
derived uncertainty ranges of both parameters\footnote{see
http://nearearthobjects.nau.edu/spitzerneos.html\label{spitzerneos}}.

Our analysis shows that both diameters and albedos derived
from the lightcurve maximum and minimum agree with the previously
derived and published\footref{spitzerneos} results at the 1$\sigma$-level in most cases, but on
the 3$\sigma$-level in all cases. Even for objects like 2015~XC and 2001~DF47, which have large
thermal flux density lightcurves, the simulated cases are well within
the reported uncertainties. This proves the conservative nature of the
thermal model uncertainties provided by the model described in
\citet{2016AJ....152..172T}. Note that this analysis does not account for
lightcurve effects in the optical counterpart - this effect will be
studied by future work (Gustafsson et al., in preparation).

\subsection{Notes on Individual NEOs}\label{objectdiscuss}

\textbf{1990 UA:}
This is an object with no previously reported rotation period. The \textit{Spitzer} 4.5 \micron\ lightcurve shows a distribution with three peaks over 315 minutes of observation. The Lomb-Scargle algorithm reports a period of 93.680 minutes, which is a solution that judges all of the peaks symmetrical, which does not appear likely to be the case. The middle peak in the curve is noticeably narrower than the other two, which seem of similar width and height. Applying the Plavchan algorithm \citep{2008ApJS..175..191P} we found that the highest peak at a period smaller than the lightcurve length gives a rotational period of 184.534 minutes. The phase plot for this solution is the one shown in Figure \ref{LSplotsFull}. Further observations of 1990~UA are necessary to unambiguously determine its rotation period.

\textbf{2005 LG8:}
Lightcurve data for 2005 LG8 were obtained by \citet{waszczak15}, who determined a period of 
4.630$\pm$0.0019~hr with an amplitude of 0.62~mag, although there is a note in the JPL Small Bodies Database that states ``Result based on less than full coverage, so that the period may be wrong by 30 percent or so''.  Our derived period in Table \ref{sinefits} (3.62~hr) is near the full \Sp\ sampling time, and there is some indication that the peaks of the lightcurves are not adequately sampled (see Figure \ref{sineplots}), so it appears that the extrapolated period from the \Sp\ data alone underestimates the period. We have performed a sine fit to the \Sp\ data, constraining the fit to a 4.63~hr period, also shown in Figure \ref{sineplots}. The $\chi^2$ value for this fit is $\sim$10\% higher than the unconstrained fit, but the data appear consistent with the curve for the longer period as reported by \citet{waszczak15}.

\textbf{2008 JM20:}
The length of time sampled for this object was just slightly longer than one full period, and this seemed to give the LS fitting some issues, with the best fit being about 120 minutes, greater than the length of time sampled. We again used the Plavchan algorithm for this object, which gave a rotational period of 208.8$\pm$18.4 minutes, consistent with the period determined from the sine fitting.

\textbf{2008 UE7:}
\citet{ye09} reported a lightcurve period of 3.25146$\pm$0.00001~hr (195.0876 minutes) based on optical photometry obtained in 2008 December. The amplitude was $\sim$0.2 mag, similar to the \Sp\ value. The optical lightcurve is double-peaked, so it appears that the \Sp\ dataset covered less than half of the period. Therefore, the \Sp-derived lower limit of 174.4 minutes is less than that derived from the optical data.

\textbf{2011 SD173:}
The \Sp\ lightcurve for 2011~SD173 has a high amplitude and  a non-symmetric double-peaked shape with a cusp-like feature at the minimum, indicating an irregular shape. 

\textbf{2011 WS2:}
Our thermal modeling based on the \Sp\ observations (see Section \ref{NEATM}) gives a diameter of 1.24$\substack{+0.75\\-0.34}$~km and an albedo of 0.104$\substack{+0.099\\-0.063}$ for this NEO. This is in agreement with previous results from {\it WISE} observations, which gave a diameter of 1.434$\pm$0.056~km \citep{mainzer11}.

\textbf{2011 XA3:}
\citet{2014AJ....147..121U} previously measured the period of this object to be 43.8$\pm$ 0.4 minutes. This compares well to the value we derived of 45.2$\pm$5.0. The \Sp\ period was not the highest peak reported by the LS algorithm, the highest periodogram peak was at 21.7$\pm$0.3 minutes, or roughly half of the value in the table. That period solution is for the case of each peak in the lightcurve being at the same phase. However, the appearance of the lightcurve indicated that there are alternating peaks of different magnitudes, therefore this solution was chosen as being more likely. This was done before comparing to previous measurements.

\citet{2014AJ....147..121U} determined that the diameter of 2011~XA3 is 255$\pm$97~m if it is S-complex, and 166$\pm$63~m if it is V-type, based on the albedo assumptions for S-complex and V-type of \citet{2012Icar..221..365P} and \citet{2013ApJ...762...56U}. Our estimate of the diameter of 2011~XA3 based on NEATM modeling using this  \textit{Spitzer} 4.5 \micron\ observation is $163\substack{+56 \\ -29}$~m, implying the object is more likely to be V-type than S-complex.

\textbf{2015 XC:}
This NEO was observed on 2015 Dec 02 by \citet{carbognani16}  who reported a period of 0.2767$\pm$0.0001 hr (16.602$\pm$0.006 minutes) and an amplitude of 0.39 mag in the $R$ band. Further observations and an analysis by Pravec  revealed that this NEO is likely a tumbler with a complex shape \citep{warner16}.  They  found periods of P1 = 0.181099 hr and P2 = 0.27998 hr, the second period being roughly consistent with the value of 16.25$\pm$0.05 minutes (0.2708$\pm$0.001 hr)  that we report here. We find the amplitude at 4.5 \micron\ of 1.566$\pm$0.066 mag is higher than seen in the optical measurements, where the maximum is $\sim$0.64 mag. Our \Sp\ lightcurve shows nonsinusoidal structure and amplitude  that varies by a factor of 4 in flux, confirming the earlier indications that this object has a complex shape.

\section{Cohesive Strength}\label{cohesive}
The minimum cohesive strength has been determined for only a small sample of NEOs to date. The elongation and rotation period of most objects is such that a minimum cohesive strength of 0 Pa is required. As a lower limit, this is not informative. For these objects, light curves alone are not enough to determine whether they are strengthless rubble piles or have some significant internal strength. Instead we look at objects with high amplitudes or very short rotation periods which require non-zero minimum strengths. The minimum cohesive strength has been studied for fast rotating objects \citep{2017Icar..297..126P} and highly elongated objects \citep{McNeill18} but the overall sample size remains small. A survey like the work presented in this paper serves to increase this population as we will incidentally identify high amplitude and fast rotating objects without the need for a targeted study.

Of the observed objects we find two with $D > 200$ m and 4.5 \micron\ lightcurves showing rotation periods shorter than the spin barrier at $P = 2.2$ h. If these bodies are rubble piles they should undergo rotational fission at their current spin rate. Instead we must assume that they have some internal cohesive strength or are monolithic in nature.
Therefore we  calculate the cohesive strength required using a simplified Drucker-Prager model \citep{2004Icar..172..272H}. 

The Drucker-Prager failure criterion models the three-dimensional stresses within a geological material at the point of critical rotation. The three orthogonal shear stresses on a body in the $xyz$ axes are dependent on the shape, density and rotational properties of the body \citep{Holsapple07b}: 

\begin{equation} 
\sigma_{x}=(\rho\omega^{2}-2\pi\rho^{2}GA_{x})\frac{a^{2}}{5}
\label{eqn:sigmax}
\end{equation}

\begin{equation} 
\sigma_{y}=(\rho\omega^{2}-2\pi\rho^{2}GA_{y})\frac{b^{2}}{5}
\label{eqn:sigmay}
\end{equation}

\begin{equation} 
\sigma_{z}=(-2\pi\rho^{2}GA_{z})\frac{c^{2}}{5} .
\label{eqn:sigmaz}
\end{equation}
where $\rho$ is the bulk density of the asteroid, $\omega$ is its rotational frequency, $G$ is the gravitational constant, and $a, b, c$ are the lengths of the semi-axes of the ellipsoidal body, in order from largest to smallest. These three $A_{i}$ functions are dimensionless parameters dependent on the axis ratios of the body:

\begin{equation} 
A_{x}=\frac{c}{a}\frac{b}{a}\int_{0}^{\infty}\frac{1}{(u+1)^{3/2}(u+\frac{b}{a}^{2})^{1/2}(u+\frac{c}{a}^{2})^{1/2}}du
\label{eqn:Ax}
\end{equation}

\begin{equation} 
A_{y}=\frac{c}{a}\frac{b}{a}\int_{0}^{\infty}\frac{1}{(u+1)^{1/2}(u+\frac{b}{a}^{2})^{3/2}(u+\frac{c}{a}^{2})^{1/2}}du
\label{eqn:Ay}
\end{equation}

\begin{equation} 
A_{z}=\frac{c}{a}\frac{b}{a}\int_{0}^{\infty}\frac{1}{(u+1)^{1/2}(u+\frac{b}{a}^{2})^{1/2}(u+\frac{c}{a}^{2})^{3/2}}du .
\label{eqn:Az}
\end{equation}

The Drucker-Prager failure criterion is the point at which the object will rotationally fission and is given by

\begin{equation} 
\frac{1}{6}[(\sigma_{x}-\sigma_{y})^{2}+(\sigma_{y}-\sigma_{z})^{2}+(\sigma_{z}-\sigma_{x})^{2}] \leq [k-s(\sigma_{x}+\sigma_{y}+\sigma_{z})]^{2}
\label{eqn:drucker}
\end{equation}

\noindent where $k$ represents the cohesive strength within the body and $s$ is a slope parameter dependent on the angle of friction, $\phi$:

\begin{equation} 
s=\frac{2\rm{sin}\phi}{\sqrt{3}(3-\rm{sin}\phi)} .
\label{eqn:slope}
\end{equation}

For these calculations we consider the value $\phi = 35^{\circ}$ corresponding to the average angle of friction from geological materials \citep{2015ApJ...798L...8H}. 

1990 MF was measured to have a full rotation period of $54.4\pm 5.4$ minutes with a lightcurve amplitude $A=0.069\pm 0.012$ mag. Scattering effects and increased shadowing at high phase angles will result in lightcurve minima appearing fainter. This causes the apparent lightcurve amplitude to be increased leading to a potential overestimation of the amplitude. We correct for this using the method of \citet{zappala90} using their derived correction coefficient for S-type asteroids, 0.03 mag deg$^{-1}$. This results in a corrected amplitude for this lightcurve of $A = 0.025 \pm 0.004$ mag. Using these parameters, its \Sp-derived diameter $D = 519^{+227}_{-116}$ m and assuming a typical S-type asteroid bulk density of $\rho = 2500$ kg m$^{-3}$ we find that a cohesive strength of 225$^{+225}_{-72}$ Pa is required for this object to resist rotational fission. This is a higher value than has been calculated for most rubble-pile asteroids and is a comparable value to the relatively large cohesion required by 2000 GD65 as calculated by \cite{2016Icar..267..243P}. Unlike the case of 2001 OE84, the cohesive strength is not so large (of order $10^3$ Pa) to be explicable only in terms of a monolithic structure \citep{2017Icar..297..126P}.

1991~BN was determined to have a rotation period $P = 114.6 \pm 8.4$ minutes, just below the spin barrier. The corrected lightcurve amplitude of this object was $A = 0.132 \pm 0.020$ mag, which results in an estimate for the required cohesive strength of $7^{+5}_{-4}$ Pa. 

These two objects were found in our relatively small sample of \Sp\ NEOs analyzed to date. The remainder of the dataset may yield many more objects where we can put limits on the cohesive strength and learn more about the internal strengths of asteroids. 

\section{Conclusions}
We have presented a sample of 38 NEO lightcurves obtained from data taken as part of the ExploreNEOs, NEO Survey, and NEO Legacy \Sp\ programs. We derived periods and amplitudes based on Lomb-Scargle or Plavchan fits for 10 objects where we appear to have complete sampling of the periods, and also present lower limits for another 28 objects based on sine fits to lightcurves shorter than or about equal to one period. Six lightcurves were fit with both periodogram and sine fits and found to have consistent periods. Enabled by the sensitivity and stability of \Sp/IRAC, the NEO surveys have observed thousands of objects where lightcurves can be extracted and periods and amplitudes can be determined or constrained by the data. Because of \Sp's current position in its orbit, it can observe NEOs that are not currently accessible by earth-based observatories. With the 4.5 \micron\ data, we can also estimate the diameter and measure albedos of the NEOs using the same observations. By analyzing the full database as we have done for this small sample, we will be able to extract lightcurves for hundreds of NEOs and determine or set limits on their periods and amplitudes.

\acknowledgements
This work is based on observations made with the \Sp\ Space Telescope, which is operated by the Jet Propulsion Laboratory, California Institute of Technology under a contract with NASA. Support for this work was provided by NASA through an award issued by JPL/Caltech.
This work is supported in part by NSF award 1229776.
IRAF is distributed by the National Optical Astronomy Observatory, which is operated by the Association of Universities for Research in Astronomy (AURA) under a cooperative agreement with the National Science Foundation.

\software{IRAF, mopex \citep{makovoz06}, IRACproc \citep{schuster06}}

\facilities{\Sp/IRAC}
\tighten

\bibliographystyle{aasjournal}
\bibliography{LCbiblio}{}

\begin{thebibliography}{}
\expandafter\ifx\csname natexlab\endcsname\relax\def\natexlab#1{#1}\fi

\bibitem[{{Bus}(1999)}]{bus99}
{Bus}, S.~J. 1999, PhD thesis, MASSACHUSETTS INSTITUTE OF TECHNOLOGY

\bibitem[{{Bus} \& {Binzel}(2002{\natexlab{a}})}]{bus02a}
{Bus}, S.~J., \& {Binzel}, R.~P. 2002{\natexlab{a}}, \icarus, 158, 146

\bibitem[{{Bus} \& {Binzel}(2002{\natexlab{b}})}]{bus02b}
---. 2002{\natexlab{b}}, \icarus, 158, 106

\bibitem[{{Carbognani} \& {Buzzi}(2016)}]{carbognani16}
{Carbognani}, A., \& {Buzzi}, L. 2016, Minor Planet Bulletin, 43, 160

\bibitem[{{DeMeo} {et~al.}(2009){DeMeo}, {Binzel}, {Slivan}, \&
  {Bus}}]{demeo09}
{DeMeo}, F.~E., {Binzel}, R.~P., {Slivan}, S.~M., \& {Bus}, S.~J. 2009,
  \icarus, 202, 160

\bibitem[{{Fazio} {et~al.}(2004){Fazio}, {Hora}, {Allen}, {Ashby}, {Barmby},
  {Deutsch}, {Huang}, {Kleiner}, {Marengo}, {Megeath}, {Melnick}, {Pahre},
  {Patten}, {Polizotti}, {Smith}, {Taylor}, {Wang}, {Willner}, {Hoffmann},
  {Pipher}, {Forrest}, {McMurty}, {McCreight}, {McKelvey}, {McMurray}, {Koch},
  {Moseley}, {Arendt}, {Mentzell}, {Marx}, {Losch}, {Mayman}, {Eichhorn},
  {Krebs}, {Jhabvala}, {Gezari}, {Fixsen}, {Flores}, {Shakoorzadeh}, {Jungo},
  {Hakun}, {Workman}, {Karpati}, {Kichak}, {Whitley}, {Mann}, {Tollestrup},
  {Eisenhardt}, {Stern}, {Gorjian}, {Bhattacharya}, {Carey}, {Nelson},
  {Glaccum}, {Lacy}, {Lowrance}, {Laine}, {Reach}, {Stauffer}, {Surace},
  {Wilson}, {Wright}, {Hoffman}, {Domingo}, \& {Cohen}}]{2004ApJS..154...10F}
{Fazio}, G.~G., {Hora}, J.~L., {Allen}, L.~E., {et~al.} 2004, \apjs, 154, 10

\bibitem[{{Giorgini} {et~al.}(1996){Giorgini}, {Yeomans}, {Chamberlin},
  {Chodas}, {Jacobson}, {Keesey}, {Lieske}, {Ostro}, {Standish}, \&
  {Wimberly}}]{giorgini96}
{Giorgini}, J.~D., {Yeomans}, D.~K., {Chamberlin}, A.~B., {et~al.} 1996, in
  Bulletin of the American Astronomical Society, Vol.~28, AAS/Division for
  Planetary Sciences Meeting Abstracts \#28, 1158

\bibitem[{{Harris}(1998)}]{1998Icar..131..291H}
{Harris}, A.~W. 1998, \icarus, 131, 291

\bibitem[{{Harris} {et~al.}(2011{\natexlab{a}}){Harris}, {Mommert}, {Hora},
  {Mueller}, {Trilling}, {Bhattacharya}, {Bottke}, {Chesley}, {Delbo}, {Emery},
  {Fazio}, {Mainzer}, {Penprase}, {Smith}, {Spahr}, {Stansberry}, \&
  {Thomas}}]{2011AJ....141...75H}
{Harris}, A.~W., {Mommert}, M., {Hora}, J.~L., {et~al.} 2011{\natexlab{a}},
  \aj, 141, 75

\bibitem[{{Harris} {et~al.}(2011{\natexlab{b}}){Harris}, {Mommert}, {Hora},
  {Mueller}, {Trilling}, {Bhattacharya}, {Bottke}, {Chesley}, {Delbo}, {Emery},
  {Fazio}, {Mainzer}, {Penprase}, {Smith}, {Spahr}, {Stansberry}, \&
  {Thomas}}]{Harris2011}
---. 2011{\natexlab{b}}, \aj, 141, 75

\bibitem[{{Hirabayashi} \& {Scheeres}(2015)}]{2015ApJ...798L...8H}
{Hirabayashi}, M., \& {Scheeres}, D.~J. 2015, \apjl, 798, L8

\bibitem[{{Holsapple}(2004)}]{2004Icar..172..272H}
{Holsapple}, K.~A. 2004, \icarus, 172, 272

\bibitem[{Holsapple(2007)}]{Holsapple07b}
Holsapple, K.~A. 2007, Icarus, 187, 500

\bibitem[{{Hora} {et~al.}(2008){Hora}, {Carey}, {Surace}, {Marengo},
  {Lowrance}, {Glaccum}, {Lacy}, {Reach}, {Hoffmann}, {Barmby}, {Willner},
  {Fazio}, {Megeath}, {Allen}, {Bhattacharya}, \& {Quijada}}]{hora08}
{Hora}, J.~L., {Carey}, S., {Surace}, J., {et~al.} 2008, \pasp, 120, 1233

\bibitem[{{Ivezi{\'c}} {et~al.}(2002){Ivezi{\'c}}, {Lupton}, {Juri{\'c}},
  {Tabachnik}, {Quinn}, {Gunn}, {Knapp}, {Rockosi}, \& {Brinkmann}}]{Juric2002}
{Ivezi{\'c}}, {\v Z}., {Lupton}, R.~H., {Juri{\'c}}, M., {et~al.} 2002, \aj,
  124, 2943

\bibitem[{{Leonard} {et~al.}(2017){Leonard}, {Christensen}, {Fuls}, {Gibbs},
  {Grauer}, {Johnson}, {Kowalski}, {Larson}, {Matheny}, {Seaman}, \&
  {Shelly}}]{leonard17}
{Leonard}, G.~J., {Christensen}, E.~J., {Fuls}, C., {et~al.} 2017, in American
  Astronomical Society, DPS meeting \#49, id.117.07, Vol.~49

\bibitem[{{Lomb}(1976)}]{1976Ap&SS..39..447L}
{Lomb}, N.~R. 1976, \apss, 39, 447

\bibitem[{{Mainzer} {et~al.}(2011{\natexlab{a}}){Mainzer}, {Grav}, {Bauer},
  {Masiero}, {McMillan}, {Cutri}, {Walker}, {Wright}, {Eisenhardt}, {Tholen},
  {Spahr}, {Jedicke}, {Denneau}, {DeBaun}, {Elsbury}, {Gautier}, {Gomillion},
  {Hand}, {Mo}, {Watkins}, {Wilkins}, {Bryngelson}, {Del Pino Molina}, {Desai},
  {G{\'o}mez Camus}, {Hidalgo}, {Konstantopoulos}, {Larsen}, {Maleszewski},
  {Malkan}, {Mauduit}, {Mullan}, {Olszewski}, {Pforr}, {Saro}, {Scotti}, \&
  {Wasserman}}]{mainzer11}
{Mainzer}, A., {Grav}, T., {Bauer}, J., {et~al.} 2011{\natexlab{a}}, \apj, 743,
  doi:10.1088/0004-637X/743/2/156

\bibitem[{{Mainzer} {et~al.}(2011{\natexlab{b}}){Mainzer}, {Grav}, {Masiero},
  {Hand}, {Bauer}, {Tholen}, {McMillan}, {Spahr}, {Cutri}, {Wright}, {Watkins},
  {Mo}, \& {Maleszewski}}]{Mainzer2011a}
{Mainzer}, A., {Grav}, T., {Masiero}, J., {et~al.} 2011{\natexlab{b}}, \apj,
  741, 90

\bibitem[{{Mainzer} {et~al.}(2014){Mainzer}, {Bauer}, {Cutri}, {Grav},
  {Masiero}, {Beck}, {Clarkson}, {Conrow}, {Dailey}, {Eisenhardt}, {Fabinsky},
  {Fajardo-Acosta}, {Fowler}, {Gelino}, {Grillmair}, {Heinrichsen}, {Kendall},
  {Kirkpatrick}, {Liu}, {Masci}, {McCallon}, {Nugent}, {Papin}, {Rice},
  {Royer}, {Ryan}, {Sevilla}, {Sonnett}, {Stevenson}, {Thompson}, {Wheelock},
  {Wiemer}, {Wittman}, {Wright}, \& {Yan}}]{mainzer14}
{Mainzer}, A., {Bauer}, J., {Cutri}, R.~M., {et~al.} 2014, \apj, 792, 30

\bibitem[{{Makovoz} {et~al.}(2006){Makovoz}, {Roby}, {Khan}, \&
  {Booth}}]{makovoz06}
{Makovoz}, D., {Roby}, T., {Khan}, I., \& {Booth}, H. 2006, in \procspie, Vol.
  6274, Society of Photo-Optical Instrumentation Engineers (SPIE) Conference
  Series, 62740C

\bibitem[{{Masiero} {et~al.}(2017){Masiero}, {Nugent}, {Mainzer}, {Wright},
  {Bauer}, {Cutri}, {Grav}, {Kramer}, \& {Sonnett}}]{masiero17}
{Masiero}, J.~R., {Nugent}, C., {Mainzer}, A.~K., {et~al.} 2017, \aj, 154, 168

\bibitem[{{McNeill} {et~al.}(2018){McNeill}, {Trilling}, \&
  {Mommert}}]{McNeill18}
{McNeill}, A., {Trilling}, D.~E., \& {Mommert}, M. 2018, \apjl, 857, L1

\bibitem[{{Mommert} {et~al.}(2014{\natexlab{a}}){Mommert}, {Hora},
  {Farnocchia}, {Chesley}, {Vokrouhlick{\'y}}, {Trilling}, {Mueller}, {Harris},
  {Smith}, \& {Fazio}}]{2014ApJ...786..148M}
{Mommert}, M., {Hora}, J.~L., {Farnocchia}, D., {et~al.} 2014{\natexlab{a}},
  \apj, 786, 148

\bibitem[{{Mommert} {et~al.}(2014{\natexlab{b}}){Mommert}, {Farnocchia},
  {Hora}, {Chesley}, {Trilling}, {Chodas}, {Mueller}, {Harris}, {Smith}, \&
  {Fazio}}]{2014ApJ...789L..22M}
{Mommert}, M., {Farnocchia}, D., {Hora}, J.~L., {et~al.} 2014{\natexlab{b}},
  \apjl, 789, L22

\bibitem[{{Mommert} {et~al.}(2014{\natexlab{c}}){Mommert}, {Hora}, {Harris},
  {Reach}, {Emery}, {Thomas}, {Mueller}, {Cruikshank}, {Trilling}, {Delbo}, \&
  {Smith}}]{2014ApJ...781...25M}
{Mommert}, M., {Hora}, J.~L., {Harris}, A.~W., {et~al.} 2014{\natexlab{c}},
  \apj, 781, 25

\bibitem[{{Mommert} {et~al.}(2015){Mommert}, {Harris}, {Mueller}, {Hora},
  {Trilling}, {Bottke}, {Thomas}, {Delbo}, {Emery}, {Fazio}, \&
  {Smith}}]{2015AJ....150..106M}
{Mommert}, M., {Harris}, A.~W., {Mueller}, M., {et~al.} 2015, \aj, 150, 106

\bibitem[{{Mueller} {et~al.}(2011){Mueller}, {Delbo'}, {Hora}, {Trilling},
  {Bhattacharya}, {Bottke}, {Chesley}, {Emery}, {Fazio}, {Harris}, {Mainzer},
  {Mommert}, {Penprase}, {Smith}, {Spahr}, {Stansberry}, \&
  {Thomas}}]{2011AJ....141..109M}
{Mueller}, M., {Delbo'}, M., {Hora}, J.~L., {et~al.} 2011, \aj, 141, 109

\bibitem[{{M{\"u}ller} {et~al.}(2017){M{\"u}ller}, {{\v D}urech}, {Ishiguro},
  {Mueller}, {Kr{\"u}hler}, {Yang}, {Kim}, {O'Rourke}, {Usui}, {Kiss},
  {Altieri}, {Carry}, {Choi}, {Delbo}, {Emery}, {Greiner}, {Hasegawa}, {Hora},
  {Knust}, {Kuroda}, {Osip}, {Rau}, {Rivkin}, {Schady}, {Thomas-Osip},
  {Trilling}, {Urakawa}, {Vilenius}, {Weissman}, \&
  {Zeidler}}]{2017A&A...599A.103M}
{M{\"u}ller}, T.~G., {{\v D}urech}, J., {Ishiguro}, M., {et~al.} 2017, \aap,
  599, A103

\bibitem[{{Perna} {et~al.}(2018){Perna}, {Barucci}, {Fulchignoni}, {Popescu},
  {Belskaya}, {Fornasier}, {Doressoundiram}, {Lantz}, \& {Merlin}}]{perna18}
{Perna}, D., {Barucci}, M.~A., {Fulchignoni}, M., {et~al.} 2018, \planss, 157,
  82

\bibitem[{{Plavchan} {et~al.}(2008){Plavchan}, {Jura}, {Kirkpatrick}, {Cutri},
  \& {Gallagher}}]{2008ApJS..175..191P}
{Plavchan}, P., {Jura}, M., {Kirkpatrick}, J.~D., {Cutri}, R.~M., \&
  {Gallagher}, S.~C. 2008, \apjs, 175, 191

\bibitem[{{Polishook} {et~al.}(2017){Polishook}, {Moskovitz}, {Thirouin},
  {Bosh}, {Levine}, {Zuluaga}, {Tegler}, \& {Aharonson}}]{2017Icar..297..126P}
{Polishook}, D., {Moskovitz}, N., {Thirouin}, A., {et~al.} 2017, \icarus, 297,
  126

\bibitem[{{Polishook} {et~al.}(2016){Polishook}, {Moskovitz}, {Binzel}, {Burt},
  {DeMeo}, {Hinkle}, {Lockhart}, {Mommert}, {Person}, {Thirouin}, {Thomas},
  {Trilling}, {Willman}, \& {Aharonson}}]{2016Icar..267..243P}
{Polishook}, D., {Moskovitz}, N., {Binzel}, R.~P., {et~al.} 2016, \icarus, 267,
  243

\bibitem[{{Pravec} {et~al.}(2012){Pravec}, {Harris}, {Ku{\v s}nir{\'a}k},
  {Gal{\'a}d}, \& {Hornoch}}]{2012Icar..221..365P}
{Pravec}, P., {Harris}, A.~W., {Ku{\v s}nir{\'a}k}, P., {Gal{\'a}d}, A., \&
  {Hornoch}, K. 2012, \icarus, 221, 365

\bibitem[{{Romanishin} \& {Tegler}(2005)}]{Romanishin2005}
{Romanishin}, W., \& {Tegler}, S.~C. 2005, \icarus, 179, 523

\bibitem[{{Scargle}(1982)}]{1982ApJ...263..835S}
{Scargle}, J.~D. 1982, \apj, 263, 835

\bibitem[{{Schuster} {et~al.}(2006){Schuster}, {Marengo}, \&
  {Patten}}]{schuster06}
{Schuster}, M.~T., {Marengo}, M., \& {Patten}, B.~M. 2006, in \procspie, Vol.
  6270, Society of Photo-Optical Instrumentation Engineers (SPIE) Conference
  Series, 627020

\bibitem[{{Sonnett} {et~al.}(2015){Sonnett}, {Mainzer}, {Grav}, {Masiero}, \&
  {Bauer}}]{sonnett15}
{Sonnett}, S., {Mainzer}, A., {Grav}, T., {Masiero}, J., \& {Bauer}, J. 2015,
  \apj, 799, 191

\bibitem[{{Thomas} {et~al.}(2014){Thomas}, {Emery}, {Trilling}, {Delb{\'o}},
  {Hora}, \& {Mueller}}]{2014Icar..228..217T}
{Thomas}, C.~A., {Emery}, J.~P., {Trilling}, D.~E., {et~al.} 2014, \icarus,
  228, 217

\bibitem[{{Thomas} {et~al.}(2011){Thomas}, {Rivkin}, {Trilling}, {Enga}, \&
  {Grier}}]{2011Icar..212..158T}
{Thomas}, C.~A., {Rivkin}, A.~S., {Trilling}, D.~E., {Enga}, M.-t., \& {Grier},
  J.~A. 2011, \icarus, 212, 158

\bibitem[{{Trilling} {et~al.}(2008){Trilling}, {Mueller}, {Hora}, {Fazio},
  {Spahr}, {Stansberry}, {Smith}, {Chesley}, \&
  {Mainzer}}]{2008ApJ...683L.199T}
{Trilling}, D.~E., {Mueller}, M., {Hora}, J.~L., {et~al.} 2008, \apjl, 683,
  L199

\bibitem[{{Trilling} {et~al.}(2010){Trilling}, {Mueller}, {Hora}, {Harris},
  {Bhattacharya}, {Bottke}, {Chesley}, {Delbo}, {Emery}, {Fazio}, {Mainzer},
  {Penprase}, {Smith}, {Spahr}, {Stansberry}, \&
  {Thomas}}]{2010AJ....140..770T}
---. 2010, \aj, 140, 770

\bibitem[{{Trilling} {et~al.}(2016){Trilling}, {Mommert}, {Hora}, {Chesley},
  {Emery}, {Fazio}, {Harris}, {Mueller}, \& {Smith}}]{2016AJ....152..172T}
{Trilling}, D.~E., {Mommert}, M., {Hora}, J., {et~al.} 2016, \aj, 152, 172

\bibitem[{{Trilling} {et~al.}(2017){Trilling}, {Hora}, {Mommert}, {Chesley},
  {Emery}, {Fazio}, {Harris}, {Mueller}, \& {Smith}}]{Trilling2017}
{Trilling}, D.~E., {Hora}, J.~L., {Mommert}, M., {et~al.} 2017, in AAS/Division
  for Planetary Sciences Meeting Abstracts, Vol.~49, AAS/Division for Planetary
  Sciences Meeting Abstracts \#49, 110.06

\bibitem[{{Urakawa} {et~al.}(2014){Urakawa}, {Ohtsuka}, {Abe}, {Ito}, \&
  {Nakamura}}]{2014AJ....147..121U}
{Urakawa}, S., {Ohtsuka}, K., {Abe}, S., {Ito}, T., \& {Nakamura}, T. 2014,
  \aj, 147, 121

\bibitem[{{Usui} {et~al.}(2013){Usui}, {Kasuga}, {Hasegawa}, {Ishiguro},
  {Kuroda}, {M{\"u}ller}, {Ootsubo}, \& {Matsuhara}}]{2013ApJ...762...56U}
{Usui}, F., {Kasuga}, T., {Hasegawa}, S., {et~al.} 2013, \apj, 762, 56

\bibitem[{{Vere{\v s}} \& {Chesley}(2017)}]{veres17}
{Vere{\v s}}, P., \& {Chesley}, S.~R. 2017, \aj, 154, 12

\bibitem[{{Vere{\v{s}}} {et~al.}(2015){Vere{\v{s}}}, {Jedicke}, {Fitzsimmons},
  {Denneau}, {Granvik}, {Bolin}, {Chastel}, {Wainscoat}, {Burgett}, {Chambers},
  {Flewelling}, {Kaiser}, {Magnier}, {Morgan}, {Price}, {Tonry}, \&
  {Waters}}]{veres15}
{Vere{\v{s}}}, P., {Jedicke}, R., {Fitzsimmons}, A., {et~al.} 2015, \icarus,
  261, 34

\bibitem[{{Warner}(2016)}]{warner16}
{Warner}, B.~D. 2016, Minor Planet Bulletin, 43, 143

\bibitem[{{Warner} {et~al.}(2009){Warner}, {Harris}, \& {Pravec}}]{warner09}
{Warner}, B.~D., {Harris}, A.~W., \& {Pravec}, P. 2009, \icarus, 202, 134

\bibitem[{{Waszczak} {et~al.}(2015){Waszczak}, {Chang}, {Ofek}, {Laher},
  {Masci}, {Levitan}, {Surace}, {Cheng}, {Ip}, {Kinoshita}, {Helou}, {Prince},
  \& {Kulkarni}}]{waszczak15}
{Waszczak}, A., {Chang}, C.-K., {Ofek}, E.~O., {et~al.} 2015, \aj, 150, 75

\bibitem[{{Werner} {et~al.}(2004){Werner}, {Roellig}, {Low}, {Rieke}, {Rieke},
  {Hoffmann}, {Young}, {Houck}, {Brandl}, {Fazio}, {Hora}, {Gehrz}, {Helou},
  {Soifer}, {Stauffer}, {Keene}, {Eisenhardt}, {Gallagher}, {Gautier}, {Irace},
  {Lawrence}, {Simmons}, {Van Cleve}, {Jura}, {Wright}, \&
  {Cruikshank}}]{2004ApJS..154....1W}
{Werner}, M.~W., {Roellig}, T.~L., {Low}, F.~J., {et~al.} 2004, \apjs, 154, 1

\bibitem[{{Wolters} {et~al.}(2008){Wolters}, {Green}, {McBride}, \&
  {Davies}}]{Wolters2008}
{Wolters}, S.~D., {Green}, S.~F., {McBride}, N., \& {Davies}, J.~K. 2008,
  \icarus, 193, 535

\bibitem[{{Wright} {et~al.}(2010){Wright}, {Eisenhardt}, {Mainzer}, {Ressler},
  {Cutri}, {Jarrett}, {Kirkpatrick}, {Padgett}, {McMillan}, {Skrutskie},
  {Stanford}, {Cohen}, {Walker}, {Mather}, {Leisawitz}, {Gautier}, {McLean},
  {Benford}, {Lonsdale}, {Blain}, {Mendez}, {Irace}, {Duval}, {Liu}, {Royer},
  {Heinrichsen}, {Howard}, {Shannon}, {Kendall}, {Walsh}, {Larsen}, {Cardon},
  {Schick}, {Schwalm}, {Abid}, {Fabinsky}, {Naes}, \& {Tsai}}]{wright10}
{Wright}, E.~L., {Eisenhardt}, P.~R.~M., {Mainzer}, A.~K., {et~al.} 2010, \aj,
  140, 1868

\bibitem[{{Ye} {et~al.}(2009){Ye}, {Shi}, {Xu}, {Lin}, \& {Zhang}}]{ye09}
{Ye}, Q., {Shi}, L., {Xu}, W., {Lin}, H.-C., \& {Zhang}, J. 2009, Minor Planet
  Bulletin, 36, 180

\bibitem[{{Zappala} {et~al.}(1990){Zappala}, {Cellino}, {Barucci},
  {Fulchignoni}, \& {Lupishko}}]{zappala90}
{Zappala}, V., {Cellino}, A., {Barucci}, A.~M., {Fulchignoni}, M., \&
  {Lupishko}, D.~F. 1990, \aap, 231, 548

\end{thebibliography}
\end{document}